\documentclass[journal]{IEEEtran}

\makeatletter
\def\ps@headings{%
\def\@oddhead{\mbox{}\scriptsize\rightmark \hfil \thepage}%
\def\@evenhead{\scriptsize\thepage \hfil \leftmark\mbox{}}%
\def\@oddfoot{}%
\def\@evenfoot{}}
\makeatother 
\usepackage{mathrsfs}
\usepackage{amsfonts}
\usepackage{array} 
\usepackage{amssymb}
\usepackage{amsmath}
\usepackage{graphicx}
\usepackage{multirow} 
\usepackage{amsthm}

\usepackage[linesnumbered,boxed]{algorithm2e}
\usepackage{epstopdf}
\usepackage{stmaryrd}
\usepackage{amssymb,amsmath}
\usepackage{multirow,url}
\usepackage{color,cite}
\usepackage{slashbox}

\newtheorem{theorem}{Theorem}
\newtheorem{definition}{Definition}
\newtheorem{proposition}{Proposition}
\newtheorem{lemma}{Lemma}

\newtheorem{property}{Property}

\newtheorem{assumption}{Assumption}

\theoremstyle{plain}

\ifodd 127212
\newcommand{\revjrr}[1]{{#1}} %JSAC revision
\newcommand{\revjr}[1]{{\color{blue}#1}} %JSAC revision
\newcommand{\revjj}[1]{{\color{blue}#1}} %JSAC submission
\newcommand{\rmkk}[1]{{\color{blue}#1}} %revise of the text
\newcommand{\rev}[1]{{\color{blue}#1}} %revise of the text
\newcommand{\revh}[1]{{\color{magenta}#1}} %revise of the text
\newcommand{\com}[1]{\textbf{\color{red} (COMMENT: #1) }} %comment of the text
\newcommand{\comg}[1]{\textbf{\color{green} (COMMENT: #1)}}
\newcommand{\response}[1]{\textbf{\color{green} (RESPONSE: #1)}} %response to comment
\else
\newcommand{\revjrr}[1]{{#1}} %JSAC revision
\newcommand{\revjr}[1]{{#1}} %JSAC revision
\newcommand{\revjj}[1]{{#1}} %JSAC submission
\newcommand{\rmkk}[1]{{#1}} %emph in equation

\newcommand{\rev}[1]{#1}
\newcommand{\revh}[1]{#1}
\newcommand{\com}[1]{}
\newcommand{\comg}[1]{}
\newcommand{\response}[1]{}
\fi
\newcommand{\tit}[1]{\textit{#1}}

\def\A{\mathcal{A}}
\def\N{\mathcal{N}}
\def\M{\mathcal{M}}

\def\S{\boldsymbol{S}}
\def\TE{\boldsymbol{\theta}}
\def\TEa{\boldsymbol{\phi}}
\def\x{\boldsymbol{x}}
\def\p{\boldsymbol{p}}

\def\z{\boldsymbol{z}}
\def\bpa{\boldsymbol{\pi}}

\def\B{\textsc{B}}
\def\C{\textsc{C}}
\def\U{\textsc{U}}
\def\V{\textsc{V}}
\def\W{\textsc{W}}
\def\R{\textsc{R}}
\def\Q{\textsc{Q}}
\def\sw{\Psi}
\def\mw{\Omega}

\def\dw{\Delta}
\def\dwc{\widetilde{\Delta}}
\def\adw{\bar{\Delta}}

\def\Ctot{\C^{\textsc{tot}}}
\def\Um{\U}
\def\Ua{\V}

\def\Wa{\W^{\textsc{ap}}}

\def\Wtot{\W^{\textsc{tot}}}

\def\pa{\pi}
\def\paxx{v}
\def\t{\xi}
\def\tl{\underline{\xi}}
\def\tu{\overline{\xi}}
\def\eq{\triangleq}
\def\eqi{=}
\def\wc{\widetilde{c}}

\def\xu{\bar{X}}

\def\st{*}
\def\stx{*}
\def\mi{{-}}
\def\ad{{+}}
\def\ei{{=}}
\def\dii{{...}}
\def\dt{;}

\def\paN{ \pa_N }

\def\paNr{ \pa_{N\mi 1} }

\def\pan{ \pa_{n} }

\def\MC{ \textsl{MC}_{\textsc{bs}} }

\def\MP{ \textsl{MP} }

\newcommand{\footnotesc}[1]{\footnote{#1}}

\makeatletter
\let\@copyrightspace\relax
\makeatother

\begin{document}

\title{Bargaining-based Mobile Data Offloading}

\author{
Lin~Gao, 
George~Iosifidis, 
Jianwei~Huang, 
Leandros~Tassiulas, and  
Duozhe~Li
\IEEEcompsocitemizethanks{
\IEEEcompsocthanksitem
This work is supported by the General Research Funds (Project Number CUHK 412713 and CUHK 412511) established under the University Grant Committee of the Hong Kong Special Administrative Region, China, and the National Natural Science Foundation of China (Project Number 61301118).
\IEEEcompsocthanksitem
The work is also supported by the project SOFON, which is implemented under the ``ARISTEIA'' Action of the ``OPERATIONAL PROGRAMME EDUCATION AND LIFELONG LEARNING'' and is co-funded by the European Social Fund (ESF) and National Resources.
\IEEEcompsocthanksitem
Lin~Gao and Jianwei~Huang {(Corresponding Author)} are with Network Communications and Economics Lab (NCEL), The Chinese University of Hong Kong,
E-mail: \{lgao, jwhuang\}@ie.cuhk.edu.hk.
\IEEEcompsocthanksitem
George~Iosifidis and Leandros~Tassiulas are with the
Department of Electrical and Computer Engineering, University of Thessaly, and CERTH, Greece, 
E-mail:
giosifid@inf.uth.gr, leandros@uth.gr.
\IEEEcompsocthanksitem
Duozhe Li is with the 
Department of Economics, The Chinese University of Hong Kong, 
E-mail: duozheli@cuhk.edu.hk.
}
\vspace{-5mm}
}

\maketitle
\begin{abstract}

The unprecedented growth of mobile data traffic challenges the performance and economic viability of today's cellular networks, and calls for novel network architectures and communication solutions. 
Data offloading through third-party WiFi or femtocell access points (APs) can effectively alleviate the cellular network congestion in a low operational and capital expenditure. 
This solution requires the cooperation and agreement of mobile cellular network operators (MNOs) and AP owners (APOs).
In this paper, we model and analyze the interaction among one MNO and multiple APOs (for the amount of MNO's offloading data and the respective APOs' compensations) by using the Nash bargaining theory. 
Specifically, we introduce a \textit{one-to-many} bargaining game among the MNO and   APOs, and analyze the bargaining solution (game equilibrium) systematically under two different bargaining protocols: (i) \emph{sequential bargaining}, where the MNO bargains with APOs sequentially, with one APO at a time, in a given order, and (ii) \emph{concurrent bargaining}, where the MNO bargains with all APOs concurrently.  
%This bargaining protocol captures the different negotiation options in this setting, and the bargaining dynamics that may impact the equilibrium. 
%Namely, the MNO may opt to bargain sequentially, with one APO at a time, or concurrently with all of them.  
We quantify the benefits for APOs when bargaining sequentially and earlier  with the MNO, and the losses for APOs when bargaining concurrently with the MNO. 
%we study the case where the APOs are organized in groups e.g., as in WiFi sharing communities, and bargain jointly with the MNO. For this scenario, 
%Moreover, we quantify the benefits for APOs when bargaining with the MNO jointly as a group, and further study whether and up to what extent the grouping of some APOs affects those APOs not in the group. 
We further study the group bargaining scenario where multiple APOs form  a group bargaining with the MNO jointly, and quantify the benefits for APOs when forming such a group. 
Interesting, our analysis indicates that grouping of APOs not only benefits the APOs in the group, but may also benefit some APOs not in the group.
%the benefits for APOs, both those participating in the groups and those acting independently, and the losses for the MNO.  
Our results shed light on the economic aspects and the possible outcomes of the MNO/APOs interactions, and can be used as a roadmap for designing policies for this promising data offloading solution.

\vspace{2mm}


\end{abstract}

\IEEEpeerreviewmaketitle

\addtolength{\abovedisplayskip}{-1mm}
\addtolength{\belowdisplayskip}{-1mm}

%!TEX root = DataOffload_main_journal.tex
%SourceDoc DataOffload_main_journal.tex

\vspace{-5mm}

\section{Introduction}

\subsection{Background and Motivations}

The global mobile data traffic is growing explosively, and it is expected that by 2018, it will reach $15.9$ exabytes per month, nearly an $11$-fold increase over 2013 \cite{cisco-2012}. 
To cope with this unprecedented traffic load, mobile network operators (MNOs) need to significantly increase their cellular network capacities. 
However, traditional methods such as acquiring more spectrum licenses, deploying new cells of small size, and upgrading technologies (e.g., from WCDMA  to LTE/LTE-A) are costly, time-consuming, and may not catch up the pace of the traffic increase. 
Clearly, MNOs must find novel methods to address this problem, and mobile data offloading appears as one of the most attractive solutions.

%\rmhocg{1. Why mobile data offloading is important (congestion, scalable solution, etc: operators cost reduction can
%be expected as fewer new macrocell
%towers are required to meet growing demand, reducing costs related to
%new equipment, site acquisition or leasing and power consumption).
%2. Why employing privately owned APs is important?
%3. Why we use the bargaining solution
%4. Why we study different bargaining protocols. (This must be stressed properly since we are the first that
%use this type of analysis for a networking problem).}

Simply speaking, \emph{mobile data offloading} is the use of complementary network technologies (such as WiFi and femtocell) for delivering the mobile data traffic originally targeted for cellular  networks.
The performance benefit of data offloading through WiFi and femtocell networks has been extensively studied in the existing literature (see, e.g., \cite{wifi-boudec,wifi-hui,wifi-aruna,wifi-lee,wifi-zhuo,femto-andrews,our3}). 
\revjj{Thus, it is not surprising that MNOs want more initiative in determining whether, when, and how much to offload their cellular traffic.\footnote{As demonstrations, MNOs have already deployed their own WiFi networks (e.g. AT\&T \cite{wifi-att}), or initiated collaborations with existing WiFi networks (as O2 did with BT \cite{O2-BTOpenzone}), to complement their cellular networks. Some MNOs have also started offering transparent cellular-WiFi services \cite{xx2}.} 
This \emph{network-initiated} offloading approach is greatly facilitated by technological advances such as the Hotspot 2.0 protocol \cite{xx1}, and the 3GPP Access Network Discovery and Selection Function (ANDSF) standard.}
In order to fully reap these benefits, it is essential to ensure that MNOs are able to offload their traffic whenever needed.
%as frequently as possible
%%anytime and anywhere
%in a flexible and scalable fashion.
%, which is rarely considered before (for example in \cite{xx2} this issue is not addressed).
%A prerequisite of this goal is obviously the 
%One way to achieve this is to 
To achieve this goal, a \emph{high coverage} of WiFi or femtocell networks is necessary. 
%Unfortunately, to achieve a ubiquitous deployment of WiFi or femtocell access points (APs) directly by MNOs is costly, and sometimes even impractical due to the limitations of additional site spaces and backhaul capacities. 
Unfortunately, the densely or ubiquitous deployment of WiFi or femtocell access points (APs) by the MNOs themselves is costly and often impractical due to the limitations of additional site spaces and backhauls. 

An alternative option for the MNOs is to \emph{employ} existing WiFi and femtocell APs already deployed by third-parties (as O2 did with BT \cite{O2-BTOpenzone}), instead of deploying their own offloading networks. 
%\revj{This approach is facilitated by the observation that there are already many APs available today, deployed by residential users, business companies or public departments, and that many of them are organized in small groups or even larger communities such as FON \cite{fon}.}
This novel network \emph{outsourcing} method is attractive due to the high population of WiFi or femtocell users \cite{jeffrey} as well as the technology innovations (e.g.,  Hotspot 2.0 protocol and  3GPP ANDSF standard) enabling such a cellular-WiFi inter-networking.
%for two main reasons.
%\revjj{First, WiFi and femtocell APs are being deployed in residencies and offices with an increasing pace \cite{jeffrey}. 
%Second, related technological advances (e.g., the Hotspot 2.0 protocol and the ANDSF standard) facilitate the integration of WiFi into cellular network, as they enable, among others, seamless hand-off and automated authentication.}
%\revjj{Second, the already increased penetration of WiFi technologies is complemented by related technical advancements \cite{europe-offloading}. (not sure what this means. Do you mean that the latest WiFi standards can support much higher data rates comparing with the cellular technology?)} 
%\footnotesc{For example, an MNO can use the femtocell APs that are already deployed by its clients to offload the traffic of other clients.} 
With this approach, MNOs can handle data offloading with a reduced capital expenditure (CAPEX) and operational expenditure (OPEX). 
Moreover, MNOs can make the offloading decisions more flexibly and efficiently, by employing APs on-demand taking into consideration the traffic dynamics. 
%\footnote{Such a leasing solution can be easily implemented. For example, consider an MNO who leases some of the APs that his subscribers have already installed in their premises for satisfying their own needs.} 
Nevertheless, without proper \emph{incentives}, the APs' owners (APOs) are expected to be reluctant to admit the cellular traffic, since offloading cellular traffic will consume their limited network capacities and increase various costs such as the energy expenditure and the backhaul cost.
This important economic incentive issue, however, is still quite under-explored in the existing literature.

\vspace{-3mm}

\subsection{Contributions}

%Therefore, the question arising in such a scenario is how much data should each APO admit to offload for each MNO, and how much should it be reimbursed by the MNO.

\begin{figure}[t] 
      \centering
    \includegraphics[scale=.27]{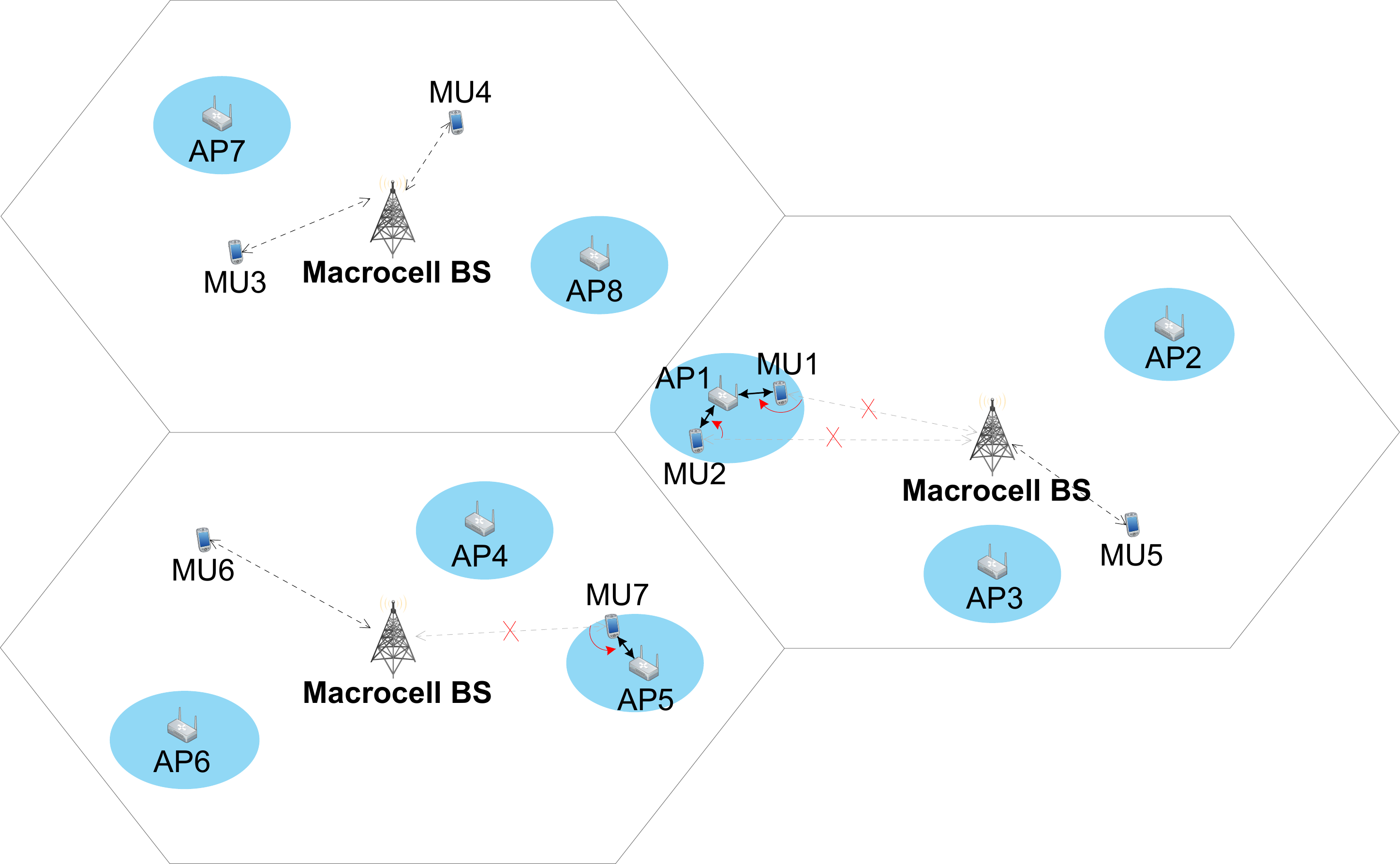}
    \caption{An instance of mobile data offloading.
The MNO can either serve an MU by its macrocell base stations (BSs) directly, e.g., MUs 3-6, or offload an MU's traffic to nearby APs, e.g., MUs 1-2 to AP 1 and MU 7 to AP 5.} \label{fig:system-illu}
   \vspace{-4mm}
\end{figure}

In this paper, we study the mobile data offloading via {third-party} WiFi and femtocell APs, 
and focus on the necessary \emph{economic incentives} that MNOs need to provide for APOs in order to achieve flexible on-demand data offloading. 
Specifically, we consider such an offloading scenario,  where \emph{one} MNO offloads its cellular traffic to a set of third-party WiFi or femtocell APs. 
The MNO serves its subscribed mobile users (MUs) in one or multiple  macrocells. 
Each AP can only offload the traffic generated by (cellular) MUs within its coverage. 
%We assume that the coverage areas of APs are non-overlapping. 
Figure \ref{fig:system-illu} illustrates such a network scenario, 
where the hexagons denote the coverage areas of the MNO's macrocells, and the blue circles denote the much smaller (non-overlapping)  coverage areas of APs.
In this example, 
the traffic of MUs 1 and 2 can be offloaded to AP 1, and the traffic of MU 7 can be offloaded to AP 5, while the rest MUs cannot take advantage of offloading, as they are not within the coverage area of any AP.
In such an offloading model, we are interested in the following technical and economical issues:
\begin{itemize}
\item \emph{Technical issue:} How to offload traffic efficiently (i.e., maximizing the social welfare)?
\item \emph{Economical issue:} How to share the offloading benefit among the MNO and APOs fairly? 
\end{itemize}
Note that the second (economical) issue is particularly important as it is closely related to the incentives of  APOs. 
More specifically, to address these issues, we need to answer the following offloading and reimbursing problem explicitly: (i) For the MNO, how much traffic should it offload to each AP and how much pay each APO? and 
(ii) For each APO, how much cellular traffic should it offload for the MNO and how much charge to the MNO? 
Clearly, the successful deployment of such a cooperative offloading architecture requires the MNO and APOs to agree on both the offloading amount and the payment.
\revjr{One proper theoretic framework to achieve this goal is the cooperative game theory.\footnote{Game theory \cite{new1} is widely used in wireless networks to model 
interactions of multiple network entities, where the actions of one entity (player) affect the payoffs of the other entities (see, e.g., \cite{add-1, add-2, add-3, add-4, add-5}). 
Cooperative game theory is usually used in situations when players have conflicting/competing interests but 
have the means (and also incentives) to coordinate and negotiate with each other to achieve a mutually beneficial outcome. 
In our model, it is natural to assume that the MNO and APOs  can get in touch and coordinate regarding  offloading and reimbursing decisions. 
%For example, an APOs may already be a subscriber of the MNO (such as an AT\&T's subscriber who owns a home WiFi). 
It is then natural to study the data offloading problem using the cooperative game theory.}} 

In this work, we model and analyze the data offloading problem by using the \emph{Nash bargaining theory} \cite{nash}, a special branch of the cooperative game theory, which is expected to yield a Pareto-efficient and fair outcome, hence self-enforcing and satisfactory for all entities. 
In this bargaining model, the MNO negotiates with each APO for the  amount of offloading data and the respective payment. 
We formulate the entire negotiation processes between the MNO and all APOs as a \emph{one-to-many} bargaining game, and study the game outcome (bargaining solution) systematically. 
There are many challenging issues arising in a one-to-many bargaining. 

\textbf{Bargaining Protocol.} 
%The basic Nash bargaining theory that has been widely applied in the  existing literature cannot capture several important aspects of our problem. 
%Namely, we need here to consider the impact of market dynamics on the bargaining outcome. For example, what is the benefits for the APO that first deploys an AP in a location where the cellular network is congested? What are the advantages if the MNO can concurrently bargain with more than one APOs? 
An important issue arising naturally in a one-to-many bargaining is the  {bargaining dynamics} (called \textit{bargaining protocol}), namely, how the MNO bargaining with multiple APOs, e.g., sequentially or concurrently? 
In this work, we will study two different {bargaining protocols} systematically: (i) \emph{sequential bargaining}, where the MNO bargains with all APOs sequentially in a predefined order, and (ii) \emph{concurrent bargaining}, where the MNO bargains with all APOs concurrently. 
There are many interesting open questions associated with this bargaining protocol. 
For example, will an APO gain certain benefit if it bargains with the MNO ahead of other APOs (in the sequential bargaining)? 
How would the MNO choose between sequential or concurrent bargaining with multiple APOs? 
%Although the study of bargaining protocols is an active research area in economics (see \cite{moresi,duozhe-li}), 
%there does not exist studies that take the unique characteristics of network into consideration. 
%and, to the best or our knowledge, have not be considered before in network resource allocation.
Although the study of bargaining theory is an active research area in economics, there is not much work analyzing this protocol comprehensively.~~~~~~~~
%Nevertheless, the consideration and analysis of these protocols is particularly important so as to understand the various aspects of the bargaining problem and the possible outcome in practice. 
%there does not exist studies that take the unique characteristics of network into consideration. 

\textbf{Grouping Effect.}
Another important issue in a one-to-many bargaining is the possibility that APOs may form groups (or larger communities) and  bargain jointly with the MNO. 
This allows individual APOs, who initially have less bargaining power than the MNO and may only have limited choices (e.g., accept or reject the terms of the MNO), to gain more market power from the larger collective coverage. 
%\footnote{Practical examples of  such groups and communities include the world-wide WiFi operator FON \cite{fon}, and other smaller groups \cite{braem}. In 2013, FON has signed an agreement with Deutsche Telekom in terms of mobile data offloading.} 
Such groups can be created within WiFi sharing communities such as FON \cite{fon}, or in the context of community networks \cite{braem}. 
%\footnote{Interestingly, FON has signed an agreement with Deutsche Telekom in terms of mobile data offloading in 2013.} 
%as their services are often complementary, to gain more benefits. 
%formed (e.g., FON \cite{fon} or smaller groups \cite{braem}) and is expected to bargain jointly with any MNO who aims to lease their equipment. 
Motivated by this, we would like to understand the impacts of the size and the structure of APO groups on the bargaining outcome. 
%We will quantify the impacts of the {APO grouping structure} under both sequential and concurrent bargaining protocols.

%It is important to note that our analysis focuses on a monopoly  scenario (i.e., one MNO), and incorporates the particularities of wireless and cellular networks (see Section \ref{sec:model-1}). 
%% which represents many actual settings. For example, it captures the bargaining of an MNO with its clients (subscribers) that have already installed femtocell APs at their premises, so as to offload cellular traffic of other mobile subscribers. Additionally, 
%This model arises naturally in today's cellular market, where a MNO has  exclusive access to the APs deployed by his subscribers (with certain cost). 
%This model can also serve as a building block for the analysis of the general competitive scenario.
%%This model can serve as a building block for the analysis of the  oligopoly competitive scenario (with multiple competitive MNOs) in the future.} 
The main contributions are are summarized as follows.

\begin{itemize}

\item 
\revjr{To the best of our knowledge, this is the first paper modeling and studying mobile data offloading using a one-to-many bargaining framework, which yields a fair, Pareto-efficient, and self-enforcing offloading solution.}~~~~~~~

\item  
We characterize the outcome of the one-to-many bargaining under different bargaining protocols and grouping structures, 
which has not yet been considered completely in the existing literature of Nash bargaining.

%Such a bargaining choice is suitable when two sides of the market have similar market powers, which is the case when APOs can form groups. 
%\revjj{Besides, bargaining is most suitable when the two sides have similar market power and are able to negotiate so as to reach an agreement.}
%The proposed scheme can be considered as an incentive structure that enables cooperative data offloading.

%\item 
%We provide a detailed framework using generic objective (cost) functions for the network operator and   APs.
%This approach renders our analysis appropriate for a variety of systems, under different modeling assumptions. 
%We calculate analytically the operator's cost reduction %benefit from offloading and the respective payments that are necessary for incentivizing the APs to admit cellular traffic.
%Our study incorporates the APs' resource constraints and needs (for serving their own traffic).

%\item
%We analytically derive the bargaining solutions under two different bargaining protocols: sequential bargaining and concurrent bargaining.

%\revjj{(The following two bullets points are too long and repetitive of our previous discussions. 
%The list of contributions need to be concise and to the point. With the previous discussions on bargaining protocol and grouping effect, readers already how the basic ideas. The list of contributions here should emphasize the key results obtained from the analysis. What is the impact of sequential and group bargaining? What is the impact of grouping to different network entities? )}

\item 
We study the impact of the bargaining protocol on the bargaining outcome comprehensively. 
%Namely, we explicitly define and analyze the concurrent and sequential bargaining protocols  in this setting.  
We quantify the benefits for APOs when bargaining sequentially and earlier with the MNO (\emph{early-mover advantage}), and the losses of APOs when bargaining concurrently with the MNO (\emph{concurrent moving tragedy}).~~~~~~~~~~~~~
%Note that bargaining protocol is an important factor as it captures the various cooperation dynamics among the MNO and APOs.
%\revjr{This analysis is very important, as it reveals the various cooperation dynamics among the MNO and APOs, and hence characterizes the possible outcomes that may arise in practice.}
%We formulate the problem as an one-to-many bargaining, and study the bargaining solutions under
%sequential and concurrent bargaining protocols systematically. 
%This aspect of bargaining theory has never been considered before, while it is very important as it models various dynamic effects, such as the appearance of new APs.
%Our analysis analytically shows that
%\begin{enumerate}
%\item
%Under sequential bargaining, APOs can secure a higher payoff by bargaining earlier with the MNO (called \emph{early-mover advantage}). We analytically calculate such a early-moving gain for APOs.
%\item
%Under concurrent bargaining, the payoff of each APO is equal to the worst-case payoff that it can ensure under sequential bargaining (called \emph{concurrent moving tragedy}).
%%\item We further study the impact of AP group structure on the bargaining outcome: (i) grouping together will always benefit the grouped APs; and (ii) it will also benefit all prior APs under sequentially protocols.
%\end{enumerate}

\item 
We study the grouping effect on the bargaining solution systematically. 
%Specifically, we analyze how grouping of multiple APOs affects the benefits of the group members, and further analyze whether and up to what extent it affects the non-group members. 
Interesting, our analysis indicates that grouping APOs not only benefits the APOs in the group (\emph{intra-grouping benefit}), but may also benefit some APOs not in the group (\emph{inter-grouping benefit}).~~~~~~

\end{itemize}

%\revjj{(This paragraph should be moved to the literature review part (Section II), and can be used to emphasize the difference between the related litearture and our work. )}
%\revj{Our analysis considerably departs from related studies which have mainly focused on the potential performance benefits of mobile data offloading without however ensuring the availability of APOs. We use cooperative game theory in order to find the self-enforcing and hence sustainable solution for the problem, and we shed light on its dependence on the bargaining protocol. This latter aspect is of particular importance since it captures the market's dynamics (the sequence of APO bargaining) and other interesting phenomena, such as group bargaining, which have been overlooked until now. The considered monopoly scenario represents many real settings such as the bargaining of a MNO with its clients (subscribers) that have installed femtocell APs, in order to offload traffic of other clients. Also, our work can be used as a building block for the analysis of the respective oligopoly market problem}.~~~~~~

The rest of this paper is organized as follows.
In Section \ref{sec:literature}, we review the literature. In Section \ref{sec:model}, we  present the system model.
In Sections \ref{sec:barg} and \ref{sec:barg-many}, we study the data offloading bargaining systematically. We provide the simulations in Section \ref{sec:simu}, and finally conclude in Section \ref{sec:conclusion}.

%!TEX root = DataOffload_main_journal.tex
%SourceDoc DataOffload_main_journal.tex

\section{Literature Review}\label{sec:literature}

\subsection{Mobile Data Offloading}

The performance benefit of mobile data offloading through WiFi networks has been studied in \cite{wifi-boudec,wifi-hui,wifi-aruna,wifi-lee,wifi-zhuo}, which showed that in urban environments, WiFi can offload about 65\% of mobile data, and save 55\% of MUs' battery energy. 
These benefits can be further enlarged if users are willing to delay their traffic \cite{our3}. 
%Obviously, ensuring a high availability of WiFi APs is the critical factor that determines the potential of this offloading method, an issue that is still open.
Another promising option for data offloading is   femtocell  \cite{femto-andrews}. 
%Today, an increasing number of subscribers install femtocell APs at their premises which can serve either only the registered users (close access mode), or offload also traffic from other mobile users (open access mode). 
The problem of incentivizing femtocell owners to admit macrocell traffic has been recently studied in \cite{femto-huang, femto-jo,femto-yun,femto-zhang,femto-pantisano}. 
However, these works studied the incentive issues using the non-cooperative game framework, which cannot capture the potential of coordination  among mobile operators and femtocell owners 
(which calls for a cooperative game approach). 
%Besides, they do not consider the MNO's explicit benefits from offloading.
%, or propose spectrum exchange schemes that raise additional implementation issues.
In \cite{femto-new}, Zhang \emph{et. al.} studied the  economic incentive issue by using the cooperative game framework (Nash bargaining) as we did in this work. 
%where the  authors study the cooperation framework for a mobile network operator (who offloads traffic to its femtocells) and a fixed-line network operator (who provides wired backhaul for femtocells) that enables the femtocell offloading. 
%where the authors use a cooperative game framework (Nash bargaining) to model the interaction between  a mobile network operator and a fixed-line network operator in the offloading process. 
%In our work, we consider a different cooperation game framework, which is not targeted for the interactions of mobile and fixed-line operators, but for those of mobile operator and WiFi/femtocell APOs. 
%Besides, our proposed bargaining framework is a one-to-many bargaining, while the bargaining framework in \cite{femto-new} is the simple one-to-one bargaining.
However, the bargaining model in \cite{femto-new} is the simple one-to-one bargaining (between one mobile operator and one fixed-line operator), while the bargaining model in our work is a more general one-to-many bargaining (between one MNO and many APOs).~~~~~~~

In our previous works, we have studied the economic incentive issue in mobile data offloading via third-party APs, by using either the non-cooperative Stackelberg game framework \cite{our1} or the  auction  framework \cite{our2}. 
However, these works can neither capture the potential of coordination among the MNOs and APOs, nor the effect of market dynamics (e.g., the bargaining protocol in our model) or  
user collusions (e.g., the grouping of APOs in our model). 

\subsection{Nash Bargaining Theory}

Nash in \cite{nash} established a basic two-person bargaining framework between two rational players,  
%and shows that \revjr{there is a unique solution (i.e., the \emph{Nash bargaining solution, NBS}) that satisfies the four axioms of Pareto efficiency, symmetry, invariance to affine transformations, and independence of irrelevant alternatives (see Section \ref{sec:theory} for details).} 
%%and suggested that we can predict the result by focusing on the properties (axioms) that we expect the bargaining outcome will exhibit (the so-called \emph{axiomatic approach} for   bargaining problems). 
%The NBS determines the portion of the jointly produced welfare each player should receive so as to agree to cooperate with the other players. 
%The last years this theory has been used to model and analyze a variety of problems in communication networks \cite{book-basar}. 
and proposed an \emph{axiomatic} solution concept---Nash Bargaining Solution (NBS), which is characterized by a set of pre-defined axioms (see Section \ref{sec:theory}), and does not rely on the detailed bargaining process of players. 
In the follow up work, Nash \cite{nash-new} and Rubinstein \cite{rubin-new} provided strategic foundations for the NBS, by analyzing specific dynamic non-cooperative bargaining processes (games) and showing that the equilibria of the  bargaining games converges to the NBS.

Since Nash's pioneering work, researchers have extended the bargaining analysis to the case of more than two players. 
In the multi-player scenario, some players may form \emph{groups} and bargain jointly in order to improve their payoff (hence the  {group bargaining}  \cite{chae}).
% A typical application of group bargaining is the labor negotiation between the {management} and a union of workers \cite{chica,dobbelaere}.
%In these settings, bargaining first takes place among the different groups and accordingly the members of each group bargain with each other in order to distribute the acquired welfare.
%The authors in \cite{chae} presented a solution for the group bargaining problems which is actually an extension of the NBS. 
In most cases, the \emph{grouping} improves the payoff of the group members (see \cite{chae2,puga2,heidhues}), as it increases their collective bargaining power. 
Interestingly, the opposite is also possible as shown by the \emph{Harsanyi bargaining paradox} \cite{harsanyi}.
\revjr{However, the above works did not consider the bargaining dynamics ({bargaining protocol}) among multiple bargainers, which arises naturally in a multi-player bargaining.}
%The bargaining outcome heavily depends on the {bargaining protocol}, e.g., bargaining concurrently or sequentially. 
\revjr{Regarding the bargaining protocol, the most relevant models are those in \cite{moresi, duozhe-li}. 
However, both papers focused only on the sequential bargaining,  using either an axiomatic approach \cite{moresi}  or a strategic approach \cite{duozhe-li}, and neither  considered the concurrent bargaining, nor the grouping effect.} 
%This aspect was studied in \cite{moresi} and \cite{duozhe-li}, where one dominant player optimally selects in each stage a weak player to bargain.
%\revjj{The above works, however, are pure economic studies that do not take into account the unique characteristics of the wireless data offloading problem. Clearly, the bargaining outcome, under any protocol or group structure, depends on the market model and also on the system model. Our work captures both   aspects.}

%Our work focuses on the one-to-many bargaining between one dominant player (the macrocell  BS) and many weaker players (APs).
%We derive the bargaining solutions based on the Nash bargaining framework, under both concurrent and sequential bargaining protocols.
%From the solutions, we show that APs have incentive to form a group bargaining jointly with the macrocell  BS, and the corresponding group bargaining solution is equivalent to that in \cite{chae} in terms of the group benefit.

%!TEX root = DataOffload_main_journal.tex
%SourceDoc DataOffload_main_journal.tex

\section{System Model}\label{sec:model}

%The mobile data traffic increases dramatically due to the explosive growth of mobile users with smartphones, laptops, and tablets. As a consequence, the Macro BS may suffer from severe network congestion, and thus the mobile operator want to offloading some traffic to complementary networks like Wi-Fi or femtocell networks, after considering certain offloading cost.

\subsection{System Description}\label{sec:model-1}

We consider one mobile network operator (MNO), operating one or multiple macrocells, wants to offload its cellular traffic to a set $\N\eq \{1,\dii,N\}$ of third-party WiFi or femtocell access points (APs).\footnotesc{In this work, we do not distinguish WiFi APs and femtocell APs, as we will model APs using generic objective (cost) functions. This renders our analysis appropriate for a variety of systems with various assumptions.}
We assume
%APs are non-overlapped in geography.
that the coverage areas of any two APs are non-overlapping.
This assumption is reasonable as the transmission range of AP is much smaller than that of the macrocell base station (BS).\footnote{\revjr{In our online technical report \cite{report}, we also  discuss  how to extend the current model to a more general model with overlapping APOs.}}
%This assumption is reasonable because the transmission range of WiFi or femtocell APs (typically tens of meters) is much smaller than that of macrocell base station (BS), which is typically hundreds of or even thousands of meters. 
Figure \ref{fig:system-illu} illustrates such a network with $ 8$ non-overlapping APs and $3$ macrocell BS.~~~~~~~~
%Notice that the discussions in this paper are general for multiple macrocell BSs. 

The MNO serves a set of macrocell mobile users (MUs) who are randomly distributed in geography. 
The traffic generated by an MU can be offloaded to an AP,  if the following conditions are all satisfied: 
%\footnotesc{The first two constraints are related to the technical issues, and ensure that the MU is able to connect to the AP. The last constraint is related to the personal attitude or policy to data offloading, and ensures that the MU is willing to offload its traffic to WiFi or femtocell APs.}
\begin{itemize}
\item
The MU is located within the coverage area of the AP (hence \emph{attainable} for the AP),
\item
The MU is equipped with the same radio frequency interface and  wireless communication protocol as the AP (hence \emph{compatible} with the AP),
\item
The MU is enabled to offload its traffic (e.g., WiFi is turned on for offloading to a WiFi AP).~~~~~~~~~~~~~~~~~~~~
\end{itemize}

%\vspace{-2mm}

Let $\M_n$ denote the set of MUs whose traffic can be offloaded to AP $n$, and $\M_0$ denote the set of MUs whose traffic cannot be offloaded to any AP.
As the APs' coverage areas are non-overlapping, we have: $\M_n \bigcap \M_{m} = \emptyset, \forall m,n\in\N$ with $m\neq n$.
%(i)
%the MU is located within the coverage area of the AP (and thus  {attainable} for the AP),
%(ii)
%the MU equips with the same wireless interface and protocol as the AP (and thus  {compatible} with the AP),
%and (iii) the MU is enabled (by himself) to offload his traffic.
In the example of Figure \ref{fig:system-illu}, the traffic of MUs 1 and 2 can be offloaded to AP 1 and the traffic of MU 7 can be offloaded to AP 5 (supposing these MUs are compatible with APs and enable WiFi), while the traffic of MUs 3-6 cannot be offloaded to any AP.~~~~~~~~~~~~~
%Mobile users (MUs) are randomly distributed within the coverage area of the BS, and have heavy data to send. Within the same area, there are $\N \eq \{1,\dii,N\}$ WiFi and femto access points (APs) owned by self-interested third parties. The goal of the operator is to employ (i.e. lease) these APs and offload a portion of his traffic to them, so as to reduce the cost of serving his subscribers (the mobile users) and to avoid network congestion.

%Each AP has a transmission range (typically tens of meters) much smaller than the range of the macrocell BS (typically hundreds or thousands of meters). Thus, we can assume that the coverage areas of the employed APs are \emph{non-overlapping}.\footnote{{Note that in practice, the operator can employ multiple APs in certain hot spots to support offload of  very high data traffic. We will show in our technical report \cite{report} that our analysis can be easily applied to such scenarios.}} The data of a user can be offloaded to a certain AP, only if this user is located within the coverage area of the AP (i.e., \emph{attainable}), and is equipped with the same wireless communication interface as the AP (i.e., \emph{compatible}). Due to the non-universal coverage of APs, a user may be located in a blank area not covered by any AP. Additionally, due to the non-overlapping coverage of APs, every user has at most one attainable AP. In Figure \ref{fig:system-illu}, AP $1$ is attainable for mobile users $1$ and $2$, while users $3, \dii, 7$ have no attainable APs.

Let $S_n$ denote the total cellular traffic that can be offloaded to AP $n$ (i.e., the total traffic generated by MUs in $\M_n$), and $S_0$ denote the total cellular  traffic that cannot be offloaded to any AP (i.e., the total traffic generated by MUs in $\M_0$). 
%\footnote{\revjr{Note that when considering the large times-scale bargaining period (e.g., when the bargaining is performed every hour or every day), $S_n$ and $S_0$ will correspond to the estimations of the \emph{average} total cellular traffic that can be offloading to AP $n$ and the \emph{average} total cellular traffic that cannot be offloaded to any AP, respectively.}}
%That is, $S_n, n=1,\dii,N$, is the total traffic generated by those MUs who are attainable for AP $n$ and meanwhile compatible with AP $n$, and $S_0$ is the total traffic generated by all other MUs.
The \emph{traffic profile} of the MNO is denoted by
$$
\S \eq(S_0, S_1, \dii, S_N).
$$
Due to the uncertainty of MUs' mobility and data usage, the value of $S_n$ for each $n$ changes randomly over time.
We consider a quasi-static network scenario,
where the values of $S_n$ for all $n$ remain unchanged within every data offloading period (e.g., one minute in our simulation).\footnote{\revjr{Note that when considering the large times-scale bargaining period (e.g., when the bargaining is performed every hour or every day), $S_n$ and $S_0$ will correspond to the estimations of the \emph{average} traffic.}}
%generated by compatible users within the coverage area of AP $n$, $n\ei 1,\dii,N$, and $S_0$ denote the aggregate traffic generated by users either lying in blank areas or being incompatible with the APs. Clearly, $S_n$ is the maximum amount of data that can be offloaded to AP $n$, while $S_0$ is the amount of data that cannot be offloaded to any AP.
%\footnote{Note that each AP also has its own traffic to send as we will discuss in details in Section \ref{sec:model:ap}.}
%Denote $\S \eq(S_0, S_1, \dii, S_N)$ the \emph{traffic profile} of the macrocell BS. Due to the uncertainty of users' activities, $\S$ changes randomly over time.
%We consider a quasi-static traffic profile, which remains unchanged in a certain period of time, called the \emph{traffic coherent time}  (e.g., one mimute in our simulation). We assume that the data offloading   is performed periodically with a period smaller than the traffic coherent time, and thus $\S$ remains unchanged in every data offloading decision period  (but can change across different decisions periods).%\footnote{We will show in our technical report \cite{report} that our solution can also be applied to the fast changing traffic profile (which changes within one data offloading period) with an acceptable performance degradation.}

We define the \emph{transmission efficiency} of a communication link (between an MU and its attached macrocell BS, or between an MU and an AP) as the average amount of data traffic (in bits) that can be delivered by one unit of spectrum resource (in Hz) per time unit (in second). 
Obviously, the transmission efficiency is closely related to the path loss and shadow fading of a link.
As a concrete example, we can compute it based on the Shannon channel capacity. 
But our discussions are general for any choice of transmission efficiencies in different communication systems. 

Let $\theta_n $ denote the  {average} transmission efficiency (in bits/Hz/s) between MUs in $\M_n$ (in AP $n$'s coverage area) and their corresponding macrocell BS,
and $\theta_0$ denote the  {average} transmission efficiency between MUs in $\M_0$ (not in any AP's coverage area) and their corresponding macrocell BS.
%\footnotesc{Here the ``average'' essentially means a weighted average. That is, $\theta_n = \frac{ \sum_{m\in \M_n} S_{n}^m \theta_{n}^m}{\sum_{m\in\M_n} S_{n}^m}$, where $S_{n}^m$ and $\theta_{n}^m$ are the traffic volume and transmission efficiency of MU $m\in \M_n$, respectively. \revjj{(Need to clarify)}}
That is, delivering one unit of traffic generated by $\M_n$ (or $\M_0$) within a single time unit, on average, consumes $\frac{1}{\theta_n}$ (or $\frac{1}{\theta_0}$) units of the MNO's   resource.
The \emph{transmission efficiency profile} of the MNO is denoted by
$$
\TE \eq (\theta_0, \theta_1, \dii, \theta_N).
$$

Let $\phi_n$ denote the  {average} transmission efficiency between   MUs in $\M_n$ and AP $n$. 
That is, offloading one unit of cellular traffic generated by $\M_n$ within a single time unit, on average, consumes $\frac{1}{\phi_n}$ units of AP $n$'s resource.
The \emph{transmission efficiency profile} of APs is denoted by
$$
\TEa \eq (\phi_1, \dii, \phi_N).
$$
We similarly assume that $\TE$ and $\TEa$ remain  unchanged within every offloading period, but may changes across periods.
Our analysis focus on the offloading solution in a single  period. 

%
%Since every AP's coverage area is small, we can reasonably assume that:
%(i) the transmission efficiency between every AP and its covered mobile users is the same maximum value, i.e., $\theta = 1$, and (ii) the transmission efficiency between the macrocell BS and all mobile users covered by the same AP (say $n$) is also the same, denoted by $\theta_n$. Depending on the location of every AP $n$, we have: $0\leq \theta_n \leq 1$.
%%Obviously, employing those APs close to the edge of the macrocell BS's coverage area is more efficient for data offloading, since this traffic consumes more spectrum resource of the macrocell BS.
%We further denote $\theta_0$ as the average transmission efficiency between the BS and all users that do not offload their data. Denote $\TE \eq (\theta_0, \theta_1, \dii, \theta_N)$ as the {transmission efficiency profile}.
%We similarly assume that $\TE$ remains unchanged in every data offloading decision period.

\subsection{MNO Modeling}

We focus on the \emph{direct} benefit for the MNO from data offloading, i.e., the \emph{serving cost reduction} due to the reduced resource consumption.\footnotesc{Some \emph{indirect} benefits include (i) the improvement of MUs' QoS and thus the increase of the average revenue per user (ARPU), (ii) the increased number of active MUs and thus the increased  total revenue, and (iii) the reduction of the network congestion and thus the saving of the MNO's reputation.}
Such a serving
cost may include the energy cost, operational cost, coordinating cost, etc.
Let $\C(b)$ denote the MNO's serving cost for $b$ units of resource consumption.
%Without loss of generality, we
%Here $\C(b)$ can be any generic function satisfying the following conditions: continues, differential, 
We will consider a generic cost function $\C(b)$ that is continuous, differentiable, 
strictly increasing, and convex, i.e., 
$\C'(b) > 0$ and $\C''(b) \geq 0$.~~~~~~~~~~~~~~~~~~~~~~~~~
%The Macro BS's (negative) \emph{payoff}, denoted by $\Um$, is directly defined as the total cost (consisting of both the serving cost and the payment), i.e.,

%Given the traffic profile $\S = (S_0, S_1, \dii, S_N)$ and transmission efficiency profile $\TE = (\theta_0, \theta_1, \dii, \theta_N)$, without data offloading, the amount of the BS's spectrum resource for delivering all traffic (by itself) is given by
%\begin{equation}
%\textstyle B_{\textsc{all}} \eq \sum_{n=0}^N \frac{S_n}{\theta_n}.
%\end{equation}

%When the macrocell BS performs data offloading, it needs to pay each collaborative AP based on the amount of data offloaded.
%When data offloading is applied, the macrocell base station offloads an amount of traffic to the APs by paying certain reimbursement prices.
Let $x_n \in [0, S_n]$ denote the traffic offloaded to AP $n$, and $z_n\geq 0$ denote the MNO's payment to the owner of AP $n$ (denoted by APO $n$).
The \textbf{traffic offloading profile} and 
\textbf{payment profile} are, respectively,
$$
\x \eq (x_1, \dii, x_N),\mbox{~~and~~}\z \eq (z_1, \dii,z_N).
$$
%Obviously, a feasible $\x$ satisfies: $x_n \leq S_n$, $\forall n $.

Given $\x$ and $\z$,
the MNO's total resource consumption for
delivering \emph{remaining} un-offloaded traffic is
\begin{equation}
\textstyle
%b(\x,\S,\TE) 
b(\x)
\eqi \frac{S_0}{\theta_0} +  \sum_{n=1}^N \frac{S_n - x_n}{\theta_n},
\end{equation}
%It is easy to see if $\x = \boldsymbol{0} \eq (0,\dii,0)$, then $\B(\x) = B_{\textsc{all}}$.
%For notation consistence, we will write $B_{\textsc{all}}$ as $\Bdir$.
and the MNO's total cost, including both the serving cost and the payment to APOs, is
\begin{equation}
\textstyle \Ctot(\x \dt \z) \eqi \C (b(\x)) + \sum_{n=1}^N z_n.
\end{equation}
%and without data offloading, the total cost is
%\begin{equation}
%\Ctot(\boldsymbol{0}\dt \boldsymbol{0}) = \C(\Bdir) + 0,
%\end{equation}
%since both the data and payments transferred to APs are $0$.
%For clarity, we will write $b(\x)$ as $b$ whenever there is no confusion caused.
%Let $z_n$ denote the Macro BS's payment to AP $n$ for data offloading.
The MNO's \emph{payoff} is defined as the \emph{total cost reduction} achieved from data offloading, 
%\footnotesc{Here $ \boldsymbol{0} \eq (0, \dii, 0)$ denotes the zero vector of size $1\times N$.} 
denoted by
\begin{equation}\label{eq:BS-payoff}
\begin{aligned}
\Um(\x\dt\z) &  \eqi  \Ctot(\boldsymbol{0}\dt \boldsymbol{0}) - \Ctot(\x\dt \z)
\\
&  \textstyle \eq \R(\x) -  \sum_{n=1}^N z_n,
\end{aligned}
\end{equation}
where $ \boldsymbol{0} \eq (0, \dii, 0)$, and $\R(\x) \eqi \C(b(\boldsymbol{0})) - \C(b(\x))$ is the MNO's serving cost reduction.
%For clarity, we will write $b(\x)$ as $b$ and $b(\boldsymbol{0})$ as $b_0$ whenever there is no confusion caused.
%, and $b_0 = b(\boldsymbol{0},\S,\TE) = \sum_{n=0}^N \frac{S_n}{\theta_n}$ is the
%MNO's total resource consumption without data offloading.
%The payoff defined in (\ref{eq:BS-payoff}) is actually the \emph{reduction} of the Macro BS's total cost (including both the serving cost and the payment) when using data offloading. Specifically, without data offloading, the total cost is $\C(\B(\boldsymbol{0})) +0 $, since the traffic volume and the payment to APs are both 0; and with data offloading, the total cost is $\C(\B(\x)) + \sum_{n=1}^N z_n$.
We refer to the MNO's payoff without data offloading as its \emph{reservation payoff}, denoted by $\Um^0 \eq \Um( \boldsymbol{0}\dt \boldsymbol{0})  = 0$. 
As we will show later, this reservation payoff serves as the disagreement point of the MNO, and plays an important role in the bargaining.

\subsection{APO Modeling}\label{sec:model:ap}

Each AP is owned by a private owner (APO), whose primary goal
is to serve its own users.
%\footnotesc{An APO can be, for example, a customer who bought a WiFi or femtocell AP, or a business like StarBucks who deployed a WiFi network.} 
Thus, each APO, when deciding whether (and how, if so) to offload traffic for the MNO, must take into consideration the demand of its own users.

Let $\t_n$ denote the APO $n$'s own resource demand (from its own users).
Due to the uncertainty of AP users' mobility and data usage, we define $\t_n$ as a random variable, falling within a certain interval $[\tl_n, \tu_n]$ and following a probability distribution function (PDF) $f_n(\t)$ and a cumulative distribution function (CDF) $F_n(\t)$. 
%\footnotesc{If APO $n$'s traffic coherent time is large enough, $\t_n$ will keep unchanged in every data offloading period.
%In this case, $\t_n$ becomes a constant in the current data offloading period, which is a special case of our model.}
We assume that  $\t_n, \forall n\in\N$, are independent of each other, but not necessarily identically distributed.
Let $B_n$ denote the total resource owned by APO $n$. 
Let $w_n$ denote the average revenue achieved from one unit of its own resource demand, and $c_n$ denote the cost for one unit of its resource consumption. %Without loss of generality, we assume that $w_n > c_n,\, \forall n\in\N$.
Then, APO $n$'s expected profit (from serving its own demand) is
%\footnotesc{Here we consider the linear serving cost for APOs, similar as the model used in \cite{femto-yun}.}
\begin{equation}
\begin{aligned}
&  \textstyle \Wa_n(B_n ) \textstyle \eqi  (w_n - c_n) \cdot \mathbf{E}_{\t_n} \min\{B_n, \t_n\}
\\
 = 
&\textstyle (w_n - c_n) \cdot  \Big( \int_{ \tl_n}^{B_n} \t   f_n(\t) \mathrm{d}\t  + \int_{B_n}^{\tu_n} B_n    f_n(\t) \mathrm{d}\t\Big).
\end{aligned}
\end{equation}

%An AP is usually very small and serves in a small area. Thus, its cost mainly includes the energy cost and the backhaul broadband cost, both are approximately linear to the spectrum resource consumption.

%Since the transmission efficiency between an AP and its covered mobile users is normalized to 1, the amount of spectrum resource the AP consumes is equivalent to the volume of the traffic it delivers.
Recall that the average transmission efficiency between AP $n$ and MUs in $\M_n$ is $\phi_n$.
If AP $n$ admits $x_n $ units of cellular traffic (generated by MUs in $\M_n$), 
the total resource consumption for the offloaded cellular traffic is $\frac{x_n}{\phi_n}$,
and thus the resource left for serving its own demand is $B_n - \frac{x_n}{\phi_n}$.
Obviously, a feasible $x_n$ must satisfy: $x_n \leq \phi_n \cdot B_n $.
Given feasible $x_n$ and $z_n$, the APO $n$'s total profit, including both the profit from serving its own demand and the profit from offloading for the MNO, is
\begin{equation}
\begin{aligned}
&\textstyle \Wtot_n(x_n\dt z_n ) \eqi \Wa_n(B_n - \frac{x_n}{\phi_n}) +
 z_n - c_n \cdot \frac{x_n}{\phi_n},
%+  \Wb(x_n\dt z_n),
\end{aligned}
\end{equation}
where $(z_n - c_n \cdot \frac{x_n}{\phi_n})$ is the profit from helping the MNO, consisting of the service income (i.e., the MNO's payment) and the serving cost.

The APO $n$'s \emph{payoff} is the \emph{profit improvement} when offloading traffic for the MNO, denoted by
\begin{equation}\label{eq:AP-payoff}
\begin{aligned}
\Ua_n (x_n\dt z_n) & \eqi \Wtot_n (x_n\dt z_n) - \Wtot_n  (0\dt 0) 
\\
& \eq \Q_n(x_n) + z_n,
\end{aligned}
\end{equation}
where $\Q_n(x_n) \eqi \Wa_n  (B_n - \frac{x_n}{\phi_n})  - \Wa_n (B_n) - c_n\cdot \frac{x_n}{\phi_n}$ is the APO $n$'s profit loss induced by data offloading.
Similarly, we refer to the APO $n$'s payoff when not offloading traffic for the MNO as its \emph{reservation payoff}, denoted by $\Ua_n^0 \eq \Ua_n (0\dt 0)  = 0$.
This reservation payoff serves as the disagreement point of APO $n$ in the bargaining.

\subsection{Social Welfare}

The \textbf{social welfare} is defined as the aggregate payoff of the MNO and all APOs, denoted by
%That is, given any feasible offloading profile $\x$ and payment profile $\z $, the social welfare is
\begin{equation}\label{eq:social-welfare}
\begin{aligned}
\sw(\x\dt \z)   \textstyle \eqi &\textstyle  \Um(\x \dt \z) + \sum_{n=1}^N \Ua_n (x_n\dt z_n)
\\ 
  =&\textstyle \R(\x) + \sum_{n=1}^N \Q_n(x_n) \eq \sw(\x).
\end{aligned}
\end{equation}
That is, the social welfare is equivalent to the sum of the MNO's serving cost reduction and the APOs' profit loss, as the payments will be canceled out. Thus, we will also write the social welfare as $\sw(\x)$.~~~~~~~~~~

\section{A Simple One-to-One Bargaining}\label{sec:barg}

%In this section, we formulate the data offloading problem as a \emph{bargaining}, and study the outcome systematically.
%based on the cooperative game theory. That is, the MNO  and APO owners bargain for the traffic offloading volume as well as the payment.

In this section, we first review the Nash bargaining theory. 
Then we consider a simple model with \emph{one} APO, and formulate the problem as a basic two-person \emph{one-to-one} bargaining. 
\revjj{We use this simple example to illustrate how to formulate and analyze a data offloading problem by using the Nash bargaining framework.
This can help us to better understand the bargaining formulation and analysis for general models with multiple APOs in Section \ref{sec:barg-many}.}

%We apply the famous Nash bargaining solution (NBS) in this work, which is based on 4 well-known axioms proposed by Nash \cite{nash}.

%The novelty of this bargaining problem is as follows.
%First, this is a \emph{one-to-many bargaining} between one dominant player (the macrocell BS) and multiple weaker players (APs).
%%\footnotesc{The MNO  is dominant, since it has more market power to determine whether to offload its traffic and how much traffic to be offloaded.}
%Second, we study two different one-to-many bargaining protocols: \emph{sequential bargaining}, and \emph{concurrent bargaining}, depending on whether the MNO  bargains with all APOs sequentially or concurrently.
%Third, we study the incentive for APOs merging together and bargaining jointly with the BS.
%%the APOs (weaker players) can either form a group jointly bargaining with the BS, or bargain independently with the BS.
%%We characterize the conditions under which APOs will form a group jointly bargaining with the macrocell BS.

 \subsection{Nash Bargaining Theory}
\label{sec:theory}

%We first briefly review the classic Nash bargaining solution (NBS) \cite{nash}.
%We first review the Nash bargaining theory basics.
In \cite{nash}, Nash established the following two-person bargaining framework.
There is a set $\N = \{1,2\}$ of two players.
The players either reach an \emph{agreement} in a set $\A$,
%\footnotesc{The set $\A$ of possible agreements is rather general in this framework. For example, an agreement in $\A$ can simply be the price in a trading activity, or the strategy profile of players in a game.}
or fail to reach agreement, in which case the disagreement event $D$ occurs.
Each Player $i\in\N$ has a preference
ordering over the set $\A\bigcup\{D\}$, represented by a utility function $U_i $ over the domain of $\A\bigcup\{D\}$.
%Given $\A$, $D$, and $\{U_i\}_{i\in\N}$, we can construct the set of all \emph{utility profiles} that can be the outcome of bargaining.
We denote such a {bargaining problem} by $\mathcal{G} \eq \left<\N,\A, D,\{U_i\}\right>$.
A \emph{bargaining solution} assigns {every} bargaining problem $\mathcal{G}$ an outcome, which can be either an agreement or the disagreement event. 
Note that an agreement outcome can be either a specific agreement in the set $\A$, or a \emph{lottery} over a set of possible agreements. 

Nash proposed four axioms that should be satisfied by a reasonable bargaining solution \cite{nash}: Pareto efficiency, symmetry, invariance to affine transformations, and independence of irrelevant alternatives. 
Nash proved that under mild technical conditions, there is a \emph{unique} bargaining solution (called Nash bargaining solution, NBS) satisfying the four axioms above. 
Moreover, the NBS has a very simple form: it corresponds to an
outcome that maximizes the product of both players' \emph{utility gains} upon the disagreement outcome.
%Specifically, consider the following bargaining game $\mathcal{G}=\left<\N,\A,\{U_i\}\right>$, where $\N\eq\{1,2,\dii ,N\}$ is the player set, $\A \eq A_1\times A_2 \times\dii \times A_N$ is the action space where $A_i$ is the set of actions available to player $i$, and $U_i(\cdot)$ or $U_i(\a)$ is the payoff of player $i$ which depends on the action profile of all players, $\a \eq (a_1,a_2, \dii ,a_N)$, where $a_i \in A_i$ is player $i$'s action.
%Consider a 2-person bargaining problem as an illustration.

Specifically, let $d_i \eq U_i(D)$ denote the utility of player $i\in\{1,2\}$ over the disagreement outcome $D$ (i.e., the \emph{reservation utility} or \emph{disagreement point} of player $i$), and $\mathcal{U} \eq \{(U_1(a), U_2(a)) \}_{a\in\A}$ denote the set of utility pairs over all possible agreements (i.e., the \emph{feasible set}). 
%\footnotesc{If it is allowed to specify a lottery over a set of possible agreements as an outcome, the feasible set is the smallest convex set that contains $\mathcal{U}$, i.e., the convex hull of $\mathcal{U}$.}
Suppose that (i) $ \mathcal{U}$ is compact (i.e. closed and bounded) and convex, and (ii) there exists an $(u_1,u_2)\in \mathcal{U}$ such that $u_i \geq d_i,  i =1,2$.

\begin{definition}[Nash Bargaining Solution -- NBS \cite{nash}]\label{def:NBS}
A pair of utilities $(u_1^*, u_2^*)
\in \mathcal{U}$ (or the associated agreement $a^*\in\A$) is an NBS (i.e., satisfying Nash's four axioms), if it solves the following  problem:
\begin{equation}\label{eq:NBS-definition}
\begin{aligned}
\max_{(u_1,u_2) \in \mathcal{U}} & \ (u_1-d_1 ) \cdot (u_2 - d_2 ) \\
\mbox{\emph{s.t.} }
& \ u_1 \geq d_1 ,\  u_2 \geq d_2 .
\end{aligned}
\end{equation}
\end{definition}

%Intuitively, an NBS maximizes the product of both players' \emph{gains} in utility over the disagreement point, and
It is easy to see that the disagreement points $d_1$ and $d_2$ play an important role in the Nash bargaining framework. 
With a higher disagreement point $d_i$, player $i$ can obtain a larger utility under the NBS.

\subsection{One-to-One Bargaining}\label{sec:barg-single}

Now we consider a simple network scenario with \emph{one} AP.
%, to illustrate the basic idea and key concepts to be used in the more general data offloading bargaining in Section \ref{sec:barg-xo}. 
In this case, the bargaining problem is a \emph{one-to-one} bargaining (one MNO and one APO). 
For notational consistence, we still denote the APO by $n$, i.e., $\N=\{n\}$.
%Obviously, in this case, there is no difference between  sequentially and concurrent bargaining protocols.
%By Definition \ref{def:NBS}, an NBS maximizes the product of both players' {payoff gain}.

Let $\mathcal{X}_n \eq [0, \min\{S_n,\phi_n B_n\}]$ and $\mathcal{Z}_n \eq [0, +\infty)$ denote the sets of   feasible $x_n$ and $z_n$, respectively.
An agreement is a feasible  tuple $(x_n, z_n)$.
% with feasible $x_n$ and $z_n$.
The agreement set is $\A \eq \{(x_n, z_n)\ | \ x_n \in \mathcal{X}_n, z_n \in \mathcal{Z}_n \}$.
The NBS is an agreement $(x_n^*, z_n^*) \in \A$ that solves the following problem:
\begin{equation}\label{eq:NBS-single}
\begin{aligned}
\max_{(x_n, z_n)\in \A} & \  \Um(x_n\dt z_n) \cdot  \Ua_n (x_n\dt z_n) \\
\mbox{s.t. } & \ \Um(x_n\dt z_n) \geq 0,\ \Ua_n (x_n\dt z_n) \geq  0.
%,\\ &\ 0\leq x_n \leq \xu_n.
\end{aligned}
\end{equation}
%where $\Um(\cdot)$ is the macrocell MNO's payoff defined in  (\ref{eq:BS-payoff}), and $\Ua_n(\cdot)$ is APO $n$'s payoff defined in (\ref{eq:AP-payoff}).
%This problem (\ref{eq:NBS-single}) is directly from the definition of NBS.
Note that in (\ref{eq:NBS-single}), both the MNO  and APO $n$ have a zero disagreement point, i.e., $\Um^0 = \Ua_n^0 = 0$.

%where $\xu_n \eq \min\{S_n, B_n\}$.
% is the upper-bound of $x_n$.
%where $\Um^0 = 0$ and $\Ua_n^0 = 0$ are the reservation payoffs (or disagreement points) of the Macro MNO  and APO $n$.
%The first two constraints state that an agreement (bargaining solution) is addressed only when every player achieves a payoff no smaller than its disagreement point.\footnotesc{In other words, an agreement is addressed only when the Macro MNO  achieves a lower total cost and the APO achieves a higher total profit with data offloading than without data offloading (disagreement).} The last constraint states that the offloaded traffic volume $x_n$ cannot be larger than $S_n$ (i.e., the Macro MNO's total traffic volume within APO $n$'s coverage area) and $B_n$ (i.e., the total resource  of APO $n$).
%Note that a bargaining solution $\{0,0\}$ is equivalent to the outcome of disagreement.

%Obviously, if the Macro MNO  and APO $n$ do not address any agreement, we have $x_n = 0$ and $z_n = 0$, and therefore
%$$
%\Um^0 \eq \Um(0,0) = 0 \mbox{ and } \Ua_n^0 \eq  \Ua_n(0, 0) = \Wa_n (0).
%$$

%Denote
%$\pa_n \eq \Ua_n (x_n\dt z_n) = \Q_n(x_n) + z_n  $ as the payoff (gain) of APO $n$.
%Then, the MNO's payoff (gain) can be written as
%$\Um(x_n\dt z_n )  = \sw(x_n) - \pa_n.$
%The problem (\ref{eq:NBS-single}) becomes

For notational convenience, we introduce a new variable $\pa_n$ to denote the APO $n$'s payoff (gain), i.e., 
$$\pa_n \eq \Ua_n (x_n\dt z_n)  = \Q_n(x_n) + z_n. $$
%\begin{equation}\label{eq:NBS-single-pa}
%\pa_n \eq \Ua_n (x_n\dt z_n) .%= \Q_n(x_n) + z_n  .
%\end{equation}
Then, the MNO's payoff (gain) can be written as $\Um(x_n\dt z_n )  = \sw(x_n) - \pa_n$
%\begin{equation}\label{eq:NBS-single-sw-pa}
%\Um(x_n\dt z_n )  = \sw(x_n) - \pa_n,
%\end{equation}
where
$ \sw(x_n ) $ is the   social welfare defined in (\ref{eq:social-welfare}). 
Substituting the above formulas to (\ref{eq:NBS-single}), we can rewrite (\ref{eq:NBS-single}) as a new optimization problem of $x_n$ and $\pa_n$, i.e.,
\begin{equation}\label{eq:NBS-single-eq}
\begin{aligned}
\max_{(x_n, \pa_n)} & \ \big( \sw(x_n) - \pa_n \big) \cdot \pa_n \\
\mbox{s.t. } & \ x_n \in \mathcal{X}_n,\ \sw(x_n) - \pa_n \geq 0,\ \pa_n \geq 0.
\end{aligned}
\end{equation}
%where $\xu_n \eq \min\{S_n, \phi_n B_n\}$.
%The problem (\ref{eq:NBS-single-eq}) has an optimal solution only if there exists a feasible $x_n \in [0, \xu_n]$ such that $\sw(x_n) \geq 0$. Otherwise, the constraint set of (\ref{eq:NBS-single-eq}) is an empty set, and thus there is no solution for (\ref{eq:NBS-single-eq}).
Note that  problems (\ref{eq:NBS-single}) and (\ref{eq:NBS-single-eq}) are   equivalent.
This implies that the bargaining for $(x_n, z_n)$ is   equivalent to the bargaining for $(x_n, \pa_n)$.
Intuitively, for any bargaining solution on $(x_n, \pa_n)$, we can compute an equivalent solution on $(x_n, z_n)$ in the following way: $  z_n = \pa_n - \Q_n(x_n) $. 
It is easy to check that (\ref{eq:NBS-single-eq}) is a convex optimization problem. Thus, we have the following NBS for this simple one-to-one bargaining problem.\footnote{We leave all of the detailed proofs in the  online technical report \cite{report}.}

\begin{lemma}[{One-to-One NBS}]\label{theorem:NBS-single}
The NBS $(x_n^*, \pa_n^*)$ for the one-to-one bargaining is
$$\textstyle
 x_n^* = x_n^o, 
\mbox{~~and~~}  \pa_n^*  =\frac{1}{2}\cdot {\sw(x_n^o)}. %\mbox{\emph{ or }} z_n^* = \frac{1}{2}\cdot {\sw(x_n^o)} - \Q_n(x_n^o),
$$
where $x_n^o = \arg \max_{x_n \in \mathcal{X}_n}   \sw(x_n) $ is the social welfare maximization offloading solution.
%(i)
%$x_n^* = x_n^o$, and (ii) $\pa_n^*  =\frac{1}{2}\cdot {\sw(x_n^o)} $.
%
%, and $z_n^* = \pa_n^* - \Q_n(x_n^*)$.
%\begin{enumerate}
%\item[\emph{(a)}] $x_n^* = x_n^o$,~~ \emph{(b)} $\pa_n^*  =\frac{1}{2}\cdot {\sw(x_n^*)} $,~~ and
%\item[\emph{(c)}] $z_n^* = \pa_n^* - \Q_n(x_n^*) =\R(x_n^*) - \pa_n^* $.
%\end{enumerate}
\end{lemma}

%\begin{proof}
%See Appendix-\ref{app:proof-one-to-one-NBS} of \cite{report}.
%\end{proof}

%Lemma \ref{theorem:NBS-single} states that the NBS maximizes the social welfare $\sw(x_n)$, and in addition, under the NBS, each player gets half of the achieved maximum social welfare.
%, i.e., $\Um(x_n^*\dt z_n^* ) = \Ua_n(x_n^*\dt z_n^* ) = \frac{1}{2}\cdot\sw(x_n^o) $.

The above lemma implies that the NBS maximizes the social welfare. 
%Under the NBS, each player gets half of the achieved maximum social welfare.
Intuitively, this is because the total generated social welfare can be freely transferred between players (through the payment $z_n$), and thus maximizing the product of their individual payoff gains can only  be achieved when maximizing the overall social welfare. 
% , they must agree on the socially optimal solution given by (\ref{eq:NBS-single-swm}).
This is a key property the bargaining problem with \emph{transferable utility}.
Note that this phenomena not only exists in a one-to-one bargaining, but also exists in the general one-to-many bargaining studied later.

\section{One-to-Many Bargaining}
\label{sec:barg-many}

{In this section, we consider a general model with multiple APOs $\N=\{1,\dii,N\}$. 
In this case, the MNO  needs to bargain with every APO $n\in \N$ for $(x_n, z_n)$ (hence a one-to-one bargaining), and thus the entire bargaining problem becomes a \emph{one-to-many}  bargaining, consisting of $N$ coupled one-to-one bargainings.
%\revjj{(connection between this section and the previous one)}
%\footnotesc{Later we will discuss the difference between one-to-man bargaining and the traditional multi-person bargaining, which we will show is not suitable for our data offloading problem.}
%\footnotesc{{Later we will study the \emph{group bargaining} where APOs form a group bargaining jointly with the BS.}}
%Since the one-to-many bargaining problem can be viewed as a combination of $N$ one-to-one bargainings, 
Accordingly, the one-to-many bargaining solution contains $N$ agreement or disagreement outcomes, each associated with a one-to-one bargaining (between the MNO and one APO). 
%\footnotesc{In this sense, the one-to-many bargaining proposed in this paper is different from the traditional multi-player bargaining.
%The latter is a direct extension of the Nash's two-person bargaining, and is essentially a single bargaining problem involving many players.}
Clearly, there are two important factors that will affect the outcome of a one-to-many bargaining:~~~~~~~~~~~~~~~~~~~~~~~
\begin{enumerate}
\item
\emph{Bargaining Protocol}:
The MNO  can either bargain with all APOs sequentially, in a predefined order, or bargain with all APOs concurrently (see Figure \ref{fig:bargaining-scheme}).
We refer to the former one as  the \emph{sequential bargaining}, and the latter one as  the \emph{concurrent bargaining}.
\item
 \emph{APO Grouping Structure}:
APOs can either bargain individually with the MNO, 
or form one or multiple groups bargaining with the MNO jointly. 
An APO group can be exogenously given (e.g., all customers of FON belong to the same group), or endogenously formed based on their instant willingnesses. 
%\footnotesc{In this paper, we consider the exogenous group. That is, we do not address the problem of group formulation,  but rather take them as given and study the bargaining solutions under the given APO grouping structures.}
\end{enumerate}

%Moreover, compared with the previous one-to-one bargaining, the one-to-many bargaining here has the following unique features.
%First, the disagreement point of the MNO  (when bargaining with a particular APO) is no longer zero, but is determined by its bargaining results with the other APOs.
%%\footnotesc{This is directly from the definition of disagreement point, which is defined as a player's payoff when \emph{not} addressing any agreement with the bargainer. In a one-to-many bargaining, when the MNO  does not address any agreement with an AP, it can still achieve certain positive payoff depending on its bargaining solutions with other APs.}
%Second, the marginal social welfare generated by an APO is related not only to its own offloading solution, but also to other APOs' offloading solutions.
%This renders the $N$ one-to-one bargainings in this one-to-many bargaining to be coupled with each other.
%%Based on these features, the solution of a particular bargaining problem (between the MNO  and a particular AP) is affected by and will affect the solutions of other bargaining problems (between the MNO  and other APs).
%%Namely, the underlying sub-problems are coupled.

In what follows, we will  study the bargaining solution of the one-to-many bargaining systematically. We will call it the one-to-many NBS, or just NBS for short. 
%Notice that the NBS specifies the bargaining solution between the MNO and any APO. 
For convenience, we denote the bargaining solution between the MNO and APO $n$ as $(x_n^*, z_n^*)$, and the one-to-many NBS as $\{\x^*, \z^*\} \eq \{(x_n^*, z_n^*)\}_{n\in\N}$ which consists of the bargaining solutions between the MNO and all APOs.
%The NBS specifies the bargaining solutions between the MNO  and all APs, i.e., $\{(x_n^*, z_n^*)\}_{n\in\N}$.
%For convenience, we will also write the NBS as $\{\x^*, \z^*\}$.

\subsection{{Traffic Offloading Profile under the NBS}}
\label{sec:barg-xo}

We first study the traffic offloading profile $\x$ under the NBS. 
Similar to that in the aforementioned one-to-one bargaining, we show that in a general one-to-many bargaining, it still maximizes the social welfare, regardless of the detailed bargaining protocol and the APO grouping structure.  Formally, 
%As mentioned before, {our model is a TU model}, and the social welfare can be transferred freely between the MNO  and APs.
%Denote $(x_n^*, z_n^*)$ as the bargaining solution between the MNO  and APO $n$.
%Then, the traffic offloading profile $\x^{\stx} \eq (x_1^{\stx}, \dii ,x_N^{\stx})$ under the NBS maximizes the social welfare.
\begin{lemma}[Traffic Offloading Profile]\label{prop:nbs-x}
The traffic offloading profile $\x^{\stx} \eq (x_1^{\stx}, \dii ,x_N^{\stx})$ under the NBS is equivalent to the socially optimal traffic offloading profile $\x^{o} \eq (x_1^{o}, \dii ,x_N^{o})$.
\end{lemma}

We present the detailed proof in \cite{report}. 
Intuitively, our bargaining model is a transferable utility model, and thus the NBS (specifying both the payment transferring and traffic offloading between the MNO and all APOs) always maximizes the social welfare. 
Besides, the payment transferring $\z$ is internal and does not affect the social welfare. 
Therefore, the traffic offloading profile $\x$ under the NBS must maximize the social welfare. 
We skip the derivation of the social welfare maximization solution $\x^o$, as it is a standard convex optimization. 
Readers can  refer to the online technical report \cite{report} for details.
%
%Next we derive the socially optimal traffic offloading profile $\x^{o} \eq (x_1^{o}, \dii ,x_N^{o})$.
%Similar to (\ref{eq:NBS-single-swm}), the social welfare maximization problem is
%\begin{equation}\label{eq:NBS-multi-swm}
%\begin{aligned}
% \max_{\x} & \  \sw(\x), \quad \mbox{s.t. } \ x_n \in \mathcal{X}_n,\ \forall n\in \N.
%\end{aligned}
%\end{equation}
%It is easy to check that (\ref{eq:NBS-multi-swm}) is a convex optimization, and thus the optimal solution satisfies the sufficient and necessary  Karush-Kuhn-Tucker (KKT) conditions. 
%\revjj{Formally,
%%\footnotesc{Due to space limit, we skip the detailed derivation here. For details, please refer to our technical report \cite{report}.}
%%\begin{theorem}[Social Optimality]\label{thrm:SWM-opt}
%if the constraints of (\ref{eq:NBS-multi-swm}) are non-binding, the socially optimal $\x^{o} \eq (x_1^{o}, \dii ,x_N^{o})$ can be solved by the first-order conditions:~~~~~~~
%\begin{equation}\label{eq:SWM-FOC-eq}
%\textstyle
%\frac{\C'(b(\x^o))}{\theta_n}   = \frac{\left(w_n  - (w_n-c_n)\cdot  F_n\left(B_n-\frac{x_n^o}{\phi_n}\right)\right)}{\phi_n}   ,\ \forall n\in \N.
%\end{equation}
%%where $b^o = b(\x^o)$ is the MNO's resource consumption under the traffic offloading profile $\x^o$.
%%\end{theorem}
%We skip the derivation of $\x^o$ in general cases (where constraints may be binding), as it is a standard convex optimization. 
%Readers can  refer to our online report \cite{report} for details.~~~~~~~~}

\subsection{{Payment Profile under the NBS}}\label{sec:payoff}

Now we study the payment profile $\z $ under the NBS. 
As discussed in (\ref{eq:NBS-single}) and (\ref{eq:NBS-single-eq}), the bargaining for $\z$ (the payments to APOs) is equivalent to the bargaining for $\boldsymbol{\pa} = \{\pan\}_{n\in\N}$ (the payoffs of APOs).
For the convenience in describing, we will present the NBS in terms of $\boldsymbol{\pa}$.

%Based on the previous discussion, the traffic offloading profile $\x^*$ under the NBS is independent to the detailed bargaining protocol and APO group structure (and always equivalent to the socially optimal profile $\x^o$).
%Unlike the traffic offloading profile $\x$,  
% which is independent to the detailed bargaining protocol and APO group structure, we will show in this subsection that 
In this subsection, we will show that the payment profile $\z$ or the APO payoff profile $\boldsymbol{\pa}$ greatly depends on the bargaining protocol, and in the next subsection (Section \ref{sec:barg:group}) we will further show that it is also affected by the APO grouping structure. 
In what follows, we   derive the NBS under sequential bargaining (in Section \ref{sec:payoff}.1) and under concurrent bargaining (in Section \ref{sec:payoff}.2) systematically.\footnote{For better understanding of the analytical bargaining solution, we also provide illustrative examples in the online technical report \cite{report}.}

%we will analytically study the bargaining for the payments $\{z_n\}_{n\in\N}$ to APs, or equivalently, the payoffs $\{\pan\}_{n\in\N}$ of APs.
%%\footnotesc{The social welfare maximization solution $\x^{o} = (x_1^{o}, \dii ,x_N^{o})$ will be present in the end of this section.}
%Notice that the bargaining for payments and payoffs are   equivalent.
%The reason is that for any bargaining solution (agreement) on payments $\z^*$ (or payoffs $\bpa^*$),
%we can directly find an equivalent solution on payoffs $\bpa^*$ (or payments $\z^*$) in the following way: $\pa_n^* = \Q_n(x_n^*) + z_n^*$ (or $z_n^* = \pa_n^* - \Q_n(x_n^*)$).

%In what follows, we first
%provide the social optimal traffic offloading profile in Section \ref{sec:barg-xo}. Then, we study two different one-to-many bargaining protocols: \emph{sequential bargaining} and \emph{concurrent bargaining}, in Sections \ref{sec:barg:seq} and \ref{sec:barg:con}, respectively.
%We further capture the incentive for APOs merging with each other under both protocols, and finally study the associated group bargaining in Section \ref{sec:barg:group}.

\subsection*{B.1)~~Sequential  Bargaining}
%\label{sec:barg:seq}

We first study the NBS under the {sequential bargaining}, where the MNO bargains with  APOs sequentially, in a predefined order (see Figure \ref{fig:bargaining-scheme} (a)).
%Later we will study the impact of the AP-order on the bargaining solution.
Without loss of generality, we assume that the MNO bargains with APOs in the order of  $1,2,\dii ,N$.
This implies that there is no APO group, i.e., each APO bargains with the MNO individually.
The impact of APO grouping will be studied in Section \ref{sec:barg:group}.~~~~~~~~~~~~~~~~~~~~~~~~~~~~~

%(The impact of APO grouping structure will be studied in Section \ref{sec:barg:group}.)

Since the underlying one-to-one bargaining problems (between the MNO and each APO) are coupled with each other,
we solve the sequential bargaining  by backward induction. 
For the convenience in writing, we introduce notations: 
$$
\x_{n} \eq (x_1,\ x_2,\ \dii , \ x_{n}),~
$$
$$
\Pi_{n} \eq \pa_1 + \pa_2 + \dii + \pa_{n},
$$
for the analysis of the sequential bargaining. 
%where $\x_{N \mi 1}^{\st} \eq (x_1^{\st}, \dii ,x_{N \mi 1}^{\st} )$, and  $\Pi_{N\mi 1} \eq
%\sum_{n=1}^{N\mi1} \pa_n^{\st} $.
%That is, we first derive the bargaining solution between the MNO and APO $N$ in the last step $N$, given the bargaining solutions between the MNO and all preceding APOs.
%Then we derive the bargaining solution between the MNO and APO $N-1$ in the step $N$, given the bargaining solutions between the MNO and all preceding APOs and the prediction of the bargaining solutions between the MNO and all posterior APOs. 
%Let $\{x_n , z_n \}$ denote the bargaining solution with APO $n$.
%Note that the bargaining solution $(0,0)$ is equivalent to the case of disagreement.

\begin{figure}[t]
    \centering
   \includegraphics[scale=.4]{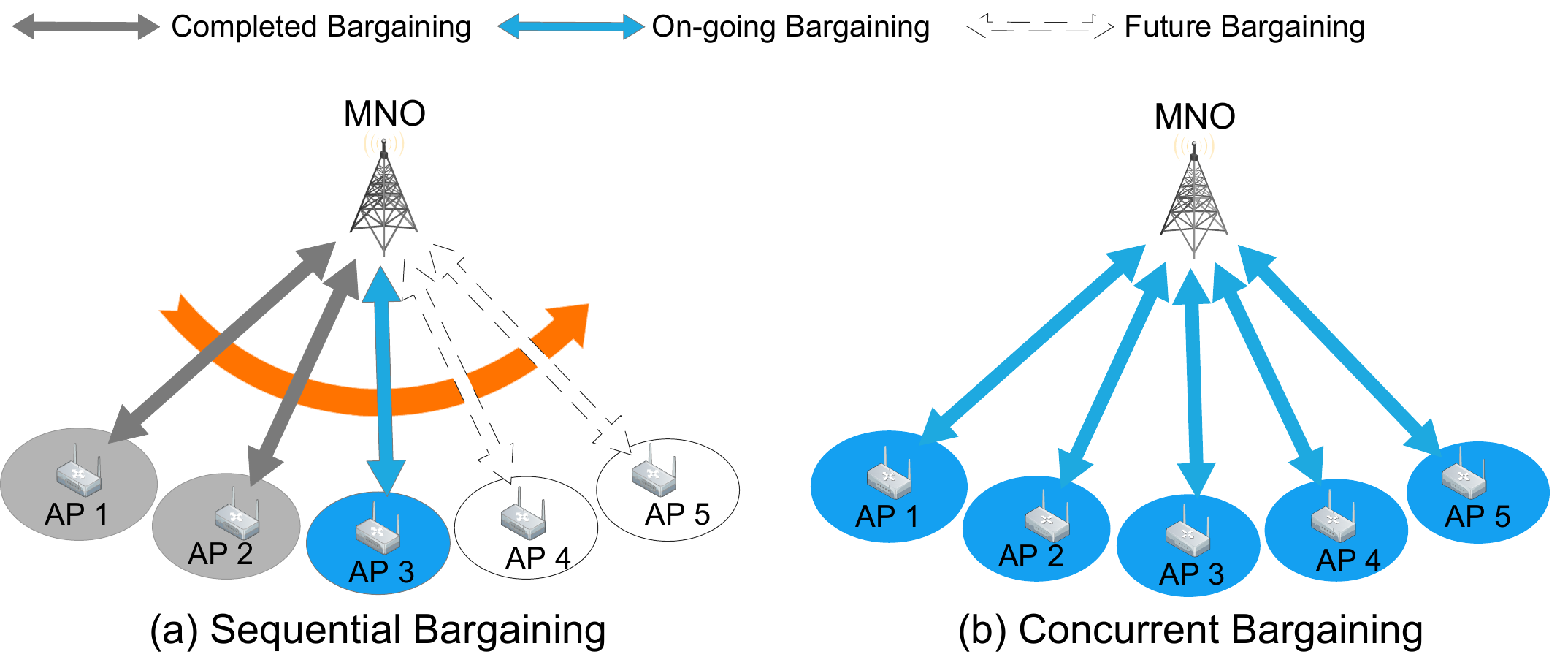}
     \vspace{-2mm}
    \caption{Illustration of bargaining protocols.} \label{fig:bargaining-scheme}
     \vspace{-2mm}
     \end{figure}

%Notice that the macrocell MNO's disagreement point when bargaining with an APO (and thus the bargaining solution with that AP) depends on the bargaining solutions with other APs. This implies that when solving a  bargaining problem with a particular AP, we need to consider not only the bargaining solutions with prior APs, but also the possible bargaining solutions with posterior APs.

\vspace{3mm}

\textbf{Step $N$.}

\vspace{1mm}

Suppose that the MNO  has finished bargaining with APOs $1$ to $N-1$, and reached bargaining solutions $\{\pa_n^{\st}\}_{n \in\{ 1,\dii ,N \mi 1\}}$.
%, with all APOs prior to $N$. 
Now it bargains with APO $N$ for $ \paN $.

%We first characterize the disagreement points of the MNO  and APO $N$.
\noindent
\underline{1.~Disagreement}:
If the MNO and APO $N$ do not reach any agreement,  
the APO $N$'s disagreement point is $0$,
and the MNO's disagreement point is its payoff achieved from all prior APOs, i.e.,\footnotesc{Here $\sw (\x_{N \mi 1}^{\stx}, 0)$ means $\sw \big(\x\big)$ with $\x = (\x_{N \mi 1}^{\stx}, 0)$, and the subscript ${[N]}$ in $\Um_{[N]}^0$ is used to indicate the stage of bargaining. Later on we will use the same form for notational convenience.}
\begin{equation*}
%\left\{
%\begin{aligned}
\textstyle \Ua_N^0  = 0,
%\\
\mbox{~~~~}
\Um^0_{[N]}  = \sw (\x_{N \mi 1}^{\stx}, 0) - \Pi_{N\mi 1}.
%\end{aligned}
%\right.
\end{equation*}
%where $\x_{N \mi 1}^{\st} \eq (x_1^{\st}, \dii ,x_{N \mi 1}^{\st} )$, and  $\Pi_{N\mi 1} \eq
%\sum_{n=1}^{N\mi1} \pa_n^{\st} $.
%\com{what is the difference between these two notations? We need to clear explain the notation of $sw$ as it keeps changing later on.}

\noindent
\underline{2.~Agreement}: If they reach an agreement $\paN = \paxx $ (and $x_N = x^*_N$), 
the  payoffs of APO $N$ and the MNO are, respectively,
\begin{equation*}
%\left\{
%\begin{aligned}
~~~~\Ua_N = \paxx,
%\\
\mbox{~~~~}
\Um_{[N]} \textstyle
 = \sw(\x_{N \mi 1}^{\stx}, \rmkk{x_N^{\stx}}) - \Pi_{N\mi 1} - \rmkk{\paxx}.
%\end{aligned}
%\right.
\end{equation*}

\noindent
\underline{3.~Payoff gain}: 
Under an agreement $\paN = \paxx $ (and $x_N = x^*_N$), the  payoff gains of APO $N$ and the MNO are, respectively,\begin{equation*}
%\left\{
%\begin{aligned}
\Ua_N - \Ua_N^0 = \paxx, 
%\\
\mbox{~~~~}
\Um_{[N]} - \Um^0_{[N]}  \textstyle = \dw_N - \paxx, 
%\end{aligned}
%\right.
\end{equation*}
where
%\begin{equation*}
%\textstyle
$\dw_N \eq \sw(\x_{N \mi 1}^{\stx}, \rmkk{x_N^{\stx}})- \sw(\x_{N \mi 1}^{\stx}, \rmkk{0})$
%\end{equation*}
denotes the  {marginal social welfare} (i.e., the {increase} of {social welfare}) generated by involving  APO $N$ in the offloading.
%(assuming that the MNO  has reached agreements with all other APs).

\noindent
\underline{4.~Bargaining solution}: 
By Definition \ref{def:NBS}, the NBS between the MNO  and the APO $N$ is given by
\begin{equation}\label{eq:NBS-seq-N}
\max_{ \paxx} \  \big(\dw_N - \paxx \big) \cdot \paxx, \quad  
\mbox{s.t.}~\dw_N  - \paxx \geq 0,\ \paxx \geq 0.
\end{equation}
%Similar to (\ref{eq:NBS-single-eq}), the objective of   (\ref{eq:NBS-seq-N}) is a quadratic function, and 
%thus the optimal solution is $\paxx^* =\frac{ \dw_N }{2}$.
Solving the above problem, we have the following NBS for the bargaining between the MNO and APO $N$.
\begin{lemma}[NBS in Step $N$]\label{lemma:NBS-seq-N}
The NBS between the MNO  and APO $N$  in Step $N$ is
%under sequential bargaining is:
\begin{equation}
\textstyle
\paN^* = \paxx^* = \frac{ \dw_N }{2}.
%= \frac{\sw(\x_{N \mi 1}^{\stx}, \rmkk{x_N^{\stx}})- \sw(\x_{N \mi 1}^{\stx}, \rmkk{0})}{2},
\end{equation}
% is the marginal social welfare generated by APO $N$.
In addition, under the NBS, the MNO's payoff is
%\footnote{Here the subscript ${[N]}$ in $\Um_{[N]}$ is used to indicate the stage of bargaining.}
\begin{equation}\label{eq:NBS-seq-N-um}
\textstyle
%\Ua_N^* = \frac{ \dw_N }{2}, 
%\mbox{~~~~}
\Um^{*}_{[N]} = U^0_{[N]} + \frac{ \dw_N }{2} = \frac{ \mw_N }{2}  - \Pi_{N\mi 1},
\end{equation}
where 
%$\dw_N= \sw(\x_{N \mi 1}^{\stx}, \rmkk{x_N^{\stx}})- \sw(\x_{N \mi 1}^{\stx}, \rmkk{0})$ and 
$\mw_N = \sw(\x_{N \mi 1}^{\stx}, \rmkk{x_N^{\stx}})+\sw(\x_{N \mi 1}^{\stx}, \rmkk{0})$.
\end{lemma}

%\begin{proof} The proof is same as that for Lemma \ref{theorem:NBS-single}.
%\end{proof}

The key insight of Lemma \ref{lemma:NBS-seq-N}
 is that the MNO  and APO $N$ equally share the marginal social welfare $\dw_N$ generated by involving APO $N$ in the offloading.
%Accordingly, the APO $N$'s payoff and the MNO's payoff under the NBS are, respectively,\footnote{Here the subscript ${[N]}$ in $\Um_{[N]}$ is used to indicate the stage of bargaining.}
%\begin{align}
%\label{eq:NBS-seq-N-pa}
%\textstyle
%%\Ua_N^* ~&\textstyle =\paN^* = \frac{ \dw_N }{2}, 
%%= \frac{ \sw(\x_{N \mi 1}^{\stx}, x_N^{\stx})- \sw(\x_{N \mi 1}^{\stx}, 0)}{2},
%\Ua_N^* ~&\textstyle 
%%= \frac{\sw(\x_{N \mi 1}^{\stx}, \rmkk{x_N^{\stx}})- \sw(\x_{N \mi 1}^{\stx}, \rmkk{0})}{2} 
%= \frac{ \dw_N }{2}, 
%\\
%\label{eq:NBS-seq-N-um}
%\textstyle
%%\Um^{*}_{[N]} & \textstyle  = \sw(\x_{N \mi 1}^{\stx}, x_N^{\stx}) - \Pi_{N\mi 1} -
%%\paN^* = \frac{ \mw_N }{2}  - \Pi_{N\mi 1},
%\Um^{*}_{[N]} & \textstyle  
%%= \frac{\sw(\x_{N \mi 1}^{\stx}, \rmkk{x_N^{\stx}}) + \sw(\x_{N \mi 1}^{\stx}, \rmkk{0})}{2} - \Pi_{N\mi 1}
%= \frac{ \mw_N }{2}  - \Pi_{N\mi 1},
%\end{align}
%\begin{equation}\label{eq:NBS-seq-N-um}
%\textstyle
%\Ua_N^* = \frac{ \dw_N }{2}, 
%\mbox{~~~~}
%\Um^{*}_{[N]} = \frac{ \mw_N }{2}  - \Pi_{N\mi 1},
%\end{equation}
%where $\mw_N \eq \sw(\x_{N \mi 1}^{\stx}, \rmkk{x_N^{\stx}})+\sw(\x_{N \mi 1}^{\stx}, \rmkk{0})$.

\vspace{3mm}

%\subsubsection*{\textsl{\bfseries  Step $N-1$}}
\textbf{Step $N-1$.}

\vspace{1mm}

Suppose  that the MNO  has reached bargaining solutions $\{\pa_n^{\st}\}_{n \in\{ 1,\dii ,N \mi 2\}}$
% ($\{x_n^{\st}\}_{n \in\{ 1,\dii ,N \mi 2\}}$) 
 with all APOs $1$ to $N-2$. Now it bargains with APO $N-1$ for $ \paNr $.

\noindent
\underline{1.~Disagreement}:
If the MNO and APO $N-1$ do not reach an agreement,
the APO's disagreement point is $0 $, and 
the MNO's disagreement point is its potential payoff \textit{after having dealt with all APOs}, i.e., $\Ua_{N\mi 1}^0 = 0$ and 
%. By (\ref{eq:NBS-seq-N-um}), we have
$$
\textstyle
\Um^0_{[N-1]}= \frac{ \sw(\x_{N \mi 2}^{\stx},\rmkk{0}, x_N^{\stx}) }{2} + \frac{ \sw(\x_{N \mi 2}^{\stx}, \rmkk{0},0)}{2}   - \Pi_{N\mi 2}, 
$$
%where $\x_{N \mi 2}^{\st} \eq (x_1^{\st}, \dii ,x_{N \mi 2}^{\st} )$, and  $\Pi_{N\mi 2} \eq
%\sum_{n=1}^{N\mi2} \pa_n^{\st} $. 
which is derived from (\ref{eq:NBS-seq-N-um}) directly, by replacing $x_{N-1}^{\st}$ and $\pa_{N-1}^{\st}$ with $0$ (i.e., not reaching   agreement).

\noindent
\underline{2.~Agreement}:
If they reach an agreement $\paNr = \paxx $ (and $x_{N-1} = x^*_{N-1}$), 
%their payoffs are, respectively,
% when an agreement $\paNr$ (and~$x^*_{N\mi1}$) is reached,  the APO $N- 1$'s payoff is
%$$
%\textstyle
%\Ua_{N\mi 1} = \paNr,
%$$
the APO's payoff is $v$, and the MNO's payoff is its potential payoff \textit{after having dealt with all APOs}, which is exactly given by (\ref{eq:NBS-seq-N-um}), i.e., $\Ua_{N\mi 1} = \paxx$ and 
%\begin{multline*}
$$\textstyle
\Um_{[N\mi 1]} = \frac{ \sw(\x_{N \mi 2}^{\stx}, \rmkk{x_{N\mi 1}^{\stx}}, x_N^{\stx})}{2}+ \frac{ \sw(\x_{N \mi 2}^{\stx}, \rmkk{x_{N\mi 1}^{\stx}},0)}{2}
- \Pi_{N\mi 2} - \rmkk{\paxx}.
$$
%\end{multline*}

%\com{use multline environment (as above) to break a long equation.}

\noindent
\underline{3.~Payoff gain}: Under an agreement $\paNr = \paxx $ (and $x_{N-1} = x^*_{N-1}$), the  payoff gains of APO $N-1$ and the MNO are, respectively, $\Ua_{N\mi 1} - \Ua_{N\mi 1}^0 = \paxx $ and
$$
\Um_{[N\mi 1]} - \Um^0_{[N\mi 1]} =\adw_{N\mi 1} - \paxx,
$$
%is $\Ua_{N\mi 1} - \Ua_{N\mi 1}^0 = \paNr $, and the macrocell MNO's payoff gain is $\Um - \Um^0 =\adw_{N\mi 1} - \paNr$, 
where $\adw_{N\mi 1} =  \frac{{\dw}_{N\mi 1}(I_N\ei1) + {\dw}_{N\mi 1}(I_N\ei0)}{2}
$, and $
\textstyle
{\dw}_{N\mi 1}(I_N ) \eq \sw(\x_{N \mi 2}^{\stx}, \rmkk{x_{N\mi 1}^{\stx}}, I_N  x_N^{\stx}) - \sw(\x_{N \mi 2}^{\stx}, \rmkk{0},I_N   x_N^{\stx}).$
%\begin{equation*}
%\begin{aligned}
%%\adw_{N\mi 1} \eq  & \textstyle  \frac{ \sw(\x_{N \mi 2}^{\stx}, \rmkk{x_{N\mi 1}^{\stx}}, x_N^{\stx})}{2} + \frac{ \sw(\x_{N \mi 2}^{\stx}, \rmkk{x_{N\mi 1}^{\stx}},0)}{2}
%%-
%%\frac{ \sw(\x_{N \mi 2}^{\stx},\rmkk{0}, x_N^{\stx}) }{2} - \frac{ \sw(\x_{N \mi 2}^{\stx}, \rmkk{0},0)}{2}
%%\\
%%=  & \textstyle  \frac{ \sw(\x_{N \mi 2}^{\stx}, \rmkk{x_{N\mi 1}^{\stx}}, x_N^{\stx}) -  \sw(\x_{N \mi 2}^{\stx},\rmkk{0}, x_N^{\stx})}{2} +\frac{\sw(\x_{N \mi 2}^{\stx}, \rmkk{x_{N\mi 1}^{\stx}},0)- \sw(\x_{N \mi 2}^{\stx}, \rmkk{0},0)}{2}
%\adw_{N\mi 1} \eq  
%%& \textstyle  \frac{ \sw(\x_{N \mi 2}^{\stx}, \rmkk{x_{N\mi 1}^{\stx}}, x_N^{\stx}) -  \sw(\x_{N \mi 2}^{\stx},\rmkk{0}, x_N^{\stx})}{2} 
%%\\
%% &
%% \textstyle +  \frac{\sw(\x_{N \mi 2}^{\stx}, \rmkk{x_{N\mi 1}^{\stx}},0)- \sw(\x_{N \mi 2}^{\stx}, \rmkk{0},0) }{2}
%%\\
%%= &
%\textstyle
%\frac{{\dw}_{N\mi 1}(I_N\ei1) + {\dw}_{N\mi 1}(I_N\ei0)}{2} ,
%\end{aligned}
%\end{equation*}
%and 
%$$ 
%\textstyle
%{\dw}_{N\mi 1}(I_N ) \eq \sw(\x_{N \mi 2}^{\stx}, \rmkk{x_{N\mi 1}^{\stx}}, I_N  x_N^{\stx}) - \sw(\x_{N \mi 2}^{\stx}, \rmkk{0},I_N   x_N^{\stx}).$$ 

Notice that ${\dw}_{N\mi 1}(I_N ) $ denotes the  {marginal social welfare} generated by involving APO $N-1$ in the offloading under a particular indicator $I_N$, 
%under the circumstance that the MNO  will ($I_N\ei1$) or not ($I_N\ei0$) reach agreement with APO $N$,
where $I_N\in \{0,1\}$ indicates the \emph{virtual possibility} of whether the MNO will reach agreement with APO $N$.
%\footnotesc{Note that  as the Step $N$ analysis shows, the MNO  will always reach an agreement with APO $N$. This means that the possibility of ``not reaching an agreement with APO $N$'' is only  a \emph{virtual} possibility, and is used to help us understand the intuitions of the NBS in Step $N-1$. 
%%and never actually realizes. 
%%This interpretation helps us to gain some intuitions of the bargaining solution.
%}
Thus, $\adw_{N\mi 1}$ can be viewed as the \emph{expected} marginal social welfare  generated by involving APO $N- 1$ in the offloading, assuming that the MNO has reached agreements with APOs $1$ to $N-2$, and will reach an agreement with APO $N$ with a probability of 0.5. We call $\adw_{N\mi 1}$ as the \textbf{virtual} marginal social welfare generated by APO $N-1$. {For a better understanding, we illustrate the structure of the virtual marginal social welfare $\adw_{n}$ in  \cite{report}.}

\noindent
\underline{4.~Bargaining solution}: 
By Definition \ref{def:NBS}, the NBS between the MNO and the APO $N-1$ is given by
\begin{equation}\label{eq:NBS-seq-N-1-xxxxaaaa}
\max_{ \paxx}  \  \big(\adw_{N\mi 1} - \paxx \big) \cdot \paxx, \quad  
\mbox{s.t.}~\adw_{N\mi 1}  - \paxx \geq 0,\ \paxx \geq 0.
\end{equation}
%Similar to the analysis in Step $N$, the agreement that MNO and AP $N-1$ will reach is $\paxx^* = \frac{\adw_{N\mi 1}}{2}$. 
%the NBS for APO $N-1$ is given by
Similarly, solving the above problem,  we have the following NBS for the bargaining between  the MNO and   APO $N-1$.~~~~~~
%\begin{equation}\label{eq:NBS-seq-N-1}
%\begin{aligned}
%\max_{ \paNr} & \textstyle \ \big( \adw_{N\mi 1} - \paNr \big) \cdot \paNr \\
%\mbox{s.t. } & \ \adw_{N\mi 1}  - \paNr \geq 0,\ \paNr \geq 0.
%\end{aligned}
%\end{equation}
%where $\adw_{N\mi 1} \eq \frac{1}{2} \cdot {\dw}_{N\mi 1}(I_N\ei1)   + \frac{1}{2}\cdot {\dw}_{N\mi 1}(I_N\ei0)   $,
%and
%${\dw}_{N\mi 1}(I_N ) \eq$ $ \sw(\x_{N \mi 2}^{\stx}, \rmkk{x_{N\mi 1}^{\stx}}, I_N  x_N^{\stx}) - \sw(\x_{N \mi 2}^{\stx}, \rmkk{0},I_N   x_N^{\stx})
%$.

%{Here $I_N\in \{0,1\}$ is used to indicate the \emph{virtual possibility} of whether the MNO  will reach an agreement with APO $N$}.\footnotesc{{Note that the possibility of ``not reaching an agreement with APO $N$'' only serves as a \emph{virtual possibility} for the current bargaining with  APO $N\mi 1$, {and helps us to gain some intuitions of the current bargaining solution.} This never actually realizes based on our analysis in Step $N$.}}
%That is, $ {\dw}_{N\mi 1}(I_N )$ is the marginal social welfare generated
%by APO $N\mi1$, suppose the MNO  will ($I_N\ei1$) or will not ($I_N\ei0$) reach an agreement with APO $N$.
%Thus, $\adw_{N\mi 1}$ can be viewed as the \textbf{virtual marginal social welfare} generated by APO $N\mi 1$ under such a situation that the MNO  and APO $N$ will reach an agreement with a probability of 0.5.

%Similarly, we have the following NBS for APO $N\mi 1$.
\begin{lemma}[NBS in Step $N - 1$]\label{lemma:NBS-seq-N-1}
The NBS between the MNO  and APO $N- 1$ in Step $N-1$ is
\begin{equation}
\textstyle
\paNr^* = \paxx^* = \frac{\adw_{N\mi 1}}{2}
%= \frac{ \adw_{N\mi 1} }{2} 
= \frac{  {\dw}_{N\mi 1}(I_N= 1)   + {\dw}_{N\mi 1}(I_N= 0)}{4}.
\end{equation}
In addition, under the NBS, the MNO's payoff is
\begin{equation}
\begin{aligned}
%\label{eq:NBS-seq-N-1-pa}
%\textstyle
%\Ua_{N\mi 1}^* ~&\textstyle =  \frac{ {\dw}_{N\mi 1}(I_N=1) + {\dw}_{N\mi 1}(I_N=0)}{4},
%\\
\label{eq:NBS-seq-N-1-um}
\textstyle
\Um^{*}_{[N\mi 1]} & \textstyle = \Um^0_{[N\mi 1]}  + \frac{\adw_{N\mi 1}}{2}
\\
&\textstyle = \frac{ \mw_{N\mi 1}(I_N=1) +  \mw_{N\mi 1}(I_N=0)}{4} - \Pi_{N\mi 2},
\end{aligned}
\end{equation}
where 
${\mw}_{N {-} 1}(I_N )  \mbox{=} \sw(\x_{N  {-} 2}^{\stx}, \rmkk{x_{N {-} 1}^{\stx}}, I_N  x_N^{\stx}) \mbox{+} \sw(\x_{N  {-} 2}^{\stx}, \rmkk{0},I_N   x_N^{\stx})$.
%${\mw}_{N\mi 1}(I_N ) $
%\\
%$
%~~~~~~~~~~\eq \sw(\x_{N \mi 2}^{\stx}, \rmkk{x_{N\mi 1}^{\stx}}, I_N  x_N^{\stx}) + \sw(\x_{N \mi 2}^{\stx}, \rmkk{0},I_N   x_N^{\stx}).
%$
\end{lemma}

%\begin{proof}
%The proof is same as that for Lemma \ref{theorem:NBS-single}.
%\end{proof}

%\revt{
Similarly, the MNO and APO $N- 1$ equally share the virtual marginal social welfare $\adw_{N\mi 1}$ generated by involving APO $N- 1$ in the offloading.

\vspace{3mm}

%\subsubsection*{\textsl{\bfseries  Step~~${n}$}}
\textbf{Step $n$,~~$\forall n\in\{1,\dii,N-2\}$.} 

\vspace{1mm}

Now we consider the bargaining between the MNO  and APO $n$ in a generic Step $n$, where the MNO has reached bargaining solutions $\{\pa_1^{\st},\dii, \pa_{n-1}^{\st}\}$ with all APOs $1 $ to $n-1$.
\revjr{By induction, we have the following NBS for the bargaining between the MNO and  an arbitrary APO $n$.}
\begin{lemma}[NBS in Step $n$]\label{lemma:NBS-seq-n}
The NBS between the MNO  and APO $n$ in Step $n$  is
%\footnote{We write $\sum_{I_{n}\in\{0,1\}}$ as $\sum_{I_{n} }$ for notational convenience.}
\begin{equation}\label{eq:NBS-seq-N-n-pa}
\textstyle
\pan^* = \frac{ \adw_{n} }{2} =   \sum_{I_{n+1}=0 }^1\dii \sum_{I_{N}=0 }^1	 \frac{{\dw}_{n}(I_{n+1};\dii ; I_N )}{ 2^{N - n+1 } },
\end{equation}
where ${\dw}_{n}(I_{n+1};\dii  ; I_N ) \mbox{=} \sw(\x_{n \mi 1}^{\stx}, \rmkk{x_{n}^{\stx}}, I_{n+ 1}  {x_{n+ 1}^{\stx}},\dii , I_N  x_N^{\stx}) 
- \sw(\x_{n \mi 1}^{\stx}, \rmkk{0}, I_{n+ 1}  {x_{n+ 1}^{\stx}},\dii , I_N  x_N^{\stx})
$.
\\
In addition, under the NBS,  the MNO's payoff is
\begin{align}
%\label{eq:NBS-seq-N-n-pa}
%\textstyle
%\Ua_n^* ~ & \textstyle =  \sum_{I_{n+1} }\dii \sum_{I_{N} }	 \frac{{\dw}_{n}(I_{n+1};\dii ; I_N )}{  2\cdot 2^{N- n}},
%\\
\label{eq:NBS-seq-N-n-um}
\textstyle
\Um^{*}_{[n]}& \textstyle 
%= \Um^0_{[n]}  + \frac{\adw_{n}}{2} 
= \sum_{I_{n+1} =0 }^1\dii \sum_{I_{N} =0 }^1	 \frac{{\mw}_{n}(I_{n+1};\dii ; I_N )}{  2^{N- n+1}} - \Pi_{n\mi 1},
\end{align}
where ${\mw}_{n}(I_{n+1};\dii  ; I_N )  \mbox{=} \sw(\x_{n \mi 1}^{\stx}, \rmkk{x_{n}^{\stx}}, I_{n+ 1}  {x_{n+ 1}^{\stx}},\dii , I_N  x_N^{\stx}) 
 + \sw(\x_{n \mi 1}^{\stx}, \rmkk{0}, I_{n+ 1}  {x_{n+ 1}^{\stx}},\dii , I_N  x_N^{\stx})
$.
%$\x_{n \mi 1}^{\st} \eq (x_1^{\st}, \dii ,x_{n \mi 1}^{\st} )$, and  $\Pi_{n\mi 1} \eq
% \pa_1^{\st}+\dii + \pa_{n\mi1}^{\st} $. 
\end{lemma}
%$$
%\pan^* =   \sum_{I_{n} \in \{0,1\}, n=n+1,\dii ,N } \frac{{\dw}_{n}(I_{n+1};\dii ; I_N )}{ 2^{N\mi n}\cdot2}
%.$$

%\revjr{
%\begin{proof}
%We can prove the lemma by induction. 
%%By induction, we can prove the lemma by showing that (i) the NBS $\paN^*$ in the last Step $N$ (for APO $N$) is characterized by (\ref{eq:NBS-seq-N-n-pa}), and (ii) if the NBS $\{\pa^*_i\}_{i=k, k+1,...,N}$  after Step $k-1$ (for APOs $k,k+1,...,N$) are all characterized by (\ref{eq:NBS-seq-N-n-pa}), then the NBS $\pa_{k-1}^*$ in Step $k-1$ (for APO $k-1$) is also characterized by (\ref{eq:NBS-seq-N-n-pa}). 
%%For the detailed and complete proof, 
%Please refer to Appendix \ref{app:proof-NBS-seq-n} of \cite{report} for the detailed proof. 
%\end{proof}
%}

Similarly,
$ {\dw}_{n}(I_{n+1};\dii ; I_N )$ denotes the  {marginal social welfare} generated
by involving APO $n$ in the offloading, under a set of indicators $I_{n+1}, ..., I_N$, each associated with an APO in $\{n+1,...,N\}$.
%under the circumstance that the MNO will ($I_i=1$) or  not ($I_i=0$) reach agreement with each later APO $i \in \{n+1,...,N\}$,
%; and
%$I_i \in \{0,1\} $
% is used to indicate the {virtual possibility} of whether the MNO  has reached an agreement with APO $i$ before this step.
%\com{I notice that you defined some new latex commands here. I changed some of them, but leave most other untouched. Although this is not essential, typically we can just the regular latex symbols as they have been optimized for different environments (texts, inline math, display math, etc.)}
Thus, $\adw_{n}$ can be viewed as the \textbf{virtual} marginal social welfare generated by involving APO $n$ in the offloading, assuming that the MNO has reached agreements with APOs $1$ to $n-1$, and will reach agreement with each APO $i \in \{n+1,...,N\}$ with a probability of 0.5.
Obviously, the MNO and APO $n$ equally share the virtual marginal social welfare $\adw_{n}$ generated by  APO $n$.

By the above analysis, we can obtain the following NBS for the sequential bargaining (denoted by S-NBS).

\begin{theorem}[{Sequential Bargaining Solution - S-NBS}]\label{theorem:NBS-seq}
The NBS 
$\{\x^*, \bpa^*\}$ under the sequential bargaining is
%(i) $x_n^* = x_n^o$, and (ii) $\pa_n^* =   \sum_{I_{n+1} }\dii \sum_{I_{N} }	 \frac{{\dw}_{n}(I_{n+1};\dii ; I_N )}{ 2^{N\mi n}\cdot2}, \forall n$=$1,\dii ,N$.
\begin{enumerate}
\item[(a)] $x_n^* = x_n^o,\ \forall n\in\N$;
\item[(b)] $\pa_n^* =   \sum_{I_{n+1} =0}^1\dii \sum_{I_{N} =0}^1	 \frac{{\dw}_{n}(I_{n+1};\dii ; I_N )}{ 2^{N- n+1}},\ \forall n\in\N$.
\end{enumerate}
\end{theorem}

%\begin{proof}
%By Lemma \ref{prop:nbs-x} and Lemma \ref{lemma:NBS-seq-n}, We can prove the theorem directly. 
%\end{proof}

%\subsubsection*{\textsl{\bfseries  Properties of S-NBS}}

Next we provide some useful properties for the S-NBS. For more detailed discussions, please refer to \cite{report}.

\begin{property}[{Early-Mover Advantage}]\label{prop:EMA-seq}
Under the sequential bargaining, 
an APO will obtain a higher payoff, 
if it bargains with the MNO earlier.
\end{property}

\begin{property}[{Invariance to APO-order Changing}]\label{prop:IOC-seq}
Under the sequential bargaining, the bargaining order of APOs does not affect the MNO's payoff.
\end{property}

%This property can be proved by (\ref{eq:NBS-seq-N-n-um}), which shows that 
%\begin{equation}\label{eq:NBS-seq-N-1-um}
%\begin{aligned}
%\Um^{*}_{[1]} &
%\textstyle = \sum_{I_{2} =0}^1\dii \sum_{I_{N} =0}^1	 \frac{{\mw}_{1}(I_{2};\dii ; I_N )}{ 2^{N}}  - \Pi_{0}
%\\
%&\textstyle = \sum_{I_{1}  =0}^1\sum_{I_{2} =0}^1\dii \sum_{I_{N}  =0}^1	 \frac{{\mw}_{0}(I_{1};\dii ; I_N )}{  2^{N}} ,
%\end{aligned}
%\end{equation}
%where
%${\mw}_{0}(I_{1};\dii  ;I_N ) \eq \sw( {I_1 {x_{1}^{\stx}}}, I_{2}  {x_{2}^{\stx}},\dii , I_N  x_N^{\stx})$. The 2nd equality follows because $\Pi_{0} = 0$ and $ {\mw}_{1}(I_{2};\dii  ;I_N ) =\sum_{I_1=0}^1 {\mw}_{0}( {I_1};I_{2}; \dii  ;I_N )$. 
%Intuitively, the MNO's payoff given in (\ref{eq:NBS-seq-N-1-um}) can be viewed as the \textbf{expected social welfare}, assuming that the MNO will reach an agreement with each APO with a probability of 0.5.
%From (\ref{eq:NBS-seq-N-1-um}), we can easily find that the bargaining order of APOs does \emph{not} affect the MNO's payoff in the S-NBS.

%The detailed  proofs and discussions can be found in \cite{report}.

\subsection*{B.2)~~Concurrent Bargaining}
%\label{sec:barg:con}

We now study the NBS under \emph{concurrent bargaining}, where the MNO bargains with APOs concurrently (see Figure \ref{fig:bargaining-scheme} (b)). Namely, $N$ one-to-one bargainings happen simultaneously.~~~~~~
%Similarly, we assume that there is no group in APOs, i.e., APOs bargain individually and concurrently with the MNO.

%Obviously, in this case, all APOs are symmetric (in bargaining order).

%\subsubsection*{\textsl{\bfseries  Bargaining with APO $n$}}

Without loss of generality, we consider the bargaining between the MNO  and an APO $n$ for $\pi_n$ (or $z_n$, equivalently). 
For the convenience in writing, we introduce notations:
$$
\x_{\mi n} \eq (x_1,\dii,x_{n\mi 1},x_{n\ad 1},\dii,x_N),
$$
$$
\bpa_{\mi n} \eq (\pa_1,\dii,\pa_{n\mi 1},\pa_{n\ad 1},\dii,\pa_N),
$$
$$
\textstyle
\Pi_{\mi n} \eq
\sum_{i\in\N,i\neq n} \pa_i.~~~~~~~~~~~~~~~~~
$$
for the analysis of the concurrent bargaining.
%Let $\bpa_{\mi n}^{\st} \eq \{\pa_i^{\st}\}_{i \in \N, i\neq n}$ and $
%\x_{\mi n}^{\stx} \eq \{x_i^{\stx}\}_{i \in \N, i\neq n}$ denote the bargaining solutions between the MNO and all APOs other than $n$.
%Let $\Pi_{\mi n} \eq
%\sum_{i\in\N,i\neq n} \pa_i^{\st} $ denote the total payment to all APOs other than $n$.
%Denote $ \x_{\mi n}^{\stx} \eq \{x_i^{\stx}\}_{i \in \N, i\neq n}$ as the traffic offloading profile.

\vspace{2mm}

\noindent
\underline{1.~Disagreement}:
If the MNO and APO $n$ do not reach an agreement,
then APO $n$'s disagreement point is $ 0 $,
and the MNO's disagreement point is its payoff after \textit{finishing all $N-1$ concurrent one-to-one bargainings  with other APOs}, i.e.,
$$ 
\textstyle \Ua_N^0  = 0,
\mbox{~~~~}
\Um^0_{[n]}= \sw(\x_{\mi n}^{\stx}, 0) - \Pi_{\mi n}.
$$
%where
%$(\x_{\mi n}^{\stx}, 0) = ( \dii ,x_{n \mi 1}^{\stx}, 0 , x_{n + 1}^{\stx},\dii  )$, and

\noindent
\underline{2.~Agreement}:
%when an agreement $\pan$ (and $x_n^{\stx}$) is reached, 
If they reach an agreement $\pan = \paxx $ (and $x_n = x^*_n$), 
then APO $n$'s payoff is $\paxx$, 
and the MNO's payoff is its payoff after \textit{finishing all concurrent one-to-one bargainings with all APOs}, i.e.,
$$
\textstyle
~~\Ua_n = \paxx,
\mbox{~~~~}
\Um_{[n]} = \sw(\x_{\mi n}^{\stx}, \rmkk{x_n^{\stx}}) - \Pi_{\mi n} - \rmkk{\paxx}.
$$

\noindent
\underline{3.~Payoff gain}: 
Under an agreement $\pan = \paxx $ (and $x_n = x^*_n$), 
the payoff gains for the MNO and APO $n$ are, respectively,
%, the APO $n$'s payoff gain and the MNO's payoff gain are, respectively,
$$
\textstyle
\Ua_n - \Ua_n^0 = \paxx,
\mbox{~~~~}
\Um_{[n]} - \Um^0_{[n]} =  \dwc_n - \paxx,
$$
%is $\Um - \Um^0 = \sw(\x_{ \mi n}^{\stx}, x_n^{\stx}) - \sw(\x_{  \mi n}^{\stx}, 0) - \pan \eq \dwc_n - \pan$, 
where
%\begin{equation*}
%\textstyle
$\dwc_n \eq \sw(\x_{ \mi n}^{\stx}, \rmkk{x_n^{\stx}}) - \sw(\x_{  \mi n}^{\stx}, \rmkk{0})$
%\end{equation*}
denotes the  {marginal social welfare} generated by involving APO $n$ into the offloading, assuming that the MNO has reached (or will reach) agreements with \emph{all} other APOs.

\noindent
\underline{4.~Bargaining solution}: 
Similar to the analysis for the sequential bargaining, the agreement that the MNO and AP $n$ will reach is $\paxx^* = \frac{\dwc_n}{2}$. 
Thus, the NBS between the MNO and APO $n$ is the following.
%
%
%by definition, we can easily obtain the NBS between the
%MNO and the APO $n$. Formally,
%the NBS with APO $n$ is given by
%\begin{equation}\label{eq:NBS-con-n}
%\begin{aligned}
%\max_{ \pan} & \ \big(\dwc_n - \pan \big) \cdot \pan \\
%\mbox{s.t. } & \ \dwc_n  - \pan \geq 0,\ \pan \geq 0.
%\end{aligned}
%\end{equation}
%where $\dwc_n \eq \sw(\x_{  \mi n}^{\stx}, \rmkk{x_n^{\stx}})- \sw(\x_{\mi n }^{\stx}, \rmkk{0})$ is the \emph{increase} of {social welfare} by offloading $ x_n^{\stx}$ units of traffic to APO $n$, assuming that the MNO  will reach agreement with all other APOs (also called the \textbf{marginal social welfare} generated by APO $n$).

%Similarly, the objective function of (\ref{eq:NBS-con-n}) is a quadratic func-tion of $\pan$. Thus, we have the following NBS for APO $n$.
\begin{lemma}[NBS with APO $n$]\label{lemma:NBS-con}
The NBS between the MNO and APO $n$ under concurrent bargaining is
\begin{equation}
\label{eq:NBS-con-n-pa}
\textstyle
\pan^* =\paxx^* = \frac{ \dwc_n }{2} = \frac{\sw(\x_{\mi n}^{\stx}, \rmkk{x_n^{\stx}}) - \sw(\x_{\mi n}^{\stx}, \rmkk{0})}{2}.
% = \frac{\sw(\x_{ \mi n}^{\stx}, x_n^{\stx}) - \sw(\x_{  \mi n}^{\stx}, 0)}{2}.
\end{equation}
%where $\dwc_n \eq \sw(\x_{\mi n}^{\stx}, \rmkk{x_n^{\stx}}) - \sw(\x_{\mi n}^{\stx}, \rmkk{0})$.
%\\
In addition, under the NBS, the MNO's payoff is
\begin{equation}
%\textstyle
%\Ua_n^*= \frac{ \dwc_n }{2},
%\mbox{~~~~}
\textstyle
\Um^{*}_{[n]} =\Um^0_{[n]} + \frac{ \dwc_n }{2} = \frac{\sw(\x_{\mi n}^{\stx}, \rmkk{x_n^{\stx}}) + \sw(\x_{\mi n}^{\stx}, \rmkk{0})}{2}  - \Pi_{\mi n} .
\end{equation}
%where $\mwc_n \eq \sw(\x_{\mi n}^{\stx}, \rmkk{x_n^{\stx}})+\sw(\x_{\mi n}^{\stx}, \rmkk{0})$.
\end{lemma}
%Accordingly, the APO $n$'s payoff and the MNO's payoff under the NBS are, respectively,
%Obviously, the MNO  and APO $n$ share equally the marginal social welfare  generated.
%\begin{align}
%\label{eq:NBS-con-n-pa}
%\textstyle
%\Ua_n^*~ &\textstyle =
%\frac{ \sw(\x_{ \mi n}^{\stx}, x_n^{\stx})- \sw(\x_{\mi n}^{\stx}, 0)}{2} = \frac{ \dwc_n }{2},
%\\
%\label{eq:NBS-con-n-um}
%\textstyle
%\Um^{*}_{[n]} &\textstyle  =  \frac{ \sw(\x_{\mi n}^{\stx}, \rmkk{x_n^{\stx}})+\sw(\x_{\mi n}^{\stx}, \rmkk{0}) }{2}  - \Pi_{\mi n}= \frac{ \mwc_n }{2}  - \Pi_{\mi n}  ,
%\end{align}
%where $\mwc_n \eq \sw(\x_{\mi n}^{\stx}, \rmkk{x_n^{\stx}})+\sw(\x_{\mi n}^{\stx}, \rmkk{0})$.

%\begin{proof}
%The proof is same as that for Lemma \ref{theorem:NBS-single}. 
%\end{proof}

%\subsubsection*{\textsl{\bfseries  Bargaining Solution: C-NBS}}
%Similarly, the MNO and APO $n$ equally share the  marginal social welfare $\dwc_n $ generated by involving APO $n$ in the offloading.
Similarly, we can obtain the following NBS for the concurrent bargaining (denoted by C-NBS).

\begin{theorem}[{Concurrent Bargaining Solution - C-NBS}]\label{theorem:NBS-con}
The NBS
$\{\x^*, \bpa^*\}$ under the concurrent  bargaining is 
\begin{enumerate}
\item[(a)] $x_n^* = x_n^o,\ \forall n= 1,... ,N$;
\item[(b)] $\pa_n^* =   \frac{ \dwc_n }{2}=
\frac{ \sw(\x_{ \mi n}^{\stx}, \rmkk{x_n^{\stx}})- \sw(\x_{\mi n}^{\stx}, \rmkk{0})}{2},\ \forall n=1,... ,N$.
\end{enumerate}
\end{theorem}
 
%\begin{proof}
%By Lemma \ref{prop:nbs-x} and Lemma \ref{lemma:NBS-con}, we can  prove the theorem  directly. 
%\end{proof}

%\subsubsection*{\textsl{\bfseries  Properties of C-NBS}}

%Next we capture some useful properties of the above concurrent bargaining solution C-NBS.
Next we provide some useful properties for the C-NBS. 
For more detailed discussions, please refer to \cite{report}.

%First, from (\ref{eq:NBS-con-n-pa}), we can easily find  that unlike the sequential bargaining,
%each APO $n$'s payoff under the concurrent bargaining is independent of its index $n$.
\begin{property}[{Invariance to AP-index Changing}]\label{prop:IIC-con}
The APO-index has no impact on the APO's payoff under the concurrent bargaining.\footnotesc{Since there is no concept of  ``order'' under the concurrent bargaining, we use the term ``index'' to distinguish APOs. Note that under the sequential bargaining, the term ``index'' is equivalent to the term ``order''.}
\end{property}

%Second, comparing equations (\ref{eq:NBS-con-n-pa}) and (\ref{eq:NBS-seq-N-n-pa}),
%we observe that the payoff of APO $n$ under the concurrent bargaining equals to its payoff under the sequential bargaining where it is the last bargainer, which, by Property \ref{prop:EMA-seq}, is the worst-case payoff that it can obtain under the sequential bargaining.
%we can further find that the last APO $N$ achieves the same payoff in the sequential bargaining and concurrent bargaining, while all APOs prior to $N$ achieve higher payoffs in the sequential bargaining.
%we can further find that each APO $n$'s payoff under the concurrent bargaining is exactly same as that under the sequential bargaining when it is the last bargainer (i.e., the worst case).
%we can further find that the payoff of each APO $n$ under the concurrent bargaining is always smaller than (or equal to) that under the sequential bargaining, regardless of the order of bargaining.
%This is because $\dwc_n \leq {\dw}_{n}(I_{n+1};\dii ; I_N )$ for any values of $I_{n+1},...,I_N$, and the equality only holds when $I_{n+1}=\dii =I_N =1$.
%Intuitively, the concurrent bargaining of APOs decrease the marginal social welfare of all APOs (comparing with the sequential bargaining), and therefore hurt all APs.
%Formally,
\begin{property}[{Concurrently Moving Tragedy}]\label{prop:CMT-con}
The payoff of   APO under the concurrent bargaining equals to the worst-case payoff that it can achieve under the sequential bargaining.
\end{property}
%
%The detailed   proofs and discussions can be found in \cite{report}.

%\revt{Third, we further notice that sequential and concurrent bargainings achieve the same maximum social welfare, while APOs can always achieve a no worse payoffs under the sequential bargaining as mentioned above. 
%This implies that the MNO can achieve a higher payoff under the concurrent bargaining.}

\subsection{Grouping Effect}\label{sec:barg:group}

So far, we have assumed that \emph{each APO bargains with the MNO  individually}. 
In practice, however, APOs may form groups and  bargain  with the MNO jointly. 
Now we study  the impact of APO grouping on the bargaining solution.\footnote{For better understanding of this grouping effect, we also provide illustrative examples 
%in Section \ref{sec:payoff}.3.
%in Appendix-\ref{app:example-group} of 
in the online technical report \cite{report}.}

% \subsection{Group Bargaining}

%It is notable that the S-NBS  and C-NBS have so far assumed that  \emph{every APO bargains with the MNO  individually}. In this section, we will study whether APOs have the incentive to merge with others (and thereby bargain jointly with the BS), and (if so) what group might emerge finally, and what is the corresponding group bargaining solution.

It is important to note that if multiple APOs form a group, they will bargain with the MNO  as a \emph{single} player. {Namely, the marginal social welfare generated by this ``player'' is the total marginal social welfare generated by all APOs in the group together; the disagreement point is the sum of all associated APOs' disagreement points.}
Thus, once the group is fixed, we can apply the results in Theorems \ref{theorem:NBS-seq} and \ref{theorem:NBS-con} directly, by viewing each APO group as a single virtual player.~~~~~~~~~~~~~~~
%To study the impact of APO group structure, we will take the previous bargaining solution (without group) as benchmark, and characterize the difference in bargaining solution

\begin{figure*}[t]
%\begin{minipage}[t]{1\linewidth}
%   \centering
%   \includegraphics[scale=.5]{Figure/offload_xn_vs_Sn}
%    \vspace{-8mm}
%    \caption{Traffic Offloading Profile vs Traffic Profile $S_n$.} \label{fig:sim-offload-xn-sn}
%    \vspace{2mm}
%\end{minipage}
\begin{minipage}[t]{0.5\linewidth}
   \centering
   \includegraphics[scale=.48]{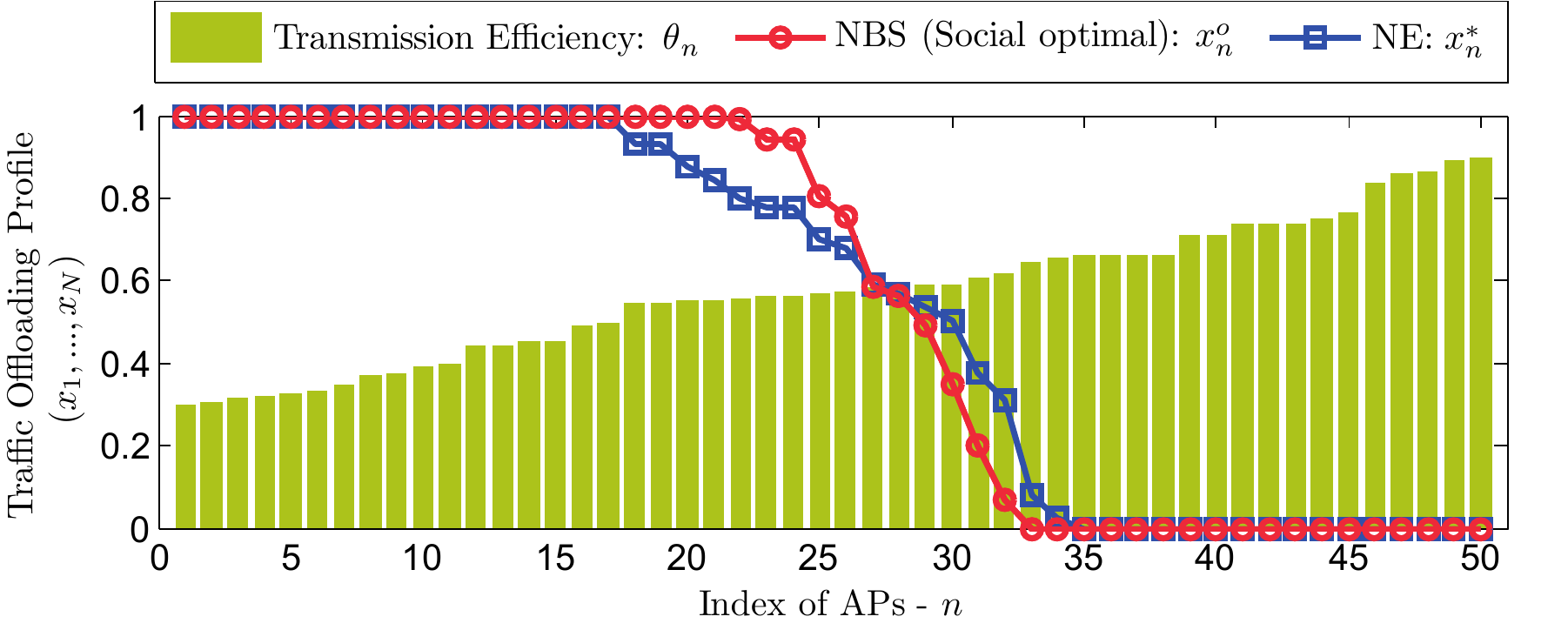}
    \vspace{-2mm}
    \caption{Traffic Offloading Profile vs Transmission Efficiency $\theta_n$.} \label{fig:sim-offload-xn-theta}
    \end{minipage}
\begin{minipage}[t]{0.5\linewidth}
   \centering
   \includegraphics[scale=.48]{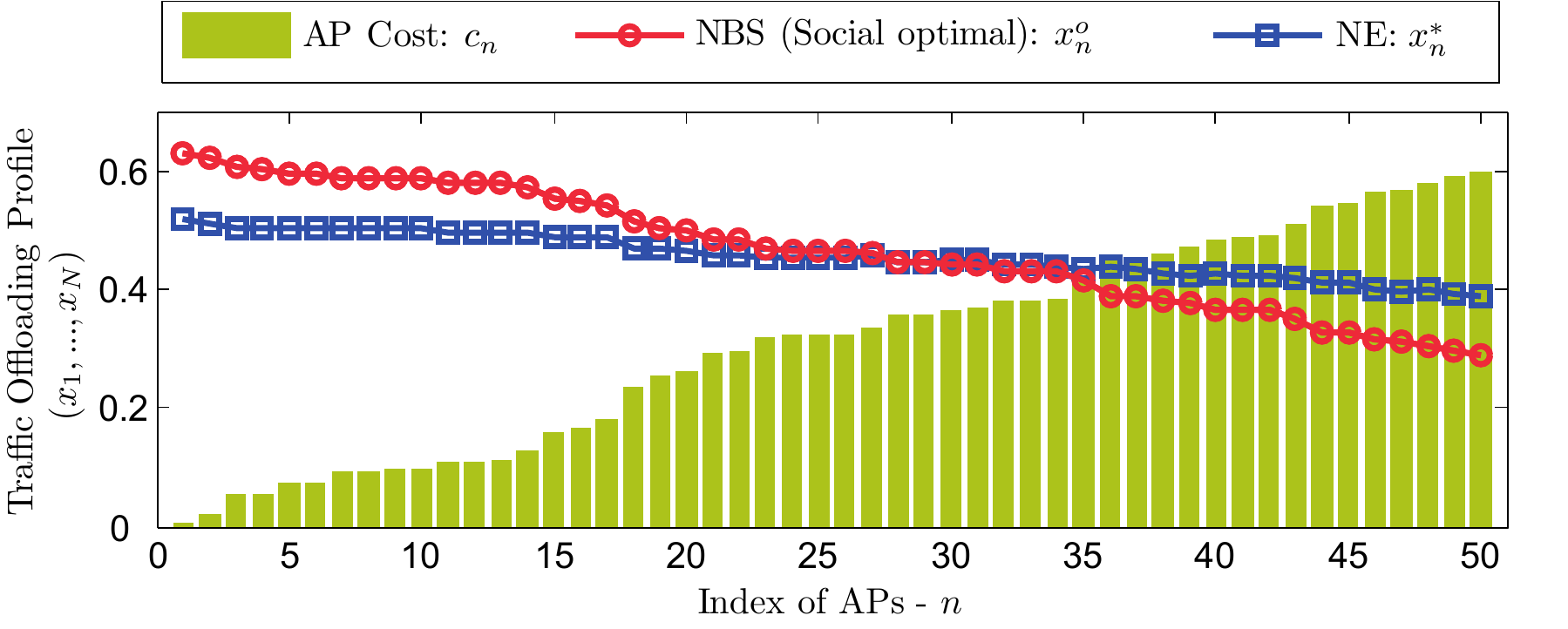}
     \vspace{-2mm}
   \caption{Traffic Offloading Profile vs AP Serving Cost $c_n$.} \label{fig:sim-offload-xn-cn}
\end{minipage}
     \vspace{-3mm}
\end{figure*}

%\subsubsection{{ Grouping Effect in Sequential  Bargaining}}

\subsection*{C.1)~~Grouping Effect in the Sequential  Bargaining}

We consider a simple, yet representative grouping scenario where two successive APOs (say $n-1$ and $n$) form a group.
%\footnotesc{Note that when studying the merge of two non-successive APs, say $n$ and $n\ad 2$, we have to consider the bargaining order of the merged group $\{n, n\ad 2\}$ and the APO between the APOs in the merged group, i.e., $n\ad 1$.}
For notational convenience,
we denote the new player (i.e., the group $\{n-1, n\}$) by $\langle{n}\rangle$.
To keep the indexes of other APOs consistent, we introduce a dummy APO  $\langle{n-1}\rangle$ before $\langle{n}\rangle$, who offloads zero traffic, and receives zero payment.~~~~~~~~~~~

By Theorem \ref{theorem:NBS-seq}, the payoff of new player $\langle{n}\rangle$ (i.e., the group of APOs $n$ and $ n\mi 1$) under the sequential bargaining is
\begin{equation}\label{eq:group-pi-n}
\textstyle
\pa_{\langle{n}\rangle}^* = \frac{ \adw_{\langle{n}\rangle} }{2} =   \sum_{I_{n+1} =0}^1\dii \sum_{I_{N} =0}^1	 \frac{{\dw}_{\langle{n}\rangle}(I_{n+1};\dii ; I_N )}{  2^{N - n +1} },
\end{equation}
where  $\textstyle {\dw}_{\langle{n}\rangle}(I_{n+1};\dii  ;I_N ) \eq $
\begin{equation*}
\begin{aligned}
  & ~~~\textstyle \sw(\x_{n \mi 2}^{\stx},  {x_{\langle{n-1}\rangle}^{\stx}}, \rmkk{ \{x_{n-1}^{\stx}, x_{n}^{\stx} \}},   I_{n+ 1}  {x_{n+ 1}^{\stx}},\dii , I_N  x_N^{\stx} )
\\
& 
\textstyle
~~~~ -  \sw(\x_{n \mi 2}^{\stx},  {x_{\langle{n-1}\rangle}^{\stx}} , \rmkk{\{0,0\}},   I_{n+ 1}  {x_{n+ 1}^{\stx}}, \dii , I_N  x_N^{\stx})
\end{aligned}
\end{equation*}
is the  {marginal social welfare} generated by APOs $n-1$ and $n$ together.
Notice  that $x_{\langle{n-1}\rangle}^{\stx} = 0$ for the dummy APO $\langle{n-1}\rangle$. Thus, we can rewrite the above marginal social welfare as~~~~~~~
\begin{equation*}
\begin{aligned}
{\dw}_{\langle{n}\rangle}(I_{n+1};\dii  ;I_N ) = & \sw(\x_{n \mi 2}^{\stx}, \rmkk{  x_{n-1}^{\stx}, x_{n}^{\stx}  }, I_{n+ 1}  {x_{n+ 1}^{\stx}},\dii , I_N  x_N^{\stx} )
\\
& -  \sw(\x_{n \mi 2}^{\stx}, \rmkk{ 0,0 }, I_{n+ 1}  {x_{n+ 1}^{\stx}}, \dii , I_N  x_N^{\stx}).
\end{aligned}
\end{equation*}
%We further notice that when APOs $n-1$ and $n$ bargain individually with the BS, their payoffs are, respectively,
%\begin{equation*}
%\begin{aligned}
%\pa_{n-1}^* & \textstyle =   \sum_{I_{n} }\dii \sum_{I_{N} }	 \frac{{\dw}_{n-1}(I_{n};\dii ; I_N )}{ 2\cdot2^{N- n+1}},
%\\
%\pa_n^* ~& \textstyle =   \sum_{I_{n+1} }\dii \sum_{I_{N} }	 \frac{{\dw}_{n}(I_{n+1};\dii ; I_N )}{ 2\cdot2^{N- n}}.
%\end{aligned}
%\end{equation*}
%It is easy to check that $
% \frac{{\dw}_{n}(I_{n+1};\dii ; I_N )}{ 2\cdot2^{N- n}} + \sum_{I_{n}}\frac{{\dw}_{n-1}(I_{n};\dii ; I_N )}{ 2\cdot2^{N- n+1}} $ $< \frac{{\dw}_{\langle{n}\rangle}(I_{n+1};\dii ; I_N )}{2 \cdot 2^{N - n} }$. This implies that
%$\pa_{n}^* + \pa_{n+1}^* <
%{\pa}_{ \langle{n}\rangle}^* ,
%$

Comparing  $\pa_{\langle{n}\rangle}^*$ in (\ref{eq:group-pi-n}) with   $\pa_{n-1}^*$ and $\pa_{n}^*$   in (\ref{eq:NBS-con-n-pa}), we can easily see  that  
APOs $ n-1 $ and $ n $ achieve a larger total payoff when forming a group. 
We can further show that this result   holds generally  in our data offloading problem. Formally,

\begin{property}[{Intra-Grouping Benefit}]\label{prop:GBS-seq}
Under the sequential bargaining, grouping of APOs always improves the payoffs of the group members.
\end{property}

By checking the payoffs of APOs other than  $n-1$ and $n$, we can further find that grouping of APOs benefits not only the group members, but also the preceding APOs, i.e., those APOs bargaining before the group. 
This means that the grouping of APOs has the \emph{positive externality} in sequential bargaining.

\begin{property}[{Inter-Grouping Benefit}]\label{prop:SPE-group}
Under the sequential bargaining, 
grouping of APOs improves the payoffs of all APOs  bargaining before the group, 
while does not affect the APOs bargaining after the group.
\end{property}

Based on the above analysis, we can find that grouping of APOs will not hurt any APO.
Since the achieved maximum social welfare does not change, the MNO will achieve a reduced payoff when APOs form groups.
%For more details and proofs, please refer to \cite{report}.

%\subsubsection{Grouping Effect in Concurrent  Bargaining}

\subsection*{C.2)~~Grouping Effect under the Concurrent Bargaining}

With a similar analysis, we can obtain the following results regarding the grouping effect  under the concurrent bargaining.
%: there is grouping benefit for group members themselves, but no externality to other APOs. 
%Here we just list the conclusions, but keep the details in \cite{report}.

\begin{property}[{Intra-Grouping Benefit}]\label{prop:GBS-con}
Under the concurrent bargaining, grouping of APOs always improves the payoffs of the group members.
\end{property}

\begin{property}[{No Inter-Group Benefit}]\label{prop:NOE-group}
Under the concurrent bargaining, grouping of APOs does not affect the APOs not in the group.
\end{property}

\noindent
We can similarly find that under the concurrent bargaining, the MNO will achieve a reduced payoff when APOs form groups.  
%Due to page limits, we put the detailed proofs and discussions in \cite{report}. 

%These two properties indicate that grouping in the concurrent bargaining will benefit the group member only.

%\subsubsection*{Discussion}
%
%
%Notice that when all APOs form a single group,
%the above group bargaining solution coincides with the NBS (Lemma \ref{theorem:NBS-single}) in an one-to-one bargaining.
%\begin{property}
%The G-NBS in the group bargaining is equivalent to the NBS in such an one-to-one bargaining that we treat the APO group as a single player.
%\end{property}
%Besides, our group bargaining coincides with the Chae's group bargaining solution \cite{chae}, in the sense that both the MNO  and the APO group achieve the same payoff  in the both solutions.
%Formally,
%\begin{property}
%The G-NBS  is equivalent to the Chae's group bargaining solution \cite{chae} in terms of the group payoff.
%\end{property}
%Intuitively, our group bargaining treats each group as a single player, and thus the G-NBS   specifies only the group payoff (total payoff of all APs), while not the payoff of each AP.
%Nevertheless, this does not affect the incentive for APOs merging together, because we can generally assume a reasonable payoff distribution among the APOs such that every APO is satisfied with what he obtains (e.g., we can distribute the payoff gain equally among the APs).
%To characterize the explicit payoff distribution among the APs, certain additional axioms are needed (e.g., RGH in \cite{chae}).

%!TEX root = DataOffload_main_journal.tex
%SourceDoc DataOffload_main_journal.tex

\begin{figure*}[t]
\begin{minipage}[t]{.38\linewidth}
   \centering
   \includegraphics[scale=.31]{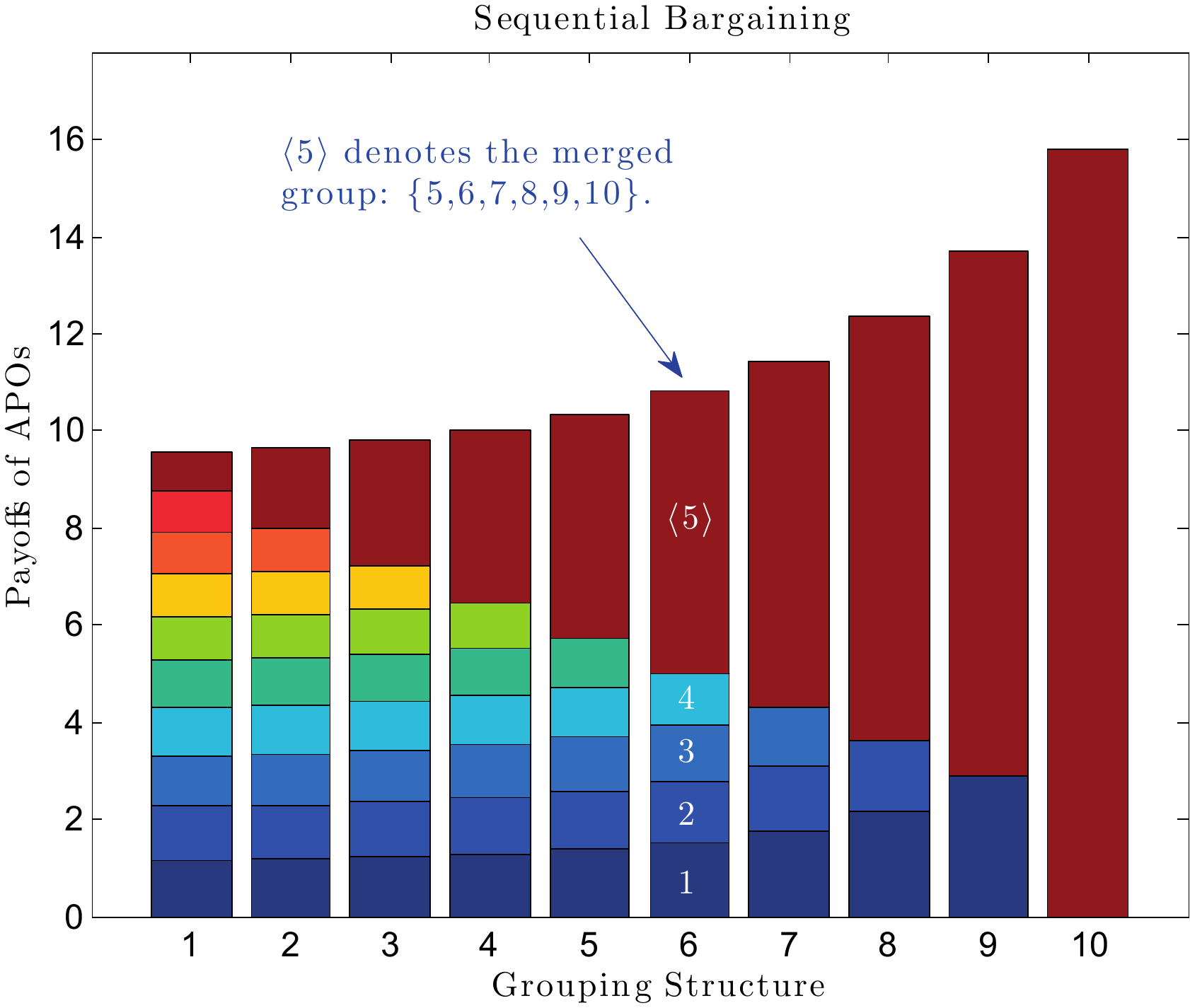}
 \end{minipage}
\begin{minipage}[t]{.38\linewidth}
   \centering
   \includegraphics[scale=.31]{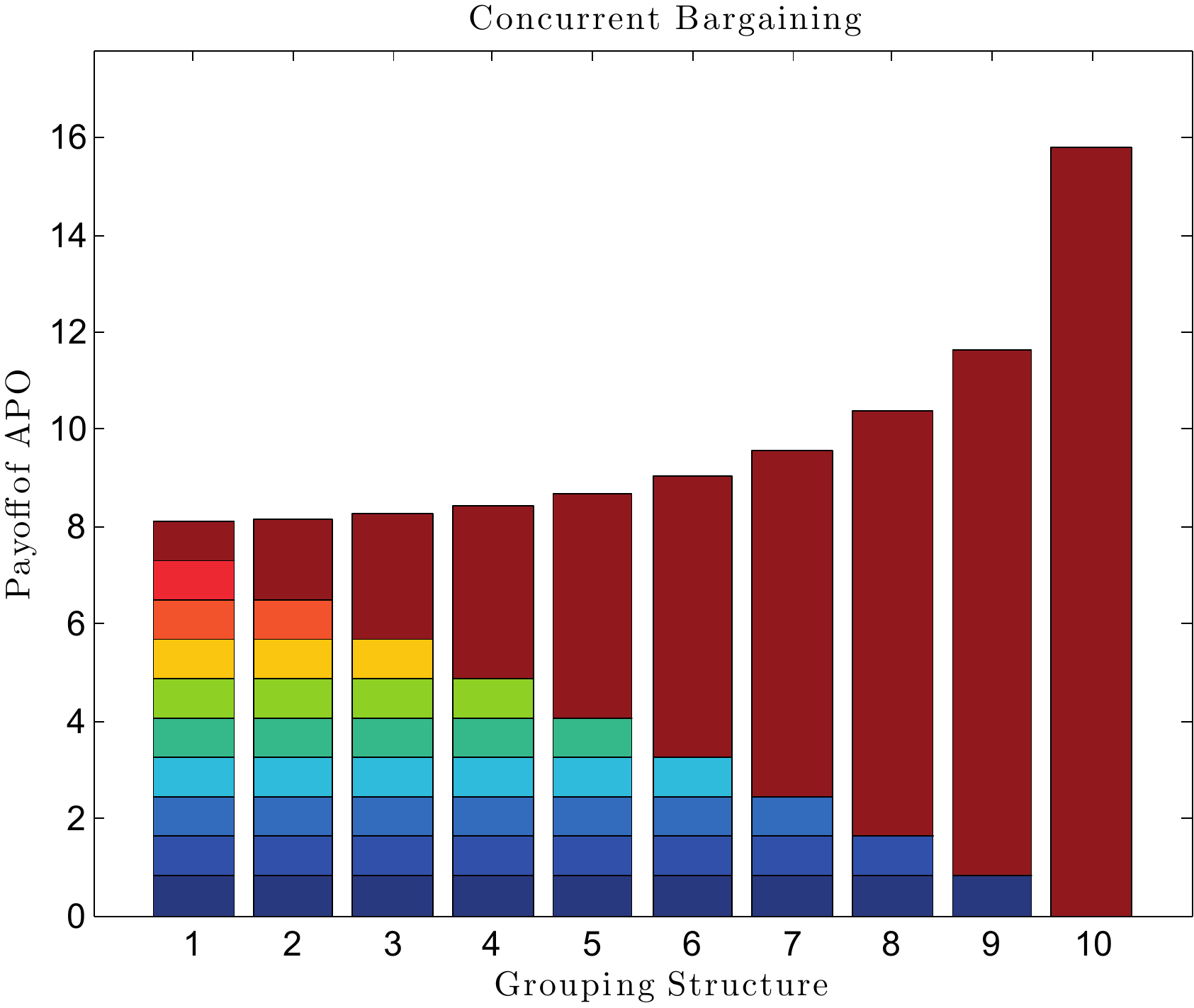}
     \end{minipage}
\begin{minipage}[t]{.19\linewidth}
   \includegraphics[scale=.36]{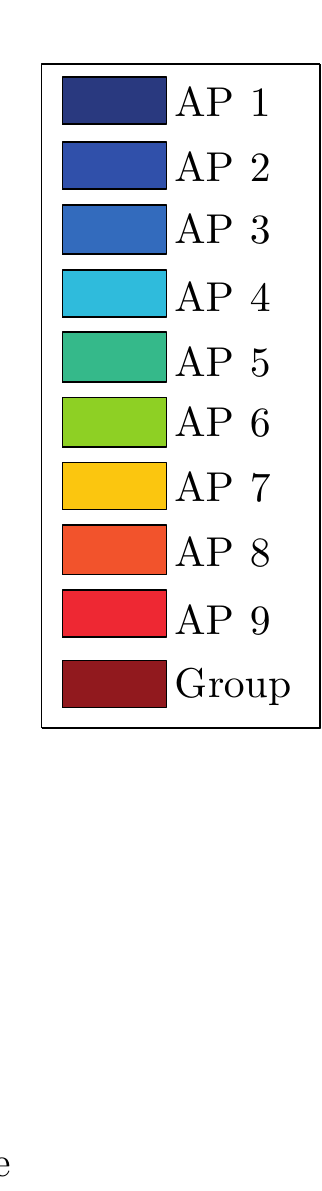}
     \end{minipage}
     \vspace{-2mm}
\caption{The payoffs of APOs under different grouping structures: (a) Sequential Bargaining, (b) Concurrent Bargaining.} \label{fig:sim-group-seq-tail}
     \vspace{-3mm}
\end{figure*}

\section{Simulations}\label{sec:simu}

In these simulations, we assume a typical 3G/4G macrocell with a transmission range of 500m, and $N=50$ WiFi APs (each operated by an APO) with a transmission range of 50m each.
The APs are located at the \emph{hot spots}, i.e., those areas with high MU densities.
The macrocell's bandwidth (resource) is 20MHz, and every AP's effective bandwidth (resource) is \revh{randomly and uniformly chosen from $\{1$, $2$, $5.5$, $11\}$MHz} (fixed within every data offloading period), depending on the interference it experiences. 
Every APO's own   demand follows a   uniform distribution in $[0, 10]$ (Mbps).

The \rev{total} MU density in hot spots is 4 times higher than that in other areas.
There are totally 250 MUs randomly distributed within the macrocell; and thus on average there are 200 MUs in the hot spots (covered by APs), and 50 MUs in areas only covered by the macrocell.
Every MU's traffic is a randomly \revh{and uniformly} selected from $\{0$, $32$, $64$, $128$, $256$, $512\}$Kbps, reflecting different types of  applications.
The MU traffic and AP resource remains unchanged within the period of data offloading (one minute in simulations), while \rev{can change} across periods.

%\begin{figure}[t]
%\begin{minipage}[t]{1\linewidth}
%   \centering
%   \includegraphics[scale=.5]{Figure/offload_xn_vs_Sn}
%    \vspace{-8mm}
%    \caption{Traffic Offloading Profile vs Traffic Profile $S_n$.} \label{fig:sim-offload-xn-sn}
%    \vspace{2mm}
%\end{minipage}
%\begin{minipage}[t]{1\linewidth}
%   \centering
%   \includegraphics[scale=.5]{Figure/offload_xn_vs_theta}
%    \vspace{-8mm}
%    \caption{Traffic Offloading Profile vs Transmission Efficiency $\theta_n$.} \label{fig:sim-offload-xn-theta}
%    \vspace{2mm}
%    \end{minipage}
%\begin{minipage}[t]{1\linewidth}
%   \centering
%   \includegraphics[scale=.5]{Figure/offload_xn_vs_cn}
%    \vspace{-8mm}
%    \caption{Traffic Offloading Profile vs AP Cost $c_n$.} \label{fig:sim-offload-xn-cn}
%\end{minipage}
%\end{figure}

\textbf{Traffic Offloading Profile.}
We first illustrate the traffic offloading profile under the NBS.
It is natural to compare the NBS with other non-cooperative game based solutions such as the Nash equilibrium (NE).
\revjj{To derive this benchmark, we formulate the problem as a Stackelberg game, where the MNO (game leader) proposes the reimbursements first, and then APOs (game followers) respond with the traffic they are willing to offload (see \cite{report} for details).}~~~~~~~~~~~~~~~~~~~~~~~~~~~~~~~~~~~~~~~~~~~

Figures \ref{fig:sim-offload-xn-theta} and \ref{fig:sim-offload-xn-cn} show the traffic offloading profiles in the NBS and the NE under different system parameters. \revh{Notice that the traffic offloading profiles under the  NBS is also the socially optimal solution (see Lemma \ref{prop:nbs-x}).}
%\com{There is one socially optimal, or there are two different ones, one under NBS and one under NE?}
%\com{need to explain what the figures mean first, and then state the insights. x-axis, y-axis.}
\revh{In both figures, the x-axis denotes the indices of APOs, and the y-axis denotes the traffic offloading to each APO.
The bar chart denote the input system parameter, representing 
%the cellular traffic volume in every AP's coverage area, 
the transmission efficiency between each AP and the macrocell BS (in Figures \ref{fig:sim-offload-xn-theta}), and the   serving cost of every AP (in Figures \ref{fig:sim-offload-xn-cn}), respectively.}
%.\com{Need more precisely explanations of inputs of all three subgraphs. }
From these figures we can see that the non-cooperative game solution \rev{(NE)} \rev{significantly} deviates from the cooperative bargaining solution (NBS) in both cases.
In Figures \ref{fig:sim-offload-xn-theta}, the weighted average  difference, i.e., $\frac{\sum_{n=1}^N |x_n^o-x_n^*|}{\sum_{n=1}^N x_n^o}$, is 6.7\%.
In   Figures \ref{fig:sim-offload-xn-cn}, the weighted average difference is 13.4\%.
%From these figures we can see that the non-cooperative game based solution \rev{(NE)} \rev{significantly} deviates from the socially optimal solution \revh{(the weighted average  difference, i.e., $\frac{\sum_{n=1}^N |x_n^o-x_n^*|}{\sum_{n=1}^N x_n^o}$, are  9.6\%, 6.7\%, and 13.4\%, respectively)}.
%\com{how significant? Give a number (percentage)? The deviation is not significant in the first subgraph?}
This implies that \emph{users' non-cooperative choices as in the NE will lead to certain social welfare loss}, which \rev{motivates our study of} the cooperative bargaining framework.~~~~~~~~~~~~~~~~~~~~~~~~~~~~~~~~~~~~~~~~~~~~~~~

Figure \ref{fig:sim-offload-xn-theta} shows that  $x_n^o$ under the NBS  
%(i) increases with the total traffic volume $S_n$, \revh{which implies that the MNO will offload more traffic to those APs in the hot spots with heavier data usage}, 
decreases with the transmission efficiency $\theta_n$, \revh{which implies that the MNO will offload more traffic to those APs farther away (as the MUs covered by such APs have a small transmission efficiency with the macrocell BS, and thus will consume more macrocell resource if not being offloaded)}.
Similarly, Figure \ref{fig:sim-offload-xn-cn} shows that $x_n^o$ decreases with the APO's serving cost $c_n$, \revh{which implies that the APO with lower cost is more likely to offload traffic for the MNO}.
%For more detailed discussion, please refer to \cite{report}.

%\com{Need to give at least one sentence of intuition to each key observation that we provide. Otherwise, it means that the observation is not important and should not be provide in the paper.}

\textbf{Payoff Division and Grouping Effect.}
Now we illustrate the payoff division under the NBS. In order to clearly show the APOs' payoff difference, we
consider a simple scenario with \revh{$N=10$ identical APOs}. \revh{Namely, they have the same cost, resource constraint, demand distribution, and transmission efficiency. Besides, the cellular traffic volumes in these APOs are also identical.}

Figure \ref{fig:sim-group-seq-tail} illustrates the payoff of every APO (group) in different grouping structures under the sequential and the concurrent bargaining.
Each bar denotes the payoffs of APOs under a particular grouping structure. 
For example, the 6th bar in both sub-figures denotes such a grouping structure: APOs 1--4 remain single, while
APOs 5--10 merge into a group $\langle5\rangle$.
\revh{Notice that the MNO's payoff equals to the maximum social welfare minus all APOs' payoffs, and later we will show that the maximum social welfare is twice the value of the last bar (i.e., twice of the total payoff of all APOs when they merge into a single group)}.
%For clarity, we focus on the bargaining with the first 10 APs in this simulation.
%
%\com{How many APs total? What is the maximum social welfare? How to derive the BS's payoff from this figure?}

Figure \ref{fig:sim-group-seq-tail} not only shows the payoff division among APOs, but also shows \rev{how grouping benefits the group members} or non-group members.
From the {left sub-figure} (corresponding to the sequential bargaining), we have the following observations. 
First, the first bar column (gropu structure 1) shows the \emph{early-mover advantage}: \rev{an earlier APO (represented by a lower block, say the dark blue one) can achieve a higher payoff than a later APO (represented by a higher block, say the brown one)}.
Second, APOs can achieve a higher total payoff as they merge into a group (e.g., the \rev{brown block} in the last column is larger than the sum of all blocks in the first column).
Third, group merging benefits all APOs bargaining before the group, e.g., \rev{APO 1's payoff increases as more APOs merge together in later columns}).
Finally, the first column corresponds to the one-to-many sequentially bargaining without any group, and the payoff division \rev{corresponds to} the one-to-many S-NBS given in Theorem \ref{theorem:NBS-seq}; the last column is essentially equivalent to a one-to-one bargaining (with all APOs forming one group), and the payoff division \rev{corresponds to} the one-to-one NBS given in Lemma \ref{theorem:NBS-single}, \revh{from which we can easily find that the maximum social welfare $\sw(\x^o)$ is twice the value of this bar (as the group gets half of the maximum social welfare).}~~~~~~~~~~~~~

\rev{The insights from  the {right sub-figure} (corresponding to the concurrent bargaining) are different.}
First, from the first column there is no \emph{early-mover advantage}, as all APOs bargaining concurrently with the MNO.
In the first column, we can see that all APOs achieve the same payoff when all of them bargain with the MNO individually. 
Second, group merging will only benefit APOs in the group, and has no impact on other APOs' payoff.
Third, \rev{notice that the last all brown column is the same as that in the left sub-figure, both representing the bargaining between the MNO and the group of all APOs.} Finally, comparing the corresponding \rev{blocks} in both sub-figures, we can observe the \emph{concurrently moving \revh{tragedy}} for APOs: all APOs achieve a lower or equal payoff under the concurrent bargaining than under the sequential bargaining  with the same grouping structure.~~~~~~~~~~~~~~~~~~~~~~~~~~~~~~~~~~~~~~~~
%For more detailed discussions and   simulations, please refer to \cite{report}.

More specifically, Figure \ref{fig:sim-group-seq-tail} shows that the generated maximum social welfare is $\sw(\x^o) = 31.8$, and the MNO obtains around $68\%$ ($75\%$, respectively) of the generated social welfare under the sequential  (concurrent, respectively) bargaining with the grouping structure 1 (all APOs bargain individually). 
This percentage decreases as more APOs form a group. For example, under the grouping structure 8 (where 8 APOs form a group), the MNO's payoff ratio decreases to $62\%$ ($67\%$, respectively). Obviously, when all APOs form a single group (grouping structure 10), the MNO can only obtain $50\%$ of the total social welfare under both bargaining protocols. 
%These simulations coincide well with our previous analysis. 
%This shows that the grouping will have a larger impact under the concurrent bargaining than the sequential bargaining.\footnote{\revjj{
%We would like to mention that our analysis assumes a \emph{fixed} grouping structure and studies the impact of the given grouping structure on the bargaining solution. 
%The group formulation process, i.e., which APOs will form a group, is a complicated issue, and may involve consideration beyond the bargaining theory (e.g., coalition game theory may be involved). We will study this issue in our future work.
%}}

%!TEX root = DataOffload_main_journal.tex
%SourceDoc DataOffload_main_journal.tex

\section{Conclusions}\label{sec:conclusion}

In this paper, we studied the economic interaction between MNO and APOs in mobile data offloading. 
We considered a monopoly setting 
which may correspond to a scenario where a MNO negotiates with its clients that have already installed femtocell APs (for their own needs), or an ISP with its clients that have installed WiFi APs. 
%First, we analytically calculated the MNO's potential cost reduction from offloading, for generic cost functions. Accordingly, 
We used Nash bargaining theory to explain how the generated benefit should be distributed among the MNO and the involved APOs, so as to ensure that all the interacting parties are satisfied and hence willing to cooperate. In this process, the bargaining protocol, i.e. the process according to which the APOs negotiate with the MNO, is of crucial importance and affects the outcome. 
%Notice that, although NBS has been used in various networking problems, 
This is the first time that this aspect is explicitly taken into account in networking problems.

This paper opens many new interesting research directions. First, it is important to study an oligopoly market where many different MNOs compete to lease the APOs. 
The monopoly scenario presented here is a prerequisite and serves as a building block for this more general analysis. 
Equally interesting is the analysis of highly dynamic systems, where MUs have to change their AP associations while offloading their data. 
More importantly, our work opens the road for a more detailed analysis of the relation between the bargaining protocol and the market outcome. 
For example, it is challenging to study the sequential bargaining scheme under imperfect knowledge about the number of the APOs or their parameters (e.g. their capacity). Similarly, one can explore the impact of competition among different APO groups.

%In this paper, we studied mobile data offloading \rev{that can effectively reduce the congestion in today's cellular networks.} \revh{We considered the scenario where a cellular operator employs WiFi and Femtocells APs that are already deployed by third parties,} and we studied the monetary incentives that must be offered \revh{to the APs in order to induce their participation and ensure the maximum possible social benefit from offloading}. Our analysis shows that the social welfare maximization can always be achieved through bargaining, and the   payoff division is affected by the bargaining protocol and the group structure of APs.
%%We further show that the APs have incentives to form a unique group and jointly bargain with the operator.
%
%%According to the current market status and the related strategic decisions of operators, data offloading
%
%\revh{Today, data offloading through privately owned APs emerges as one of the most important solutions for addressing the explosion of mobile data traffic. We believe that our study constitutes a solid first step towards understanding how this promising method will succeed in providing the anticipated social benefits. There are many directions for future work. For example, it is interesting to analyze an \emph{oligopoly market} where multiple operators aim to lease a common set of APs, and study the impact of competition on the social welfare. Another challenging direction is to consider highly dynamic systems, where users need to change AP association while offloading their data.}

\vspace{-3mm}

%!TEX root = DataOffload_main_journal.tex
%SourceDoc DataOffload_main_journal.tex

%!TEX root = DataOffload_main_journal.tex
%SourceDoc DataOffload_main_journal.tex

\vspace{-5mm}

\begin{biography}[{\includegraphics[width=1in,clip,keepaspectratio]{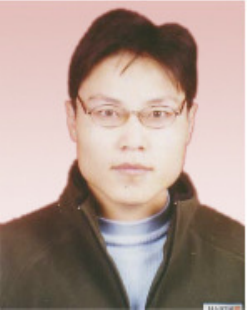}}]{Lin Gao}
is a Postdoctoral Researcher in the Department of Information Engineering at the Chinese University of Hong Kong.
He received the M.S. and Ph.D. degrees in Electronic Engineering from Shanghai Jiao Tong University (China) in 2006 and 2010, respectively.
%Prior to that, he received the B.E. degree in Information Engineering from Nanjing University of Posts and Telecommunications (China) in 2002.
%His research interests lie in wireless communication and network technologies, including dynamic spectrum access, network resource optimization, economic incentive, secondary spectrum market modeling, and game theoretic modeling.
His research interests lie in the field of wireless communications and networking with emphasis on the economic incentives in various communication and network scenarios, including cooperative communications, dynamic spectrum access, cognitive radio networks, TV
white space networks, cellular-WiFi internetworks, and user-provided networks.
\end{biography}

\begin{biography}[{\includegraphics[width=1in,clip,keepaspectratio]{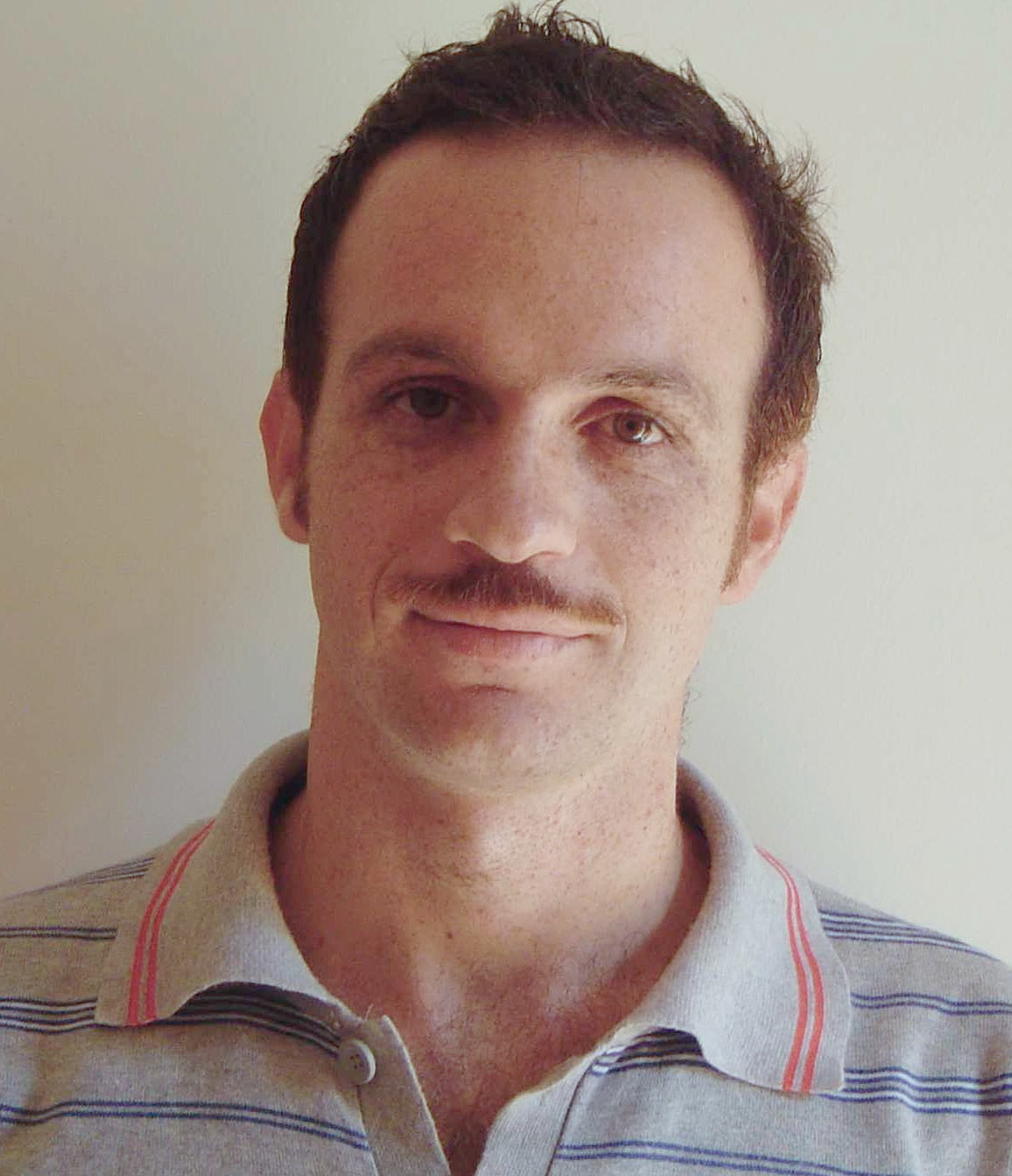}}]{George Iosifidis}
obtained the Diploma in Electronics and Telecommunications Engineering from the Greek Air Force Academy, in 2000, and the M.S. and Ph.D. degrees in Electrical Engineering from the University of Thessaly, Greece, in 2007 and 2012, respectively. Currently he is a Post-doc Researcher at the University of Thessaly and the Center for Research and Technology Hellas (CERTH), Greece. His research interests lie in the broad area of network optimization and network economics.
\end{biography}

\begin{biography}[{\includegraphics[width=1in,clip,keepaspectratio]{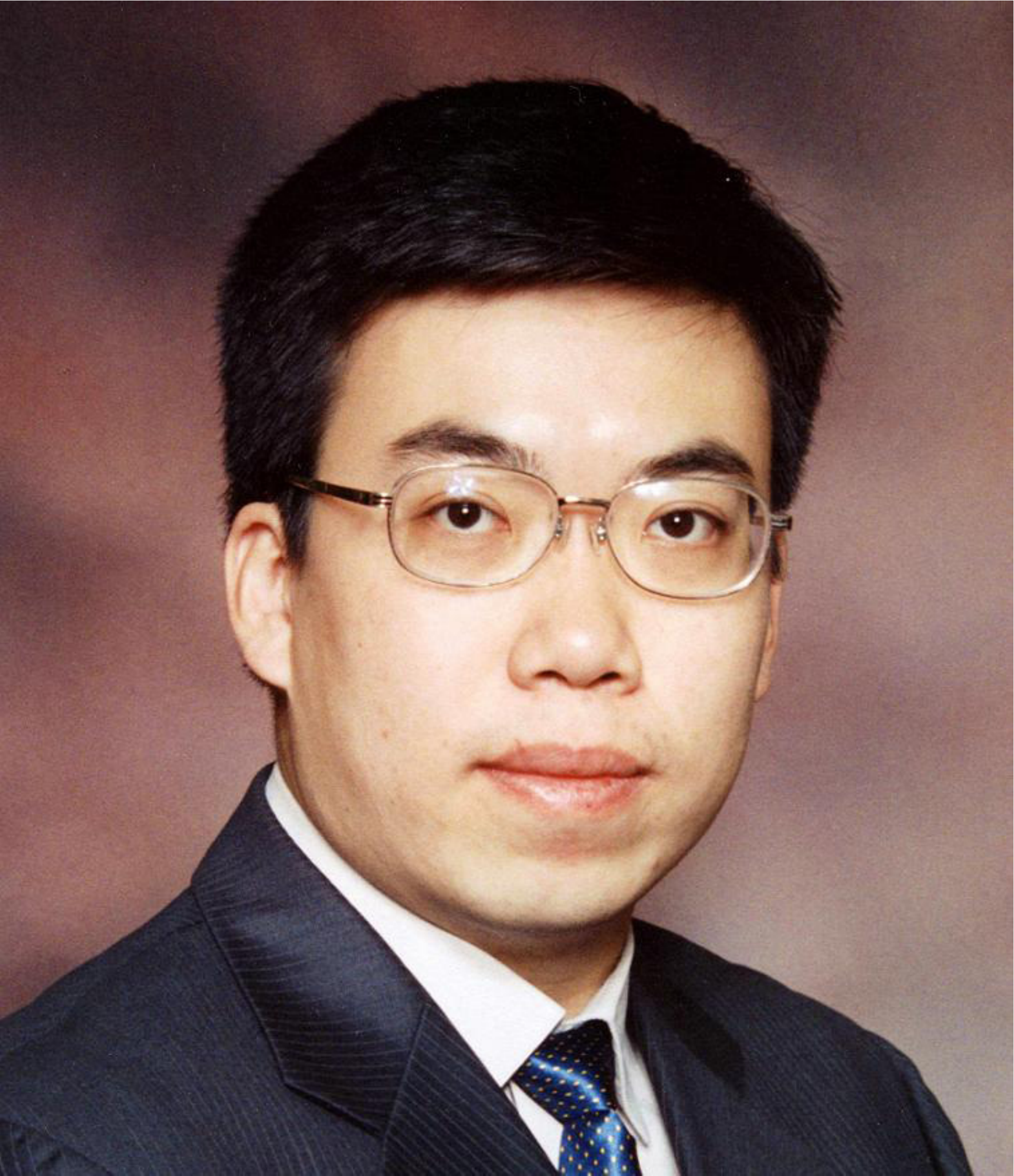}}]{Jianwei Huang}
 (S'01-M'06-SM'11) is an Associate Professor in the Department of Information Engineering at the Chinese University of Hong Kong. He is the recipient of 7 Best Paper Awards in leading international journal and conferences, including the 2011 IEEE Marconi Prize Paper Award in Wireless Communications. He is the co-author of three recent monographs: ``Wireless Network Pricing'', ``Monotonic Optimization in Communication and Networking Systems'', and ``Cognitive Mobile Virtual Network Operator Games''. He is the Editor of IEEE Journal on Selected Areas in Communications--Cognitive Radio Series and IEEE Transactions on Wireless Communications, and Chair of IEEE Communications Society Multimedia Communications Technical Committee.
\end{biography}

\begin{biography}[{\includegraphics[width=1in,clip,keepaspectratio]{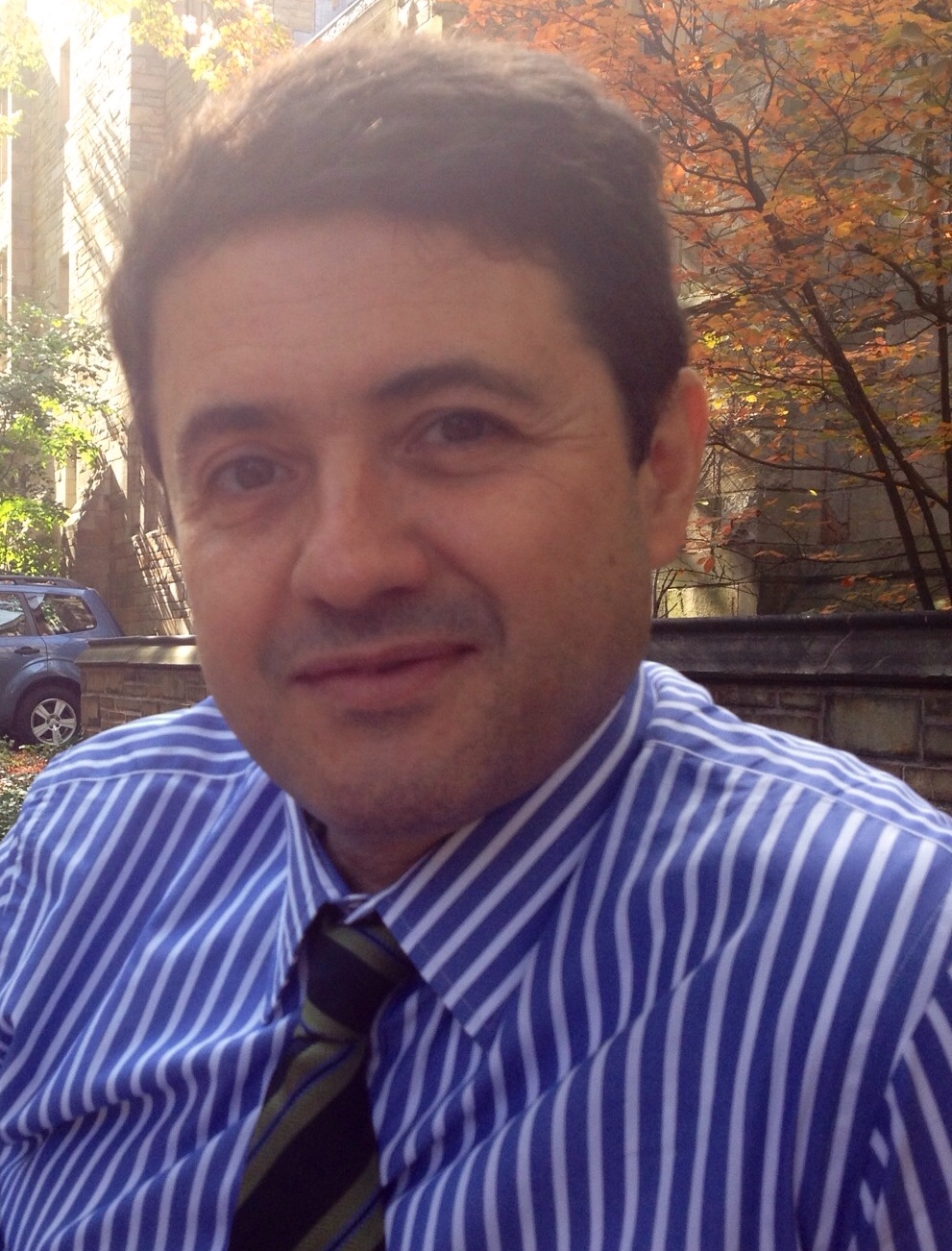}}]{Leandros Tassiulas}
(S'89-M'91-SM/06-F07) obtained the Diploma in Electrical Engineering from the Aristotelian University of Thessaloniki, Thessaloniki, Greece in 1987, and the M.S. and Ph.D. degrees in Electrical Engineering from the University of Maryland, College Park in 1989 and 1991, respectively. He is Professor in the Dept. of Computer and Telecommunications Engineering, University of Thessaly, since 2002. He has held positions as Assistant Professor at Polytechnic University New York (1991-1995), Assistant and Associate Professor University of Maryland College Park (1995-2001) and Professor University of Ioannina Greece (1999-2001). His research interests are in the field of computer and communication networks with emphasis on fundamental mathematical models, architectures and protocols of wireless systems, sensor networks, high-speed internet and
satellite communications. Dr. Tassiulas is a Fellow of IEEE. He received a National Science Foundation (NSF) Research Initiation Award in 1992, an NSF CAREER Award in 1995 an Office of Naval Research, Young Investigator Award in 1997 and a Bodosaki Foundation award in 1999. He also received the INFOCOM 1994 best paper award and the INFOCOM 2007 achievement award.
\end{biography}

\begin{biography}[{\includegraphics[width=1in,clip,keepaspectratio]{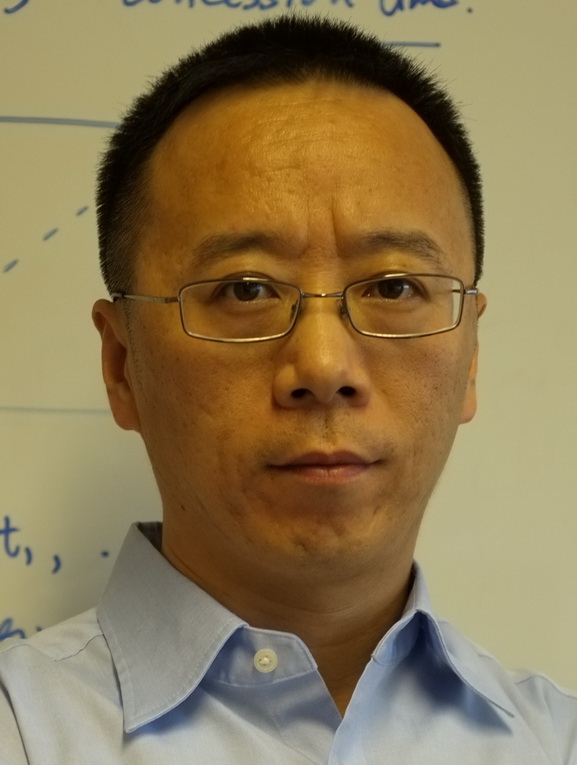}}]{Duozhe Li}
 is an Associate Professor in the Department of Economics at the Chinese University of Hong Kong. 
 Dr. Li received B.S. in Physics and M.A. in Economics from Fudan University, 
 and received Ph.D. in Economics from Boston University in 2005. 
 Dr. Li's primary research interest is game theory with the emphasis on noncooperative bargaining theory and its applications in microeconomics. His work has appeared in Journal of Economic Theory, Journal of Economic Behavior \& Organization, Economics Letters, etc.
\end{biography}

\newpage
%!TEX root = DataOffload_main_journal.tex
%SourceDoc DataOffload_main_journal.tex

\appendix

\begin{center}
\Large
\textbf{Technical Report}
\end{center}

\noindent
\textbf{Title}: Bargaining-Based Mobile Data Offloading

\noindent
\textbf{Authors}: Lin Gao, George Iosifidis, Jianwei Huang, Leandros Tassiulas, and Duozhe Li

\vspace{3mm}

\noindent
{\small
Note: The original version is published in IEEE Journal on Selected Areas in Communications (JSAC) Special Issue in 5G Communication Systems, 2014. 
}

\begin{center}
\Large
\textbf{Outline of This Technical Report}
\end{center}

\begin{itemize}
\item[]
\hspace{-8mm}
\textbf{(I) Model Discussion}

\item \textit{\ref{app:model-extension}. Model Extension}

\item[] 
\hspace{-8mm}
\textbf{(II) Illustration and Example}

\item \textit{\ref{app:illu-virtual-marginal}. Illustration of Virtual Marginal Social Welfare $\adw_{n}$}
\item \textit{\ref{app:example-nbs}. Examples of Nash Bargaining Solutions}
\item \textit{\ref{app:example-group}. Examples of Grouping Effect}

\item[] 
\hspace{-8mm} 
\textbf{(III) Proofs}

\item \textit{\ref{app:proof-one-to-one-NBS}. Proof for Lemma \ref{theorem:NBS-single} in Section \ref{sec:barg}}
\item \textit{\ref{app:proof-nbs-x}. Proof for Lemma \ref{prop:nbs-x} in Section \ref{sec:barg-xo}}
\item \textit{\ref{app:proof-NBS-seq-N}. Proof for Lemma \ref{lemma:NBS-seq-N} in Section \ref{sec:payoff}}
\item \textit{\ref{app:proof-NBS-seq-N-1}. Proof for Lemma \ref{lemma:NBS-seq-N-1} in Section \ref{sec:payoff}}
\item \textit{\ref{app:proof-NBS-seq-n}. Proof for Lemma \ref{lemma:NBS-seq-n} in Section \ref{sec:payoff}}
\item \textit{\ref{app:proof-theorem-NBS-seq}. Proof for Theorem \ref{theorem:NBS-seq} in Section \ref{sec:payoff}} 
%\item[] ~~ -- Sequential Bargaining Solution (S-NBS)
\item \textit{\ref{app:proof-property-seq}. Proofs for Properties \ref{prop:EMA-seq} and \ref{prop:IOC-seq} in Section \ref{sec:payoff}}
\item \textit{\ref{app:proof-NBS-con}. Proof for Lemma \ref{lemma:NBS-con} in Section \ref{sec:payoff}}
\item \textit{\ref{app:proof-theorem-NBS-con}. Proof for Theorem \ref{theorem:NBS-con} in Section \ref{sec:payoff}}
%\item[] ~~ -- Concurrent Bargaining Solution (C-NBS)
\item \textit{\ref{app:proof-property-con}. Proofs for Properties \ref{prop:IIC-con} and \ref{prop:CMT-con} in Section \ref{sec:payoff}}
\item \textit{\ref{app:proof-prop:GBS-seq}. Proof for Property \ref{prop:GBS-seq} in Section \ref{sec:barg:group}}
\item \textit{\ref{app:proof-prop:SPE-group}. Proof for Property \ref{prop:SPE-group} in Section \ref{sec:barg:group}}
\item \textit{\ref{app:proof-prop:GBS-con}. Proof for Property \ref{prop:GBS-con} in Section \ref{sec:barg:group}}
\item \textit{\ref{app:proof-prop:NOE-group}. Proof for Property \ref{prop:NOE-group} in Section \ref{sec:barg:group}}

\item[] 
\hspace{-8mm}
\textbf{(IV) An Alternative Modeling Approach}

\item \textit{\ref{app:game}. Non-cooperative Game Formulation and Analysis} 
\end{itemize}

\begin{figure*}[t]
    \centering
   \includegraphics[scale=.8]{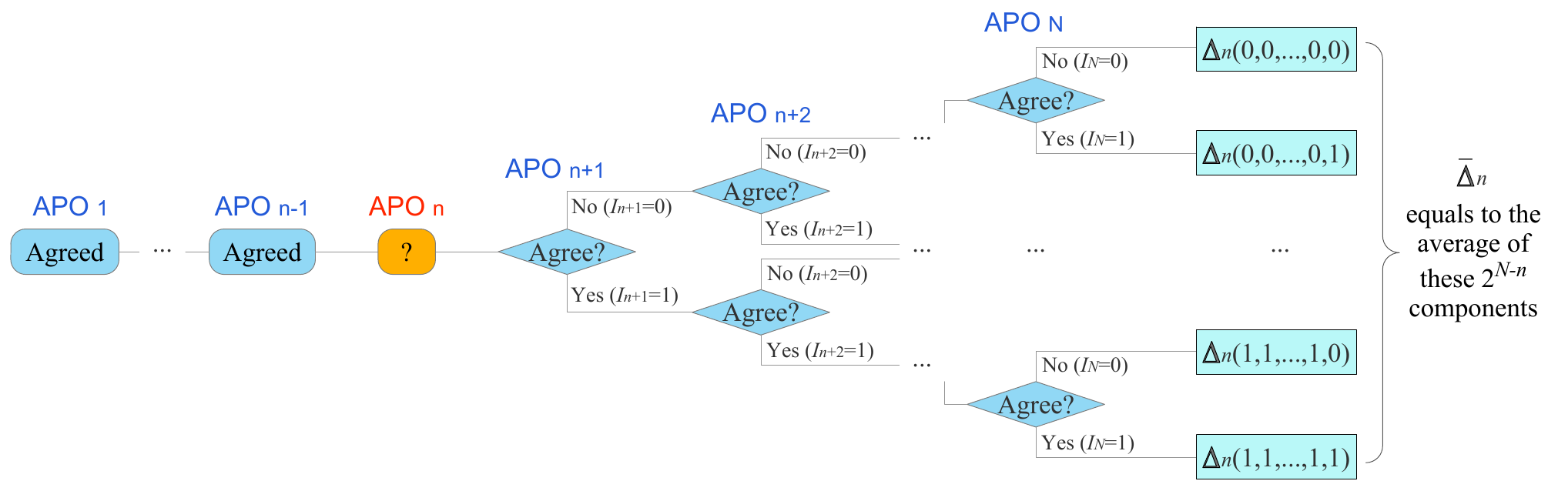}
    \caption{Illustration of the virtual marginal social welfare $\adw_n$ generated by APO $n$.}
%$I_k \in \{0,1\}, k\in\{n+1,...,N\}$, indicates whether the MNO will reach an agreement with APO $k$.
\label{fig:average-msw}
\end{figure*}

 \begin{center}
\Large\textbf{(I) Model Discussion}
\end{center}

\vspace{-5mm}

%!TEX root = DataOffload_main_journal.tex
%SourceDoc DataOffload_main_journal.tex

\subsection{Model Extension}\label{app:model-extension}

\revjr{
Now we discuss how to extend the current model (with non-overlapping APOs) to a new model with overlapping APOs. Specifically, we first show that in the new model (with overlapping APOs), the key challenges include (i) modeling the overlap relationship of APOs, and (ii) solving the
optimal offloading solution (even in the centralized manner). 
Then we propose a different modeling method, which can model the data offloading problem more effectively. 
It is important to note that as long as the optimal offloading solution is obtained, all of the bargaining analysis (regarding the welfare division) in this paper can be directly applied to the new model. 

We first discuss  the challenges in modeling and solving the offloading problem with overlapping APOs.

\noindent
\emph{1) Modeling the overlap relationship of APOs:} 

To characterize the overlap relationship of APOs, we need to define the overlapping area of any 2 APOs (hence a maximum of $\frac{N\cdot (N-1)}{2\cdot 1}$ areas), the overlapping area of any 3 APOs (hence a maximum of $\frac{N\cdot (N-1)\cdot (N-2)}{3\cdot2\cdot 1}$ areas), ..., the overlapping area of any $N-1$ APOs (hence a maximum of $\frac{N\cdot (N-1)\cdot ...\cdot 2}{(N-1)\cdot...\cdot 2\cdot 1} = N$ areas), and finally, the overlapping area of all APOs. Thus, for a network of $N$ APOs, we need to define a maximum of $K$ areas, where
$$
\textstyle
K =  K_1 + K_2 + ... + K_N = \sum_{n=1}^N 
\frac{N\cdot...\cdot (N-n+1)}{n\cdot ...\cdot 1}, 
$$
and $K_1 = N$ is the number of areas covered by a single APO, $K_n = \frac{N\cdot...\cdot (N-n+1)}{n\cdot ...\cdot 1}, n\geq 2, $ is the number of overlapping areas covered by $n$ APOs jointly.  
Accordingly, we need to define the MNO's traffic distribution in a maximum of $K+1$ areas (including the above $K$ areas and the blank area not covered by any APO). 
Obviously, it is challenging to model the offloading problem using the above method as $K$ increases exponentially with $N$.~~~~~~~~

\noindent  
\emph{2) Solving the optimal offloading solution:}

Note that even if we model the problem in the above way (i.e., dividing the whole area into $K+1$ parts), finding the optimal offloading solution (even in the centralized manner) is still challenging, as it requires us to solve a matching problem which is usually NP-hard. Specifically, for any traffic within any area covered by multiple APOs, we need to determine which APOs are actually scheduled to offload it. Therefore, the whole data offloading problem is essentially a matching problem (between the traffic in $K$ areas and $N$ APOs). 
Solving a matching problem is usually time consuming, especially when the matching size $K$ or $N$ is large. 

Now we propose a different modeling method to model the data offloading problem more effectively. 
%Note that although the modeling method is different, the essential models under different methods are equivalent.
The key idea is as follows. 
First, we divide the whole area into $I$ small areas, and each can be a square or hexagon, with a small size (e.g., $10$ meters). 
Let $S_i$ denote the traffic within the $i$th small areas, $i=1,...,I$. 
Let $a_{n,i} \in\{0, 1\}$ denote whether the $i$th area is covered by APO $n$. 
Then, the traffic $S_i$ can be offloaded to an APO $n$ with $a_{n,i} = 1$ (and there can be multiple of such APOs, each offloading a fraction of $S_i$).
It is easy to check that the model under this new modeling method is equivalent to the original model (based on the overlapping areas among APOs), but it can avoid the complicated characterization of overlap relationships among APOs. 

Certainly, with this new modeling method, solving the optimal offloading problem is a matching problem (between the traffic in $I$ areas and $N$ APOs) and hence is still challenging. 
Nevertheless, many classic algorithms or approximate algorithms can be used to solve a matching problem (Interested readers can refer to the book ``\emph{Algorithm Design (Pearson Education, 2006)}'' by Eva Tardos and Jon Kleinberg). 
Notice that in the original modeling method,  $K$ increases exponentially with the number of APOs $N$, while in the new modeling method, $I$ is independent of $N$. 
Therefore, 
the new modeling method is more efficient, especially in the scenarios with a large number of APOs.
}

 \begin{center}
\Large\textbf{(II) Illustration and Example}
\end{center}

\vspace{-5mm}

%!TEX root = DataOffload_main_journal.tex
%SourceDoc DataOffload_main_journal.tex

\subsection{Illustration of Virtual Marginal Social Welfare $\adw_{n}$}
\label{app:illu-virtual-marginal}

Lemma \ref{lemma:NBS-seq-n} shows that under the sequential bargaining solution (S-NBS), every APO $n$ (bargaining in Step $n$) achieves half of the virtual marginal social welfare $\adw_{n}$ it generates. In addition, the virtual marginal social welfare $\adw_{n}$ is given by
$$
\textstyle
\adw_{n} =   \sum_{I_{n+1}=0 }^1\dii \sum_{I_{N}=0 }^1	 \frac{{\dw}_{n}(I_{n+1};\dii ; I_N )}{ 2^{N - n } },
$$
where ${\dw}_{n}(I_{n+1};\dii  ; I_N ) \mbox{=} \sw(\x_{n \mi 1}^{\stx}, \rmkk{x_{n}^{\stx}}, I_{n+ 1}  {x_{n+ 1}^{\stx}},\dii , I_N  x_N^{\stx}) 
- \sw(\x_{n \mi 1}^{\stx}, \rmkk{0}, I_{n+ 1}  {x_{n+ 1}^{\stx}},\dii , I_N  x_N^{\stx})
$.

For a better understanding, we illustrate the structure of the virtual marginal social welfare $\adw_{n}$ in Figure \ref{fig:average-msw}. 
Intuitively, it equals to the average of the marginal social welfares generated by APO $n$, under the conditions that the MNO has reached agreements with each APO in $\{1,...,n-1\}$ (before APO $n$), and will reach agreements with each APO in $\{n+1,...,N\}$ (after APO $n$) with a probability of 0.5.

%!TEX root = DataOffload_main_journal.tex
%SourceDoc DataOffload_main_journal.tex

%\subsection*{B.3)~~Examples}
%\label{sec:barg:seq}

\subsection{Examples of Nash Bargaining Solutions}\label{app:example-nbs}

Now we provide examples to illustrate the NBS under both the sequential bargaining and the concurrent bargaining. 

Consider the following example: (i) $N=4$ APOs, (ii) the socially optimal   offloading solution is $x_n^o= 1, n\in\{1,2,3,4\}$, and (iii) the social welfare function $\sw(\x)$ is a concave  function $\psi(\cdot)$ of 
the total offloaded amount, i.e., $\sw(\x) \eq \psi(\sum_{n=1}^4 x_n)$.\footnote{\revjj{This implies that, from the social perspective, offloading one unit of traffic by an APO $n_1$ is totally same as by another APO $n_2$. 
This social welfare  function may correspond to such a network scenario where all APOs are symmetric.}
%This social welfare function is suitable for an offloading model without considering the transmission efficiency and the differences of APOs' cost functions.
}
By Lemma \ref{prop:nbs-x}, the traffic offloading profiles under both sequential and concurrent bargainings are $x_n^* =x_n^o= 1, n\in \{1,2,3,4\}$. 
Next, we illustrate the payoff profiles under different bargaining protocols.

\vspace{3mm}

\noindent
\begin{tabular}{m{0.95\linewidth}}
\textbf{Example: Sequential Bargaining}\\
\hline
\end{tabular}
 
\vspace{1mm}

In Step $4$, the disagreement points (D) of APO $4$ and the MNO, and their payoffs (A) and payoff gains (G) if they reach an agreement $\pa_4 = \paxx $ (and $x_4 = x_4^*  = 1$) are 
\begin{equation*}
\left\{
\begin{aligned}
&
\mbox{(D)}
&& 
\textstyle \Ua_4^0  = 0,
&&
\textstyle
\Um^0_{[4]}  = \psi (3) - \Pi_{3},
\\
&
\mbox{(A)}
&& 
\textstyle 
\Ua_4  = \paxx, 
&&
\textstyle
\Um_{[4]}  \textstyle = \psi (4)- \Pi_{3} - \paxx,
\\
&
\mbox{(G)}
&& 
\textstyle 
\Ua_4 -\Ua_4^0  = \paxx, 
&&
\textstyle
\Um_{[4]} - \Um^0_{[4]}   \textstyle = \psi (4)  - \psi (3) - \paxx.
\end{aligned}
\right.
\end{equation*}
Then, the NBS in Step $4$ (i.e., the APO $4$'s payoff), and the MNO's payoff under the NBS are, respectively,
\begin{equation}\label{ex1:step4}
\begin{aligned}
\pa_4^* ~& \textstyle
  = \frac{\dw_4}{2} = \frac{\psi (4)  - \psi (3)}{2},
\\
\Um_{[4]}^* & \textstyle = \Um^0_{[4]}+ \frac{\dw_4}{2}=  \frac{\psi (4)  + \psi (3)}{2} - \Pi_{3},
\end{aligned}
\end{equation}
where $\psi (4)  - \psi (3) \eq \dw_4$ is the marginal social generated by APO $4$.
Obviously, both the APO $4$ and the MNO get half of the marginal social  ${\dw_4}$ generated by APO $4$.
%, and $\psi (4)  + \psi (3) \eq \mw_4$.

In Step $3$, the disagreement points (D) of APO $3$ and the MNO, and their payoffs (A) and payoff gains (G) if they reach an agreement $\pa_3 = \paxx $ (and $x_3 = x_3^*  = 1$) are
\begin{equation*}
\left\{
\begin{aligned}
&
\mbox{(D)}
&& 
\textstyle \Ua_{3}^0  = 0,
&&
\textstyle
\Um^0_{[3]}  = \frac{\psi (3)  + \psi (2)}{2} - \Pi_{2},
\\
&
\mbox{(A)}
&& 
\textstyle 
\Ua_{3}   = \paxx, 
&&
\textstyle
\Um_{[3]}  \textstyle = \frac{\psi (4)  + \psi (3)}{2} - \Pi_{2}  - \paxx,
\\
&
\mbox{(G)}
&& 
\textstyle 
\Ua_3 -\Ua_3^0  = \paxx, 
&&
\textstyle
\Um_{[3]} - \Um^0_{[3]}   \textstyle = 
\frac{\psi (4)  - \psi (2)}{2}
 - \paxx,
\end{aligned}
\right.
\end{equation*}
where $\Um_{[3]}^0$ and $\Um_{[3]}$ are derived from $\Um_{[4]}^*$ in Step $4$, denoting the MNO's potential payoff after having dealt with all APOs.

Thus, the NBS in Step $3$ (i.e., the APO $3$'s payoff), and the MNO's payoff under the NBS are, respectively,
\begin{equation}\label{ex1:step3}
\begin{aligned}
\pa_3^* ~& \textstyle = \frac{\adw_3}{2} = \frac{\psi (4) -  \psi (3)}{4} + \frac{\psi (3)  - \psi (2)}{4},
\\
\Um_{[3]}^* & \textstyle=  \frac{\psi (4) + \psi (3)}{4} + \frac{\psi (3)  + \psi (2)}{4}
 - \Pi_{2},
\end{aligned}
\end{equation}
where (i) $\psi (4)  - \psi (3) \eq \dw_3(I_4=1)$ 
is the marginal social welfare generated by APO $3$, assuming that the MNO will reach an agreement with APO $4$, 
(ii) $\psi (3)  - \psi (2) \eq \dw_3(I_4=0)$ is the marginal social welfare generated by APO $3$, assuming that the MNO will \textbf{not} reach an agreement with APO $4$, 
(iii) ${\adw_3} = \frac{\sum_{I_4=0}^1\dw_3(I_4 )}{2}  $ is the virtual marginal social welfare generated by APO $3$, assuming that the MNO will reach an agreement with APO $3$ with a probability of 0.5. 
Both the APO 3 and the MNO get half of the virtual marginal social  ${\adw_3}$ generated by APO $3$.

In Step $2$, the disagreement point for the MNO is 
$\Um^0_{[2]}  = \frac{\psi (3) + \psi (2)}{4} + \frac{\psi (2)  + \psi (1)}{4} - \Pi_{1}$, which is directly obtained from $\Um_{[3]}^*$ in (\ref{ex1:step3}). 
Then, with a similar analysis, we can derive the NBS in Step $2$ (i.e., the APO $2$'s payoff) and the MNO's payoff under this NBS as follows.
%In Step $2$, the disagreement points (D) of APO $2$ and the MNO, and their payoffs (A) and payoff gains (G) if they reach an agreement $\pa_2 = \paxx $ (and $x_2 = x_2^*  = 1$) are
%\begin{equation*}
%\left\{
%\begin{aligned}
%&
%\mbox{(D)}
%&& 
%\textstyle \Ua_{2}^0  = 0,
%&&
%\textstyle
%\Um^0_{[2]}  = \frac{\psi (3) + \psi (2)}{4} + \frac{\psi (2)  + \psi (1)}{4} - \Pi_{1},
%\\
%&
%\mbox{(A)}
%&& 
%\textstyle 
%\Ua_{2}   = \paxx, 
%&&
%\textstyle
%\Um_{[2]}  \textstyle = \frac{\psi (4) + \psi (3)}{4} + \frac{\psi (3)  + \psi (2)}{4} - \Pi_1 - \paxx,
%\\
%&
%\mbox{(G)}
%&& 
%\textstyle 
%\Ua_2 -\Ua_2^0  = \paxx, 
%&&
%\textstyle
%\Um_{[2]} - \Um^0_{[2]}   \textstyle = 
%\frac{\psi (4) -  \psi (3)}{4} +
%\frac{\psi (3) -  \psi (2)}{4}\cdot 2 + 
%\frac{\psi (2) -  \psi (1)}{4} 
% - \paxx,
%\end{aligned}
%\right.
%\end{equation*}
%where $\Um_{[2]}^0$ and $\Um_{[2]}$ are derived from $\Um_{[3]}^*$ in Step $3$.
%The NBS in Step $2$, and the MNO's payoff under this NBS are
\begin{equation}\label{ex1:step2}
\begin{aligned}
\pa_2^* ~& \textstyle  = \frac{\adw_2}{2} = 
\frac{\psi (4) -  \psi (3)}{8} +
\frac{\psi (3) -  \psi (2)}{8}\cdot 2 + 
\frac{\psi (2) -  \psi (1)}{8} ,
\\
\Um_{[2]}^* & \textstyle =  \frac{\psi (4) + \psi (3)}{8} +
\frac{\psi (3) +  \psi (2)}{8}\cdot 2 + 
\frac{\psi (2) +  \psi (1)}{8} 
 - \Pi_{1},
\end{aligned}
\end{equation}
where (i) $\psi (4)  - \psi (3) \eq \dw_2(I_3=I_4=1)$ 
is the marginal social welfare generated by APO $2$, assuming that the MNO will reach agreements with both APOs $3$ and $4$, 
(ii) $\psi (3)  - \psi (2) \eq \dw_2(I_3=0,I_4=1)\eq \dw_2(I_3=1,I_4=0)$ is the marginal social welfare generated by APO $2$, assuming that the MNO will reach an agreement with one of APOs $3$ and $4$,
 (iii) $\psi (2)  - \psi (1) \eq \dw_2(I_3=I_4=0)$ 
is the marginal social welfare generated by APO $2$, assuming that the MNO will \textbf{not} reach an agreement with any  of APOs $3$ and $4$, 
and
(iv) $ 
%\frac{\dw_2(I_3=I_4=1) + \dw_2(I_3=0,I_4=1) + \dw_2(I_3=1,I_4=0) + \dw_2(I_3=I_4=0)}{4} 
 {\adw_2} = \frac{\sum_{I_3=0}^1 \sum_{I_4=0}^1 \dw_2(I_3,I_4)}{4}
 $ is the virtual marginal social welfare generated by APO $2$ assuming that the MNO will reach an agreement with each APO in $\{3, 4\}$ with a probability of 0.5.
 Both the APO $2$ and the MNO get half of the virtual marginal social welfare ${\adw_2}$ generated by APO $2$.

In Step 1, 
%We skip the detailed derivation here due to space limit, and readers can refer to \cite{report} for details.  
we can similarly  derive the NBS, and the MNO's payoff under this NBS as follows.
\begin{equation}\label{ex1:step1}
\begin{aligned}
\pa_1^* ~&\textstyle
 = 
\frac{\adw_1}{2} = 
\frac{\psi (4) -  \psi (3)}{16} +
\frac{\psi (3) -  \psi (2)}{16}\cdot 3 + 
\frac{\psi (2)  - \psi (1)}{16} \cdot 3 +
\\
& \textstyle \qquad \qquad
\frac{\psi (1) -  \psi (0)}{16} 
,
\\
\Um_{[1]}^*&\textstyle =  
\frac{\psi (4) +  \psi (3)}{16} +
\frac{\psi (3) +  \psi (2)}{16}\cdot 3 + 
\frac{\psi (2) + \psi (1)}{16} \cdot 3 +
\frac{\psi (1) +  \psi (0)}{16} ,
\end{aligned}
\end{equation}
where 
%(i) $\psi (4)  - \psi (3) \eq \dw_1(I_2=I3=I_4=1)$, 
%(ii) $\psi (3)  - \psi (2) \eq \dw_1(I_2=0,I_3=I_4=1)\eq \dw_1(I_2=1,I_3=0,I_4=1)\eq \dw_1(I_2=I_3=1,I_4=0)$,
%(iii) $\psi (2)  - \psi (1) \eq \dw_1(I_2=1,I_3=I_4=0)\eq \dw_1(I_2=0,I_3=1,I_4=0)\eq \dw_1(I_2=I_3=0,I_4=1)$,
%and (iv) $\psi (1)  - \psi (0) \eq \dw_1(I_2=I3=I_4=0)$, 
%are marginal social welfares generated by APO $1$ under different possibilities, 
%and
%(v) 
$ 
{\adw_1} \eq \frac{\sum_{I_2=0}^1 \sum_{I_3=0}^1 \sum_{I_4=0}^1 \dw_1(I_2, I_3,I_4)}{8}
 $ is the virtual marginal social welfare generated by APO $1$, assuming that the MNO will reach an agreement with each APO in $\{2, 3, 4\}$ with a probability of 0.5.
 Both the APO $1$ and the MNO get half of the virtual marginal social welfare ${\adw_1}$ generated by APO $1$.

We summarize the APOs' payoffs (i.e., $\pa_n^*,\ n=1,2,3,4$) under the above sequential bargaining solution in Figure \ref{fig:example}, where $w_n \eq \frac{\psi (n+1) - \psi (n)}{2}$, denoting a half of the marginal social welfare generated by a new APO when there are $n$ other APOs reaching agreements with the MNO, 
and $w_3<w_2<w_1<w_0$ by the concavity of $\psi(\cdot)$.
These results can be easily extended to a more general case with $N$ APOs.~~~~~~~~~~~~~~~~~~
%  The first line in the table states that the payoff of APO $4$ is $w_3$, the second line states that the payoff of APO $3$ is $\frac{w_3}{2} + \frac{w_2}{2}$, and so on.

\begin{figure}[t]
    \centering
   \includegraphics[scale=.8]{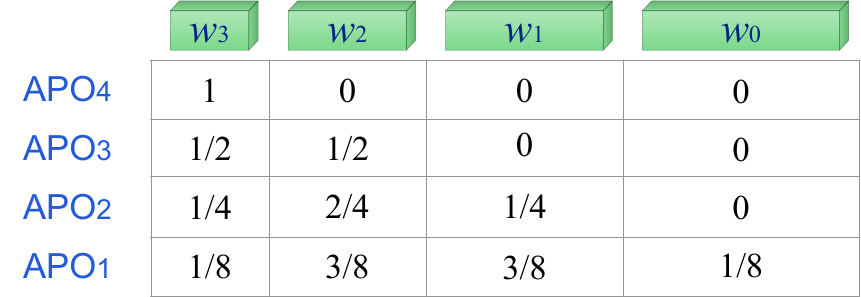}
    \caption{APOs' payoffs under the sequential bargaining solution, where $w_n = \frac{\psi (n+1) - \psi (n)}{2}$. The payoff of APO 4 is $w_3$, the payoff of APO 3 is $
\frac{1}{2}(w_3+w_2) $, the payoff of APO 2 is $\frac{1}{4} (w_3 +2 w_2 + w_1) $, and the payoff of APO 1 is $\frac{1}{8}( w_3 +3 w_2 +3 w_1+ w_0)$.} \label{fig:example}
\end{figure}

\textit{Verification of  Property \ref{prop:EMA-seq} 
(Early-Mover Advantage):}
From Eqs. (\ref{ex1:step4})-(\ref{ex1:step1}) or from Figure \ref{fig:example} we can easily find that 
$$\pa_1^* > \pa_2^* > \pa_3^* > \pa_4^*,$$
as $w_3<w_2<w_1<w_0$ by the concavity of $\psi(\cdot)$.

\textit{Verification of  Property \ref{prop:IOC-seq}
(Invariance to APO-order Changing):}
Notice that the MNO's payoff given in (\ref{ex1:step1})
can be rewritten as $\Um_{[1]}^* = \frac{\psi (4) + 4 \psi (3) +  6 \psi (2) + 4 \psi (1) +  \psi (0)}{16}
= \frac{\sum_{I_1=0}^1\sum_{I_2=0}^1 \sum_{I_3=0}^1 \sum_{I_4=0}^1 \sw(I_1, I_2, I_3,I_4)}{16}
$, which is exactly the expected social welfare when the MNO reaches agreement with each APO with a probability of 0.5.  Obviously, changing the order of APOs does not affect the MNO's payoff. $\hfill \blacksquare$

\vspace{3mm}

\noindent
\begin{tabular}{m{0.95\linewidth}}
\textbf{Example: Concurrent Bargaining.}\\
\hline
\end{tabular}

\vspace{1mm}

Consider the bargaining between the MNO and an arbitrary APO $n\in\{1,2,3,4\}$.
The disagreement points (D) of APO $n$ and the MNO, and their payoffs (A) and payoff gains (G) if they reach an agreement $\pa_n = \paxx $ (and $x_n = x_n^*  = 1$) are
\begin{equation*}
\left\{
\begin{aligned}
&
\mbox{(D)}
&& 
\textstyle \Ua_n^0  = 0,
&&
\textstyle
\Um^0_{[n]}  = \psi (3) - \Pi_{-n},
\\
&
\mbox{(A)}
&& 
\textstyle 
\Ua_n  = \paxx, 
&&
\textstyle
\Um_{[n]}  \textstyle = \psi (4)- \Pi_{-n} - \paxx,
\\
&
\mbox{(G)}
&& 
\textstyle 
\Ua_n -\Ua_n^0  = \paxx, 
&&
\textstyle
\Um_{[n]} - \Um^0_{[n]}   \textstyle = \psi (4)  - \psi (3) - \paxx,
\end{aligned}
\right.
\end{equation*}
where $\Um^0_{[n]}$ and $\Um_{[n]}$ are based on the expectation that the MNO will  reach agreements with all other APOs bargaining concurrently. 
Then, the NBS with APO $n$ (i.e., the APO $n$'s payoff), and the MNO's payoff under this NBS are
\begin{equation}\label{ex2:n}
\begin{aligned}
\pa_n^*  ~ &\textstyle = \frac{\dwc_n}{2} = \frac{\psi (4)  - \psi (3)}{2},
\\
\Um_{[n]}^* &\textstyle =   \frac{\psi (4)  + \psi (3)}{2} - \Pi_{-n},
\end{aligned}
\end{equation}
where $\psi (4)  - \psi (3) \eq \dwc_n$ is the marginal social welfare generated by APO $n$ (assuming the MNO will reach agreements with all other APOs).
%Both the APO $n$ and the MNO get half of the marginal social $\dwc_n$ generated by APO $n$.
%, and $\psi (4)  + \psi (3) \eq \mw_4$.

\textit{Verification  of Property \ref{prop:IIC-con} (Invariance to AP-index Changing):}
Due to the symmetry of APOs in this example, we have: 
$$
\textstyle
\pa_n^* = \frac{\psi (4)  - \psi (3)}{2},\quad \forall n\in\{1,2,3,4\}.
$$ 
That is, the APO's payoff is independent of its index.

\textit{Verification  of Property \ref{prop:CMT-con} (Concurrently Moving Tragedy):}
It is easy to see that each APO $n$'s payoff in (\ref{ex2:n}) is equal to the worst APO's payoff under the sequential bargaining (i.e., the payoff of the last bargainer at Step 4). 
%That is,  the worst-case payoff that it can obtain under the sequential bargaining. 

By Property \ref{prop:IIC-con}, we further have: $\Pi_{-n} = \sum_{i\neq n} \pa_i^* =3 \cdot \frac{\psi (4)  - \psi (3)}{2}$. Thus, the MNO's payoff can be written as: $ \Um_{[n]}^* = 2 \cdot  \psi (3)  - \psi (4) $.
Comparing it with (\ref{ex1:step1}), we can easily find that the MNO can achieve a higher payoff under the concurrent bargaining. $\hfill \blacksquare$

%!TEX root = DataOffload_main_journal.tex
%SourceDoc DataOffload_main_journal.tex

%\subsection*{B.3)~~Examples}
%\label{sec:barg:seq}

\subsection{Examples of Grouping Effect}\label{app:example-group}

%\subsection*{C.3)~~Examples}
Now we use the   example in Appendix \ref{app:example-nbs} to illustrate the grouping effect. 
For a better illustration, we consider that APOs $2$ and $3$ form a new group, denoted by $\langle{3}\rangle \eq \{2,3\}$, while APOs $1$ and $4$ bargain individually.
A dummy APO $\langle{2}\rangle$ is introduced for the notational consistence. 
With this APO grouping structure, the bargaining order under the sequential bargaining is  $\{1\}$, $\{2,3\}$,  $\{4\}$.
By Lemma \ref{prop:nbs-x}, the traffic offloading profiles are still $x_n^* =x_n^o= 1, n\in \{1,2,3,4\}$ with this APO grouping structure. 
Next, we illustrate the payoff profiles
under this APO grouping structure.

\vspace{3mm}

\noindent
\begin{tabular}{m{0.95\linewidth}}
\textbf{Example: Grouping Effect in Sequential Bargaining.}
\\
\hline
\end{tabular}

\vspace{1mm}

In Step $4$, the MNO has reached agreements with APO $1$ and APO group $\{2,3\}$, and now is bargaining with APO $4$. 
With a similar analysis in Appendix \ref{app:example-nbs}, we can obtain the NBS in Step $4$ (i.e., the APO $4$'s payoff) and the MNO's payoff under the NBS as follows.
\begin{equation}\label{ex3:step4}
\begin{aligned}
\pa_4^*  ~ & \textstyle = \frac{\dw_4}{2} = \frac{\psi (4)  - \psi (3)}{2},
\\
\Um_{[4]}^*  & \textstyle =   \frac{\psi (4)  + \psi (3)}{2} - \Pi_{3}.
\end{aligned}
\end{equation}

In Step $3$, the MNO has reached agreements with APO $1$, and now is bargaining with the APO group $\{2,3\}$). 
The disagreement points (D) of APO group $\{2,3\}$ and the MNO, and their payoffs (A) and payoff gains (G) if they reach an agreement $\pa_2 + \pa_3 = \paxx $ (and $x_2 = x_2^*=1$, $x_3 =x_3^* = 1$) are
\begin{equation*}
\left\{
\begin{aligned}
&
\mbox{(D)}
&& 
\textstyle \Ua_{\langle{3}\rangle}^0  = 0,
&&
\textstyle
\Um^0_{[3]}  = \frac{\psi (2)  + \psi (1)}{2} - \Pi_{1},
\\
&
\mbox{(A)}
&& 
\textstyle 
\Ua_{\langle{3}\rangle}   = \paxx, 
&&
\textstyle
\Um_{[3]}  \textstyle = \frac{\psi (4)  + \psi (3)}{2} - \Pi_{1}  - \paxx,
\\
&
\mbox{(G)}
&& 
\textstyle 
\Ua_{\langle{3}\rangle} -\Ua_{\langle{3}\rangle}^0  = \paxx, 
&&
\textstyle
\Um_{[3]} - \Um^0_{[3]}   \textstyle = 
\frac{\psi (4) -  \psi (2) + \psi (3)  - \psi (1)}{2}
 - \paxx,
\end{aligned}
\right.
\end{equation*}
where $\Um_{[3]}^0$ and $\Um_{[3]}$ are derived from $\Um_{[4]}^*$ in Step $4$. 
Note that $\Um_{[3]}^0$ here is different from that in Appendix \ref{app:example-nbs}, as in this new grouping structure, the MNO will not reach an agreement with any APO in $\{2,3\}$ under the disagreement outcome. 
Then, the NBS in Step $3$ (i.e., the total payoff of APOs $2$ and $3$), and the MNO's payoff under this NBS are
\begin{equation}\label{ex3:step3}
\begin{aligned}
\pa_{\langle{3}\rangle}^* & \textstyle = \frac{\adw_{\langle{3}\rangle}}{2} = \frac{\psi (4) -  \psi (2)}{4} + \frac{\psi (3)  - \psi (1)}{4},
\\
\Um_{[3]}^* & \textstyle =  \frac{\psi (4) + \psi (2)}{4} + \frac{\psi (3)  + \psi (1)}{4}
 - \Pi_{1},
\end{aligned}\end{equation}
where ${\adw_{\langle{3}\rangle}}\eq \frac{\sum_{I_4=0}^1\dw_{\langle{3}\rangle}(I_4 )}{2}  $ is the virtual marginal social welfare generated by APOs $\{2,3\}$, assuming that the MNO will reach agreement with APO $4$ with a probability of 0.5.~~~~~~

In Step $2$, the MNO bargains with the dummy APO $\langle{2}\rangle$), and thus the  bargaining result is straightforward: 
\begin{equation}\label{ex3:step2}
\begin{aligned}
\pa_{\langle{2}\rangle}^*   &\textstyle
 = 0,
\\
\Um_{[2]}^* &\textstyle =\Um_{[3]}^*=  \frac{\psi (4) + \psi (2)}{4} + \frac{\psi (3)  + \psi (1)}{4}
 - \Pi_{1}.
\end{aligned}
\end{equation}

In Step $1$, the MNO bargains with APO $1$.  
with a similar analysis in Appendix \ref{app:example-nbs}, 
we can obtain the NBS in Step $1$ (i.e., the APO $1$'s payoff) and the MNO's payoff under this NBS as follows.
\begin{equation}\label{ex3:step1}
\begin{aligned}
\pa_1^* ~& \textstyle = \frac{\adw_1}{2} = 
\frac{\psi (4) -  \psi (3)}{8} +
\frac{\psi (2) -  \psi (1)}{8}+ 
\frac{\psi (3) -  \psi (2)}{8}+ 
\frac{\psi (1) -  \psi (0)}{8} 
\\
\Um_{[1]}^*& \textstyle =  
\frac{\psi (4) +  \psi (3)}{8} +
\frac{\psi (2) +  \psi (1)}{8}+ 
\frac{\psi (3) +  \psi (2)}{8}+ 
\frac{\psi (1) +  \psi (0)}{8} .
\end{aligned}
\end{equation}

\textit{Verification of Property \ref{prop:GBS-seq} (Intra-Grouping Benefit):}
Comparing (\ref{ex3:step3}) with (\ref{ex1:step3}) and (\ref{ex1:step2}), we can find that the total payoff of APOs $2$ and $3$ increases if they form a group.

\textit{Verification of Property \ref{prop:SPE-group} (Inter-Grouping Benefit (Positive Externality)):}
Comparing (\ref{ex3:step1}) with (\ref{ex1:step1}), we can find that the payoff of APO $1$ also increases if APOs $2$ and $3$ form a group.  
Furthermore, comparing (\ref{ex3:step4}) with (\ref{ex1:step4}), the payoff of APO $4$ does not change.
$\hfill\blacksquare$

\vspace{3mm}

\noindent
\begin{tabular}{m{0.95\linewidth}}
\textbf{Example: Grouping Effect in Concurrent Bargaining.}
\\
\hline
\end{tabular}

\vspace{1mm}

Consider the bargaining between the MNO and an APO $n \notin \{2,3\}$ (i.e., those not in the group).
With a similar analysis in Appendix \ref{app:example-nbs},  
we can obtain the NBS and the MNO's payoff under this NBS as follows.
\begin{equation}\label{ex4:n14}
\begin{aligned}
\pa_n^*  ~ &\textstyle = \frac{\dwc_n}{2} = \frac{\psi (4)  - \psi (3)}{2},
\\
\Um_{[n]}^*  &\textstyle =   \frac{\psi (4)  + \psi (3)}{2} - \Pi_{-n}.
\end{aligned}
\end{equation}

Consider the bargaining between the MNO and the APO group $ \langle{3}\rangle \eq \{2,3\}$.
With a similar analysis, 
we can obtain the NBS with the APO group $\langle{3}\rangle$ and the MNO's payoff under this NBS as follows.
\begin{equation}\label{ex4:n23}
\begin{aligned}
\pa_{\langle{3}\rangle}^*~&\textstyle = \frac{\dwc_{\langle{3}\rangle}}{2} = \frac{\psi (4)  - \psi (2)}{2},
\\
\Um_{[\langle{3}\rangle]}^* &\textstyle=   \frac{\psi (4)  + \psi (2)}{2} - \Pi_{-\langle{3}\rangle}.
\end{aligned}
\end{equation}

\textit{Verification of Property \ref{prop:GBS-con} (Intra-Grouping Benefit):}
Comparing (\ref{ex4:n23}) with (\ref{ex2:n}), we can find that the total payoff of APOs $2$ and $3$ increases if they form a group (as $\frac{\psi (4)  - \psi (2)}{2} = \frac{\psi (4)  - \psi (3)}{2} + \frac{\psi (3)  - \psi (2)}{2} > \frac{\psi (4)  - \psi (3)}{2}\cdot 2$).

\textit{Verification of Property \ref{prop:NOE-group} (No Inter-Group Benefit (Non-Externality)):}
Comparing (\ref{ex4:n14}) with (\ref{ex2:n}), we can find that the payoff of each APO $n\notin \{2,3\}$ does not change under the new APO grouping structure.  $\hfill \blacksquare$

\begin{center}
\Large\textbf{(III) Proofs}
\end{center}

\vspace{-5mm}

%!TEX root = DataOffload_main_journal.tex
%SourceDoc DataOffload_main_journal.tex

\subsection{Proof for Lemma \ref{theorem:NBS-single} in Section \ref{sec:barg}}\label{app:proof-one-to-one-NBS}

\begin{proof}
To prove this lemma, we only need to prove that the NBS $\{x_n^*, z_n^*\}$ or $\{x_n^*, \pa_n^*\}$ given in this lemma uniquely solves the problem (\ref{eq:NBS-single}) or (\ref{eq:NBS-single-eq}).
Since (\ref{eq:NBS-single-eq}) is a strictly convex optimization problem, it must has a unique solution. Next we solve (\ref{eq:NBS-single-eq}) by sequential optimization on each variable. 
Specifically, we divide the derivation into two sequential steps: Step-I, finding the optimal $\pa_n^*$ under any feasible $x_n$; and Step-II, finding the optimal $x_n^*$ by substituting the optimal $\pa_n^*$ into problem (\ref{eq:NBS-single-eq}). 
The social optimality of the above sequential optimization method is guaranteed by the facts that both sub-problems in the above two steps are convex optimization.~~~~~~~~~~

\emph{Step-I: Finding the optimal $\pa_n^*$.}
Given any feasible $x_n$, the optimal $\pa_n^*$ is given by the following optimization problem:
\begin{equation}\label{eq:NBS-single-opt-pa}
\begin{aligned}
\max_{\pa_n} \ & \big[ \sw(x_n) - \pa_n \big] \cdot \pa_n 
\\
 \quad \mbox{s.t. } \ &  \sw(x_n) - \pa_n \geq 0,\ \pa_n \geq 0.
\end{aligned}
\end{equation}
The objective function of (\ref{eq:NBS-single-opt-pa}) is a quadratic function of $\pa_n$, and therefore the problem (\ref{eq:NBS-single-opt-pa}) is convex optimization.
Thus, we have the following optimal $\pa_n^*$ under any feasible $x_n$:
\begin{equation}
\textstyle
\pa_n^* = \frac{1}{2}\cdot \sw(x_n).
\end{equation}

\emph{Step-II: Finding the optimal $x_n^*$.}
Substitute the above optimal $\pa_n^*$ into (\ref{eq:NBS-single-eq}), we can find that the optimal $x_n^*$ for problem (\ref{eq:NBS-single-eq}) solves the following problem
\begin{equation}\label{eq:NBS-single-opt-xn}
\begin{aligned}
\max_{x_n}\  & \textstyle \  \frac{1}{4}\cdot	\sw(x_n)  \cdot  \sw(x_n) 
\\
\mbox{s.t. } \  & \sw(x_n) \geq 0,\  x_n \in[0, \xu_n].
\end{aligned}
\end{equation}
It is easy to see that $x_n^*$ equals to the social welfare maximization solution $x_n^o$ given by
\begin{equation}\label{eq:NBS-single-opt-xn-swm}
\begin{aligned}
x_n^o \triangleq \arg \max_{x_n} \ & \sw(x_n), 
\\
 \mbox{s.t. } \ & x_n \in[0, \xu_n].
\end{aligned}
\end{equation}
\end{proof}

%!TEX root = DataOffload_main_journal.tex
%SourceDoc DataOffload_main_journal.tex

\subsection{Proof for Lemma \ref{prop:nbs-x} in Section \ref{sec:barg-xo}}\label{app:proof-nbs-x}

\emph{Proof:} 
We first show that for any one-to-one bargaining with transferable utility, the disagreement points of bargainers will not affect the achieved social welfare, but only affect the welfare \emph{division} among bargainers. 
Then, the bargaining solution must maximize the social welfare, regardless of the disagreement points of bargainers (this result is analytically shown in Section IV.B). 

Take the one-to-one bargaining between the MNO and APO $n$ as an example. Let $\Um^0$ and $\Ua_n^0$ denote the disagreement points of the MNO and the APO $n$, respectively. 
%Denote $(\boldsymbol{x}_{-n}^* = \{x_i^*, \forall i\neq n\},\ \boldsymbol{z}_{-n}^* = \{z_i^*, \forall i\neq n\})$ as the bargaining solutions between the MNO and all APOs other than $n$. 
Then, the NBS $(x_n^*, z_n^*)$ between the MNO and APO $n$ is given by 
\begin{equation}\label{eq:NBS-single-hk}
\begin{aligned}
\max_{(x_n, z_n)\in \A}\  & \ \big( \Um(x_n, \boldsymbol{x}_{-n}^*\dt z_n, \boldsymbol{z}_{-n}^*) - \Um^0 \big) \cdot \big( \Ua_n (x_n\dt z_n) - \Ua_n^0 \big)
\\
\mbox{s.t. } \ & \ \Um(x_n, \boldsymbol{x}_{-n}^*\dt z_n, \boldsymbol{z}_{-n}^*) - \Um^0 \geq 0,
\\
&  \ \Ua_n (x_n\dt z_n) - \Ua_n^0 \geq 0, 
%,\\ &\ 0\leq x_n \leq \xu_n.
\end{aligned}
\end{equation}
where $\boldsymbol{x}_{-n}^* = \{x_i^*, \forall i\neq n\}$, $ \boldsymbol{z}_{-n}^* = \{z_i^*, \forall i\neq n\}$, and $(x_i^*, z_i^*)$ is the NBS between the MNO and   other APO $i \neq n$. 
We further notice that 
\begin{equation*}
\begin{aligned}
& \textstyle \Um(x_n, \boldsymbol{x}_{-n}^*\dt z_n, \boldsymbol{z}_{-n}^*)   =  \R(x_n, \boldsymbol{x}_{-n}^*) -  \sum_{i\neq n} z_i^* - z_n,
\\
 & \textstyle\Ua_n (x_n\dt z_n)  =  \Q_n(x_n) + z_n.
\end{aligned}
\end{equation*}
Then, the above optimization problem can be rewritten as
\begin{equation}\label{eq:NBS-single-xx}
\begin{aligned}
\max_{(x_n, z_n)\in \A} & \ \big( A (x_n) - z_n \big) \cdot \big( B(x_n) + z_n \big)\\
\mbox{s.t. } & \ A (x_n) - z_n \geq 0,\ B (x_n) + z_n \geq 0, 
%,\\ &\ 0\leq x_n \leq \xu_n.
\end{aligned}
\end{equation}
where $A(x_n) = \R(x_n, \boldsymbol{x}_{-n}^*) -  \sum_{i\neq n} z_i^* - \Um^0$, and $B(x_n) = \Q_n(x_n) - \Ua_n^0$. 

It is easy to obtain the following optimal solution for the above problem: (i) $z_n^* = \frac{A(x_n^*) - B(x_n^*)}{2}$, and (ii) $x_n^*$ is the solution of the following optimization problem
\begin{equation}\label{eq:NBS-single-yy}
\begin{aligned}
\max_{x_n \in \mathcal{X}_n} & \  A (x_n) +  B(x_n). \\ 
%,\\ &\ 0\leq x_n \leq \xu_n.
\end{aligned}
\end{equation}
Notice that  $A(x_n)  +B(x_n) = \R(x_n, \boldsymbol{x}_{-n}^*) -  \sum_{i\neq n} z_i^* - \Um^0 +  \Q_n(x_n) - \Ua_n^0$. We further notice that the terms  $ \Um^0$, $ \Ua^0$, and $\sum_{i\neq n} z_i^* $ are independent of $x_n$. 
Thus, the above optimization problem is equivalent to the following social welfare maximization problem 
\begin{equation}\label{eq:NBS-single-zz}
\begin{aligned}
\max_{x_n \in \mathcal{X}_n} & \ \R(x_n, \boldsymbol{x}_{-n}^*) + \Q_n(x_n) \eq \sw(x_n, \boldsymbol{x}_{-n}^*). \\ 
%,\\ &\ 0\leq x_n \leq \xu_n.
\end{aligned}
\end{equation}
That is, the NBS (or $x_n^*$ in the NBS) between the MNO and APO $n$ always maximizes the conditional social welfare, regardless of their disagreement points. 
Besides, the disagreement points will affect the payment $z_n^* = \frac{A(x_n^*) - B(x_n^*)}{2}$ in the NBS, as both $A(x_n^*)$ and $B(x_n^*)$ rely on the disagreement points. 

Based on the above discussion, we have the following important proposition. 
\begin{proposition}\label{prop:xx}
Given the NBS  $(x_i^*, z_i^*)$ between the MNO and every APO $i \neq n$, the NBS $(x_n^*, z_n^*)$ between the MNO and the APO $n$ always maximizes the social welfare $\sw(x_n, \boldsymbol{x}_{-n}^*)$, regardless of the disagreement points of the MNO and the APO $n$. That is,
\begin{equation}
x_n^* = \arg \max_{x_n \in \mathcal{X}_n} \ \sw(x_n, \boldsymbol{x}_{-n}^*). 
\end{equation}
\end{proposition}

{Next we show that under mild conditions, there is a unique bargaining solution for the entire one-to-many bargaining, and such a solution maximizes the overall social welfare.} 

Notice that the one-to-many bargaining consists of $N$ one-to-one bargaining, each corresponding to the bargaining between the MNO and a particular APO. 
Let $(x_n^*, z_n^*)$ denote the NBS between the MNO and every APO $n \in \N$, and $(\x^*, \z^*)$ denote the NBS of the entire one-to-many bargaining, where $\x^* = \{x_n^*, \forall n\in\N\}$ and $\z^* = \{z_n^*, \forall n\in\N\}$. 
Then, we need to show that the NBS $(\x^*, \z^*)$ is unique, and solves the  social welfare maximization problem
\begin{equation}\label{eq:NBS-multi-swm}
\begin{aligned}
\x^* = \arg \max_{\x} & \  \sw(\x), 
\\
 \mbox{s.t. } & \ x_n \in \mathcal{X}_n,\ \forall n\in \N.
\end{aligned}
\end{equation}

Consider the bargaining between the MNO and a particular APO $n \in \N$. 
By Proposition \ref{prop:xx}, the NBS (or $x_n^*$ in the NBS) between the MNO and the APO $n$ satisfies:
\begin{equation}
x_n^* = \arg \max_{x_n \in \mathcal{X}_n} \ \sw(x_n, \boldsymbol{x}_{-n}^*). 
\end{equation} 
Thus, the NBS (or $\x^*$ in the NBS) of the entire one-to-many bargaining satisfies: 
\begin{equation}\label{eq:fset}
\left\{
\begin{aligned}
x_1^* & = \arg \max_{x_1 \in \mathcal{X}_1} \ \sw(x_1, \boldsymbol{x}_{-1}^*) 
\\ 
x_2^* & = \arg \max_{x_2 \in \mathcal{X}_2} \ \sw(x_2, \boldsymbol{x}_{-2}^*) 
\\
... & ...
\\
x_N^* & = \arg \max_{x_N \in \mathcal{X}_N} \ \sw(x_N, \boldsymbol{x}_{-N}^*) 
\end{aligned}
\right.
\end{equation} 
Obviously, the social welfare maximization solution $\x^o$ must be a solution of the above equations, since $x_n^o= \arg \max_{x_n\in\mathcal{X}_n} \sw(x_n, \x_{-n}^o)$ for every $n\in\N$. 
Thus, if there is a unique solution for (\ref{eq:fset}), then it must be  $\x^o$. 

In general, however, there may have multiple solutions for (\ref{eq:fset}), depending on the form of $\sw(\x)$. 
To avoid this (multi-solution) situation, we introduce the following assumption:
\begin{assumption}
The MNO's serving cost $\C(\cdot)$ is an additive function. That is, $\C(x+y) = \C(x) + \C(y)$.
\end{assumption}
Let $b_n$ denote the MNO's resource consumption for delivering the traffic within the coverage area of AP $n$, and $b_0$ denote the MNO's resource consumption for delivering the traffic not within the coverage area of any AP. Obviously, the total resource consumption is $b = b_0 + \sum_{n\in\N} b_n$. The above assumption implies that 
\begin{equation}
\textstyle
\C(b) = \C(b_0) + \sum_{n\in\N} \C(b_n).
\end{equation}
That is, the total serving cost is the summation of the serving costs in all different areas. 
\revjrr{Notice that in cellular networks, the total serving area is divided into small areas (called \emph{cells}), and each cell is usually served by a particular base station. Thus, the actual total serving cost of the MNO can be viewed as the summation of the serving costs in all cells. 
Therefore, the above additive serving cost can be a good approximation to the actual serving cost when the cell size is small enough (hence each cell will not cover many APs), which will become more and more common given the current trend of reducing the cell size to increase the cellular capacity.}

Based on this assumption, the MNO's total serving cost without data offloading is 
\begin{equation}
\textstyle
\C(b(\boldsymbol{0})) = \C\left(\frac{S_0}{\theta_0}\right) + \sum_{n\in\N} \C\left(\frac{S_n}{\theta_n}\right).
\end{equation}
With data offloading, the MNO's total serving cost under the offloading profile $\x$ is
\begin{equation}
\textstyle
\C(b(\x)) = \C\left(\frac{S_0}{\theta_0}\right) + \sum_{n\in\N} \C\left(\frac{S_n-x_n}{\theta_n}\right).
\end{equation}
Thus, the serving cost reduction can be written as
\begin{equation}
\begin{aligned}
\R(\x)&\textstyle \eqi \C(b(\boldsymbol{0})) - \C(b(\x)) 
\\
&\textstyle= \sum_{n\in\N} \left( 
\C\left(\frac{S_n-x_n}{\theta_n}\right) - \C\left(\frac{S_n}{\theta_n}\right) \right)
\\
&\textstyle \eq \sum_{n\in\N} \R_n(x_n),
\end{aligned}
\end{equation}
where $\R_n(x_n) = \C\left(\frac{S_n-x_n}{\theta_n}\right) - \C\left(\frac{S_n}{\theta_n}\right)$. That is, $\R(\x)$ is also an additive function. 
Based on the above, we can further rewrite the social welfare $\sw(\x)$ as
\begin{equation}\label{eq:social-welfare-ap}
\begin{aligned} 
\sw(\x) & \textstyle= \sum_{n\in\N}\R_n(x_n) + \sum_{n\in \N} \Q_n(x_n) 
\\
& \textstyle\eq \sum_{n\in\N} \sw_n(x_n), 
\end{aligned}
\end{equation}
where $ \sw_n(x_n) = \R_n(x_n) + \Q_n(x_n)$. That is, $\sw(\x)$
is also an additive function.

Notice that $\sw_n(x_n)$ depends only on $x_n$, while not on $x_i$, $\forall i\neq n$. Thus, we can rewrite the function set (\ref{eq:fset}) as the following equivalent function set. 
\begin{equation}\label{eq:fset-2}
\left\{
\begin{aligned}
x_1^* & = \arg \max_{x_1 \in \mathcal{X}_1} \ \sw_1(x_1) 
%\\ 
%x_2^* & = \arg \max_{x_2 \in \mathcal{X}_2} \ \sw_2(x_2) 
\\
... & ...
\\
x_N^* & = \arg \max_{x_N \in \mathcal{X}_N} \ \sw_N(x_N) 
\end{aligned}
\right.
\end{equation} 
Obviously, the above function set has a unique solution, since all functions in (\ref{eq:fset-2}) are decoupled, each having a unique solution (as it is a strictly convex optimization problem). 

Denote $\x^* = (x_n^*)_{n=1,...,N}$ as the solution of (\ref{eq:fset-2}). Then, 
$$
\sw_n(x_n^* ) \geq \sw_n(x_n), \quad \forall x_n \neq x_n^*, \ \forall n\in\N.
$$
Thus, we have: for any $ \x \neq \x^*$, 
$$
\textstyle
 \sum_{n\in\N} \sw_n(x_n^* )\eq \sw(\x^*)  \geq \sw_n(x_n) \eq \sum_{n\in\N}  \sw(\x).
$$
That is, the solution $\x^*$ of (\ref{eq:fset-2}) maximizes the social welfare. 

%Formally, we summarize the above result in the following proposition. 
%\begin{proposition}
Based on the above analysis, we can easily obtain the result in Lemma 2.
That is, if the MNO's serving cost $\C(\cdot)$ is an additive function, then the NBS $(\x^*, \z^*)$ of the one-to-many bargaining is unique and maximizes the social welfare $\sw(\x)$.
%, i.e., 
%\begin{equation}
%\x^* = \arg \max_{\x} \  \sw(\x), \quad \mbox{s.t. } \ x_n \in \mathcal{X}_n,\ \forall n\in \N.
%\end{equation}

$\hfill \blacksquare$

%!TEX root = DataOffload_main_journal.tex
%SourceDoc DataOffload_main_journal.tex

%%%%%%%%%%%%%%%%%%%%%%%%%%%%%%%%%%%%%%%%%%%%%%%%%%%
%%%%%%%%%%%%%%%%%%%%%%%%%%%%%%%%%%%%%%%%%%%%%%%%%%%

\subsection{Proof for Lemma \ref{lemma:NBS-seq-N} in Section \ref{sec:payoff}}\label{app:proof-NBS-seq-N}

\begin{proof}
We first notice that the objective function of   (\ref{eq:NBS-seq-N}) is a quadratic function of $\paN$. 
Thus, we have the following optimal solution for (\ref{eq:NBS-seq-N}) when there is no constraint in (\ref{eq:NBS-seq-N}):
$$\textstyle
\paN^* = \frac{\dw_N}{2}.
$$

We next show that the above optimal $\paN^*$ is located in the feasible set of (\ref{eq:NBS-seq-N}), that is, it satisfies the constraints of (\ref{eq:NBS-seq-N}).
Recall that $\x^{\stx}$ is equivalent to the social welfare maximization solution. Thus, we have:
$$
\sw(\x_{N \mi 1}^{\stx}, \rmkk{x_N^{\stx}}) \geq \sw(\x_{N \mi 1}^{\stx}, \rmkk{x_N} ),\quad \forall x_N \neq x_N^{\stx}.
$$
This implies that $\dw_N = \sw(\x_{N \mi 1}^{\stx}, \rmkk{x_N^{\stx}})- \sw(\x_{N \mi 1}^{\stx}, \rmkk{0}) \geq 0$, and thus both constraints of (\ref{eq:NBS-seq-N}) are satisfied under the optimal $\paN^*$.
By independence of irrelevant alternatives (IIA), the above $\paN^*$ is also the optimal solution of 
(\ref{eq:NBS-seq-N}) with constraints.
\end{proof}

%!TEX root = DataOffload_main_journal.tex
%SourceDoc DataOffload_main_journal.tex

%%%%%%%%%%%%%%%%%%%%%%%%%%%%%%%%%%%%%%%%%%%%%%%%%%%
%%%%%%%%%%%%%%%%%%%%%%%%%%%%%%%%%%%%%%%%%%%%%%%%%%%

\subsection{Proof for Lemma \ref{lemma:NBS-seq-N-1} in Section \ref{sec:payoff}}\label{app:proof-NBS-seq-N-1}

\begin{proof}
Similar to Lemma \ref{lemma:NBS-seq-N}, we have the following optimal solution for (\ref{eq:NBS-seq-N-1-xxxxaaaa}) when there is no constraint in (\ref{eq:NBS-seq-N-1-xxxxaaaa}):~~~~~~~~~
$$\textstyle
\paNr^* = \frac{\adw_{N\mi 1}}{2}.
$$
Thus, we only need to prove that the above optimal 
$\paNr^*$ satisfies the constraints of (\ref{eq:NBS-seq-N-1-xxxxaaaa}). 
Similarly, we first have:
$$\textstyle
{\dw}_{N\mi 1}(I_N \ei 1) = \sw(\x_{N \mi 2}^{\stx}, \rmkk{x_{N\mi 1}^{\stx}}, x_N^{\stx}) - \sw(\x_{N \mi 2}^{\stx}, \rmkk{0},  x_N^{\stx}) \geq 0,
$$
since $\x^{\stx}$ is the social welfare maximization solution. 
We further notice that
\begin{equation}\label{eq:app-second-derivative}
\textstyle
\frac{\partial( \frac{\partial \sw(\x) }{\partial x_n} )}{\partial x_m}
=
\frac{\partial^2 \sw(\x) }{\partial x_m \partial x_n}
=- \frac{\C''(\B(\x))}{\theta_m  \theta_n} \leq 0,\ \forall m\neq n,
\end{equation}
which implies that the more traffic offloaded to other APs,
the less marginal welfare generated by an AP $n$ (with the same traffic offloading volume $x_n$). 
%Intuitively, this is because the BS's serving cost is a convex function of resource consumption, and thus the less resource consumed (i.e., the more traffic already offloaded to other APs), the less benefit (serving cost reduction) generated by AP $n$'s offloading. 
By (\ref{eq:app-second-derivative}), we have:
\begin{equation*}
\begin{aligned}
\textstyle
{\dw}_{N\mi 1} & \textstyle(I_N \ei 0) =  \sw(\x_{N \mi 2}^{\stx}, \rmkk{x_{N\mi 1}^{\stx}}, 0) - \sw(\x_{N \mi 2}^{\stx}, \rmkk{0},  0) \\
& \textstyle \geq \sw(\x_{N \mi 2}^{\stx}, \rmkk{x_{N\mi 1}^{\stx}}, x_N^{\stx}) - \sw(\x_{N \mi 2}^{\stx}, \rmkk{0},  x_N^{\stx})  \geq 0.
\end{aligned}
\end{equation*}
Based on above, we immediately have: 
$$\textstyle
\adw_{N\mi 1} = \frac{1}{2} \cdot {\dw}_{N\mi 1}(I_N\ei1)   + \frac{1}{2}\cdot {\dw}_{N\mi 1}(I_N\ei0)   \geq 0.
$$
Thus, both constraints of (\ref{eq:NBS-seq-N-1-xxxxaaaa}) are satisfied under the optimal $\paNr^*$ given above.
\end{proof}

%!TEX root = DataOffload_main_journal.tex
%SourceDoc DataOffload_main_journal.tex

%%%%%%%%%%%%%%%%%%%%%%%%%%%%%%%%%%%%%%%%%%%%%%%%%%%
%%%%%%%%%%%%%%%%%%%%%%%%%%%%%%%%%%%%%%%%%%%%%%%%%%%

\subsection{Proof for Lemma \ref{lemma:NBS-seq-n} in Section \ref{sec:payoff}}\label{app:proof-NBS-seq-n}

\revjr{
\begin{proof}
%We prove the lemma by induction. Specifically, if we prove the following two statements, then by induction, we can show that the NBS $\pan^*$ at any Step $n$ (between the MNO and any APO $n$) is  characterized by (\ref{eq:NBS-seq-N-n-pa}). 
\revjrr{We prove the lemma by induction. Namely, we can prove the lemma by proving the following two statements:
\begin{itemize}
\item
\emph{Statement 1:} 
The NBS $\paN^*$ in the last Step $N$ (for APO $N$) is characterized by (\ref{eq:NBS-seq-N-n-pa});
\item 
\emph{Statement 2:} 
If the NBS  $\{\pa^*_i\}_{i=k, k+1,...,N}$ after Step $k-1$ (i.e., for APOs $k,k+1,...,N$) are all characterized  by (\ref{eq:NBS-seq-N-n-pa}), then the NBS $\pa_{k-1}^*$ in Step $k-1$ (i.e., for APO $k-1$) is also characterized by (\ref{eq:NBS-seq-N-n-pa}). 
\end{itemize}}

%We prove the lemma by induction. 
%Specifically, we show that (i) the NBS $\paN^*$ in the last Step $N$ (between the MNO and APO $N$) is given by (\ref{eq:NBS-seq-N-n-pa}), and (ii) if the NBS after Step $k$ (between the MNO and APOs $k+1,...,N$), i.e.,  $\{\pa^*_i\}_{i=k+1,...,N}$ are all given by (\ref{eq:NBS-seq-N-n-pa}), then the NBS $\pa_k^*$ in Step $k$ (between the MNO and APO $k$) is also given by (\ref{eq:NBS-seq-N-n-pa}). 
%Then, by induction, the NBS $\pan^*$ between the MNO and any APO $n$ is given by (\ref{eq:NBS-seq-N-n-pa}). 

\emph{Proof for Statement 1:} 
By Lemma \ref{lemma:NBS-seq-N}, we can easily find that $\paN^* = \frac{ \dw_N }{2}$, which is characterized by (\ref{eq:NBS-seq-N-n-pa}). 
Besides, the MNO's payoff is $\Um^{*}_{[N]} = \frac{ \mw_N }{2}  - \Pi_{N\mi 1}$, which is characterized by (\ref{eq:NBS-seq-N-n-um}).

\emph{Proof for Statement 2:} 
Suppose that the NBS $\{\pa^*_i\}_{i=k,k+1...,N}$ after Step $k-1$ (i.e., for APOs $k, k+1,...,N$) are all characterized by (\ref{eq:NBS-seq-N-n-pa}). 
Accordingly, the MNO's payoff after Step $k-1$ can be characterize by (\ref{eq:NBS-seq-N-n-um}). 
Now we prove that the NBS $\pa_{k-1}^*$ in Step $k-1$ (i.e., for APO $k-1$) is also characterized by (\ref{eq:NBS-seq-N-n-pa}). 

Since the MNO's payoff in Step $k$ can be characterize by (\ref{eq:NBS-seq-N-n-um}), we can easily find that when bargaining with APO $k-1$ in Step $k-1$, the MNO's 
disagreement point is
\begin{equation*}
\begin{aligned}
\textstyle
\Um^0_{[k-1]} = \sum_{I_{k+1} =0 }^1\dii \sum_{I_{N} =0 }^1	
\left( \frac{  \sw(\x_{k \mi 2}^{\stx}, 0, \rmkk{x_{k}^{\stx}}, I_{k+1}  {x_{k+ 1}^{\stx}},\dii , I_N  x_N^{\stx})}{2^{N- k+1}}\right.
\\
\textstyle
\left. + \frac{\sw(\x_{k \mi 2}^{\stx}, 0, \rmkk{0}, I_{k+ 1}  {x_{k+ 1}^{\stx}},\dii , I_N  x_N^{\stx})
           }{  2^{N- k+1}} \right) - \Pi_{k\mi 2},
\end{aligned}
\end{equation*}
and the MNO's payoff, if reaching an agreement $\pa_{k-1} = \paxx $ with APO $k-1$, is 
\begin{equation*}
\begin{aligned}
\textstyle
\Um_{[k-1]} = \sum_{I_{k+1} =0 }^1\dii \sum_{I_{N} =0 }^1	 \left(\frac{  \sw(\x_{k \mi 2}^{\stx}, \rmkk{x_{k-1}^{\stx}}, \rmkk{x_{k}^{\stx}}, I_{k+1}  {x_{k+ 1}^{\stx}},\dii , I_N  x_N^{\stx})}{2^{N- k+1}}\right.
\\ 
\textstyle
\left. + \frac{\sw(\x_{k \mi 2}^{\stx}, \rmkk{x_{k-1}^{\stx}}, \rmkk{0}, I_{k+ 1}  {x_{k+ 1}^{\stx}},\dii , I_N  x_N^{\stx})
           }{  2^{N- k+1}}\right) - \Pi_{k\mi 2} - \paxx.
\end{aligned}
\end{equation*}
Thus, the NBS $\pa_{k-1}^*$ between the MNO and APO $k-1$ is given by the following optimization problem
\begin{equation}\label{eq:NBS-seq-n-xxx}
\begin{aligned}
\max_{ v} & \textstyle \ \big[ \adw_{k-1} - v \big] \cdot v \\
\mbox{s.t. } & \ \adw_{k-1}  - v \geq 0,\ v \geq 0, 
\end{aligned}
\end{equation}
where $\adw_{k-1} = \Um_{[k-1]} - \Um^0_{[k-1]}$. 

Solving the above problem, we can obtain the NBS $\pa_{k-1}^*$ in Step $k-1$, i.e., 
\begin{equation*}
\begin{aligned}
\textstyle
\pa_{k-1}^* = \frac{\adw_{k-1}}{2} = 
\sum_{I_{k} =0 }^1\dii \sum_{I_{N} =0 }^1	\left( \frac{  \sw(\x_{k \mi 2}^{\stx}, \rmkk{x_{k-1}^{\stx}}, I_{k}  {x_{k}^{\stx}},\dii , I_N  x_N^{\stx})  }{ 2\cdot 2^{N- k+1}} \right.
\\
\textstyle
\left.
 +\frac{\sw(\x_{k \mi 2}^{\stx}, 0,  I_{k}  {x_{k}^{\stx}},\dii , I_N  x_N^{\stx})
           }{ 2\cdot 2^{N- k+1}} \right) - \Pi_{k\mi 2},
\end{aligned}
\end{equation*}
which is exactly characterized by (\ref{eq:NBS-seq-N-n-pa}). Accordingly, we can easily check that the MNO's payoff in Step $k-1$ is also characterized by (\ref{eq:NBS-seq-N-n-um}).  
\end{proof}}

%!TEX root = DataOffload_main_journal.tex
%SourceDoc DataOffload_main_journal.tex

%%%%%%%%%%%%%%%%%%%%%%%%%%%%%%%%%%%%%%%%%%%%%%%%%%%
%%%%%%%%%%%%%%%%%%%%%%%%%%%%%%%%%%%%%%%%%%%%%%%%%%%

\subsection{Proof for Theorem \ref{theorem:NBS-seq} in Section \ref{sec:payoff}}\label{app:proof-theorem-NBS-seq}

\begin{proof}
By Lemma \ref{prop:nbs-x} and Lemma \ref{lemma:NBS-seq-n}, we can prove the theorem directly.
\end{proof}

%!TEX root = DataOffload_main_journal.tex
%SourceDoc DataOffload_main_journal.tex

%%%%%%%%%%%%%%%%%%%%%%%%%%%%%%%%%%%%%%%%%%%%%%%%%%%
%%%%%%%%%%%%%%%%%%%%%%%%%%%%%%%%%%%%%%%%%%%%%%%%%%%

\subsection{Proofs for Properties \ref{prop:EMA-seq} and \ref{prop:IOC-seq}  in Section \ref{sec:payoff}}\label{app:proof-property-seq}

\begin{proof}
We first prove Property \ref{prop:IOC-seq} (Invariance to AP-order Changing). 
By (\ref{eq:NBS-seq-N-n-um}), the MNO's payoff can be written as
\begin{equation}\label{eq:NBS-seq-N-1-um}
\begin{aligned}
\Um^{*}_{[1]} & \textstyle = \sum_{I_{2} }\dii \sum_{I_{N} }	 \frac{{\mw}_{1}(I_{2};\dii ; I_N )}{ 2^{N\mi 1}\cdot2} - \Pi_{0}
\\
&\textstyle = \sum_{I_{1} }\sum_{I_{2} } \dii \sum_{I_{N}}	\frac{{\mw}_{0}(I_{1};\dii ; I_N )}{ 2^{N\mi 1}\cdot2} ,
\end{aligned}
\end{equation}
where
${\mw}_{0}(I_{1};\dii  ;I_N ) \eq \sw(\rmkk{I_1 {x_{1}^{\stx}}}, I_{2}  {x_{2}^{\stx}},\dii , I_N  x_N^{\stx})$. The second line follows because $ {\mw}_{1}(I_{2};\dii  ;I_N ) = {\mw}_{0}(\rmkk{I_1 \ei{1}};I_{2}; \dii  ;I_N ) + {\mw}_{0}(\rmkk{I_1\ei{0}};I_{2}; \dii  ;I_N ) $ and $\Pi_{0} = 0$. 

Intuitively, the above MNO's payoff (\ref{eq:NBS-seq-N-1-um}) can be viewed as the \emph{average social welfare} under such a situation that the MNO and every APO will reach agreement with a probability of 0.5. 
By (\ref{eq:NBS-seq-N-1-um}), we can easily find that the AP-order does \emph{not} affect  the MNO's payoff in the S-NBS.

We then prove Property \ref{prop:EMA-seq} (Early-Mover Advantage). 
Take an arbitrary APO $n$ as an example. 
By Lemma \ref{lemma:NBS-seq-n}, its payoff is 
\begin{equation*}
\begin{aligned}
&\textstyle \qquad \pan^* 
 = \frac{ \adw_{n} }{2} =   \sum_{I_{n+1} }\dii \sum_{I_{N} }	 \frac{{\dw}_{n}(I_{n+1};\dii ; I_N )}{ 2^{N\mi n}\cdot2}
\\
& \textstyle =\sum_{I_{n+2} }\dii \sum_{I_{N} } \frac{{\dw}_{n}(I_{n+1}\ei 0;I_{n+2};\dii ; I_N )  
+ {\dw}_{n}(I_{n+1} \ei 1;I_{n+2};\dii ; I_N ) }{ 2^{N\mi n}\cdot2},
\end{aligned}
\end{equation*}
where
\begin{equation*}
\begin{aligned}
{\dw}_{n}(I_{n+1};\dii  ;I_N ) = & \sw(\x_{n \mi 1}^{\stx}, \rmkk{x_{n}^{\stx}}, I_{n+ 1}  {x_{n+ 1}^{\stx}},\dii , I_N  x_N^{\stx})
\\
& - \sw(\x_{n \mi 1}^{\stx}, \rmkk{0}, I_{n+ 1}  {x_{n+ 1}^{\stx}},\dii , I_N  x_N^{\stx}).
\end{aligned}
\end{equation*}

Now suppose APO $n$ moves backward by one step. That is,  APO $n$ becomes $n\ad 1$ (denoted by $\langle{n\ad 1}\rangle$ to avoid confusion), and the original APO $n\ad 1$ becomes $n$ (denoted by $\langle{n}\rangle$) in the new bargaining sequence.
Then, by Lemma \ref{lemma:NBS-seq-n}, the APO $n$'s payoff in the new bargaining sequence is 
\begin{equation*}
\begin{aligned}
\pa_{\langle{n+ 1}\rangle}^* & \textstyle
 = \frac{ \adw_{\langle{n+ 1}\rangle} }{2} =   \sum_{I_{n+2} }\dii \sum_{I_{N} }	 \frac{{\dw}_{\langle{n+1}\rangle}(I_{n+2};\dii ; I_N )}{ 2^{N\mi n \mi 1}\cdot2},
\end{aligned}
\end{equation*}
where ${\dw}_{\langle{n+ 1}\rangle}(I_{n+2};\dii  ;I_N ) $
\begin{equation*}
\begin{aligned}
= & \sw(\x_{n\mi 1}^{\stx}, \rmkk{x_{\langle{n}\rangle}^{\stx}, x_{\langle{n+1}\rangle}^{\stx}}, I_{n+ 2}  {x_{n+ 2}^{\stx}},\dii , I_N  x_N^{\stx}) \quad\quad
\\
 & - \sw(\x_{n \mi 1}^{\stx}, \rmkk{x_{\langle{n}\rangle}^{\stx},  0}, I_{n+ 2}  {x_{n+ 2}^{\stx}},\dii , I_N  x_N^{\stx})
\\
= & \sw(\x_{n\mi 1}^{\stx}, \rmkk{x_{n}^{\stx}, x_{ n+1 }^{\stx}}, I_{n+ 2}  {x_{n+ 2}^{\stx}},\dii , I_N  x_N^{\stx}) \quad\quad
\\
 & - \sw(\x_{n \mi 1}^{\stx}, \rmkk{0, x_{n+1}^{\stx}}, I_{n+ 2}  {x_{n+ 2}^{\stx}},\dii , I_N  x_N^{\stx}).
\end{aligned}
\end{equation*}
The last line follows because $x_{\langle{n}\rangle}^{\stx} = x_{n+1}^{\stx}$ and $x_{\langle{n+1}\rangle}^{\stx} = x_{n}^{\stx}$. Here we impliedly use the fact that the social optimal traffic offload profile $\x^o$ are identical under any bargaining sequence.

Based on above, we can easily find that 
$$
{\dw}_{n}(I_{n+1}\ei 1; I_{n+2};\dii  ;I_N ) =
{\dw}_{\langle{n+ 1}\rangle}(I_{n+2};\dii  ;I_N ). 
$$
By the concavity of $\sw(\cdot)$, we further have:
$$
{\dw}_{n}(I_{n+1}\ei 0; I_{n+2};\dii  ;I_N ) \geq
{\dw}_{\langle{n+ 1}\rangle}(I_{n+2};\dii  ;I_N ). 
$$
Therefore, we have $\pan^* \geq \pa_{\langle{n+ 1}\rangle}^*$, that is, APO $n$ can achieve a higher payoff when bargaining earlier with the BS.

Intuitively, from (\ref{eq:NBS-seq-N-n-pa}) we can view the APO $n$'s payoff as (half of) the \emph{average marginal social welfare} generated by APO $n$ under such a situation that all APOs prior to $n$ always reach agreements with the MNO while every posterior APO reaches agreement with the MNO or disagrees  with a probability of 0.5. 
Furthermore, the concavity of $\sw(\cdot)$ implies that the more APOs accept the bargaining solution,
the less marginal social welfare generated by an additional APO (under the same traffic offloading volume). 
Thus, we can immediately find that the APOs bargaining earlier with the MNO is more likely to generate larger average marginal social welfare, and therefore get higher payoff. 
\end{proof}

%!TEX root = DataOffload_main_journal.tex
%SourceDoc DataOffload_main_journal.tex

%%%%%%%%%%%%%%%%%%%%%%%%%%%%%%%%%%%%%%%%%%%%%%%%%%%
%%%%%%%%%%%%%%%%%%%%%%%%%%%%%%%%%%%%%%%%%%%%%%%%%%%

\subsection{Proof for Lemma \ref{lemma:NBS-con} in Section \ref{sec:payoff}}
\label{app:proof-NBS-con}

\begin{proof}
By definition, we can easily find that the NBS between the
MNO and the APO $n$ is given by
\begin{equation}\label{eq:NBS-con-n}
\begin{aligned}
\max_{ \pan} & \ \big(\dwc_n - \pan \big) \cdot \pan \\
\mbox{s.t. } & \ \dwc_n  - \pan \geq 0,\ \pan \geq 0, 
\end{aligned}
\end{equation}
where $\dwc_n \eq \sw(\x_{  \mi n}^{\stx}, \rmkk{x_n^{\stx}})- \sw(\x_{\mi n }^{\stx}, \rmkk{0})$.

Similar to (\ref{eq:NBS-seq-N}), the objective function of   (\ref{eq:NBS-con-n}) is a quadratic function of $\pan$. 
Thus, we have the following optimal solution for (\ref{eq:NBS-con-n}) when there is no constraint in (\ref{eq:NBS-con-n}):
$$\textstyle
\pan^* = \frac{\dwc_n}{2}.
$$
Thus, we only need to prove that the above optimal 
$\pan^*$ satisfies the constraints of (\ref{eq:NBS-con-n}). We can easily obtain that 
$$\textstyle
 \sw(\x_{  \mi n}^{\stx}, \rmkk{x_n^{\stx}})- \sw(\x_{\mi n }^{\stx}, \rmkk{0}) \geq 0,
$$
since $\x^{\stx}$ is the social welfare maximization solution. 
This implies that $\dwc_n =\sw(\x_{  \mi n}^{\stx}, \rmkk{x_n^{\stx}})- \sw(\x_{\mi n }^{\stx}, \rmkk{0}) \geq 0$, and thus both constraints of (\ref{eq:NBS-con-n}) are satisfied under the optimal $\pan^*$.
\end{proof}

%%%%%%%%%%%%%%%%%%%%%%%%%%%%%%%%%%%%%%%%%%%%%%%%%%%
%%%%%%%%%%%%%%%%%%%%%%%%%%%%%%%%%%%%%%%%%%%%%%%%%%%

\subsection{Proof for Theorem \ref{theorem:NBS-con} in Section \ref{sec:payoff}}\label{app:proof-theorem-NBS-con}

\begin{proof}
By Lemma \ref{prop:nbs-x} and Lemma \ref{lemma:NBS-con}, we can prove the theorem directly.
\end{proof}

%!TEX root = DataOffload_main_journal.tex
%SourceDoc DataOffload_main_journal.tex

%%%%%%%%%%%%%%%%%%%%%%%%%%%%%%%%%%%%%%%%%%%%%%%%%%%
%%%%%%%%%%%%%%%%%%%%%%%%%%%%%%%%%%%%%%%%%%%%%%%%%%%

\subsection{Proofs for Properties \ref{prop:IIC-con} and \ref{prop:CMT-con} in Section \ref{sec:payoff}}\label{app:proof-property-con}

\begin{proof}
By (\ref{eq:NBS-con-n-pa}), we can easily prove Properties \ref{prop:IIC-con}  (Invariance to AP-index Changing).
Intuitively, this is because all APOs are symmetric (in terms of the bargaining order) in the concurrent bargaining, and thus the AP-index has no impact on the APO's payoff.

Compare (\ref{eq:NBS-con-n-pa}) and (\ref{eq:NBS-seq-N-n-pa}), we can further find that in the concurrent bargaining, every APO $n$ achieves a payoff equal to its payoff in the sequential bargaining when it bargains with the MNO in the last step. 
By Property \ref{prop:EMA-seq}, this is exactly the worst payoff that it would achieve in the sequential bargaining.
\end{proof}

%!TEX root = DataOffload_main_journal.tex
%SourceDoc DataOffload_main_journal.tex

%%%%%%%%%%%%%%%%%%%%%%%%%%%%%%%%%%%%%%%%%%%%%%%%%%%
%%%%%%%%%%%%%%%%%%%%%%%%%%%%%%%%%%%%%%%%%%%%%%%%%%%

\subsection{Proof for Property \ref{prop:GBS-seq} in Section \ref{sec:barg:group}}\label{app:proof-prop:GBS-seq}

\begin{proof}
For convenience, we focus only on the merge of two successive APOs, say $n$ and $n\ad 1$.\footnote{Note that when studying the merge of two non-successive APs, say $n$ and $n\ad 2$, we have to consider the bargaining order of the merged group $\{n, n\ad 2\}$ and the APO between the APOs in the merged group, i.e., $n\ad 1$.} 
Later we will show that such a discussion is sufficient, since it leads to the unique outcome where all APOs form a single group.

For notation convenience, 
we denote the new player (i.e., the merged group $\{n, n\ad 1\}$) by $\langle{n}\rangle$.
To keep the indexes of APOs $n\ad 2, \dii, N$, we introduce a dummy APO $\langle{n\ad 1}\rangle$, that is, APO $\langle{n\ad 1}\rangle$ offers zero resource for data offloading, and receives zero payoff from the MNO.
By Lemma \ref{lemma:NBS-seq-n}, the payoff of the new player $\langle{n}\rangle$, i.e., the total payoff of APOs $n$ and $ n\ad 1$, is 
$$\textstyle
\pa_{\langle{n}\rangle}^* = \frac{ \adw_{\langle{n}\rangle} }{2} =   \sum_{I_{n+1} }\dii \sum_{I_{N} }	 \frac{{\dw}_{\langle{n}\rangle}(I_{n+1};\dii ; I_N )}{ 2^{N\mi n}\cdot2},
$$
where ${\dw}_{\langle{n}\rangle}(I_{n+1};\dii  ;I_N ) =$
\begin{equation*}
\begin{aligned}
 & \sw(\x_{n \mi 1}^{\stx}, \rmkk{ \{x_{n}^{\stx}, x_{n+1}^{\stx} \}}, I_{n+ 1}  {x_{\langle{n+1}\rangle}^{\stx}}, I_{n+ 2}  {x_{n+ 2}^{\stx}},\dii , I_N  x_N^{\stx} ) 
\\
&-  \sw(\x_{n \mi 1}^{\stx}, \rmkk{\{0,0\}}, I_{n+ 1}  {x_{\langle{n+1}\rangle}^{\stx}} , I_{n+ 2}  {x_{n+ 2}^{\stx}}, \dii , I_N  x_N^{\stx}).
\end{aligned}
\end{equation*}
Notice that $x_{\langle{n+1}\rangle}^{\stx} = 0$ for dummy AP. Thus, we further have:
\begin{equation*}
\begin{aligned}
{\dw}_{\langle{n}\rangle}(I_{n+1};\dii  ;I_N ) = & \sw(\x_{n \mi 1}^{\stx}, \rmkk{  x_{n}^{\stx}, x_{n+1}^{\stx}  }, I_{n+ 2}  {x_{n+ 2}^{\stx}},\dii , I_N  x_N^{\stx} ) 
\\
&-  \sw(\x_{n \mi 1}^{\stx}, \rmkk{ 0,0 }, I_{n+ 2}  {x_{n+ 2}^{\stx}}, \dii , I_N  x_N^{\stx}).
\end{aligned}
\end{equation*}
Intuitively, ${\dw}_{\langle{n}\rangle}(I_{n+1};\dii  ;I_N )$ is the marginal welfare generated by both APOs $n$ and $n \ad 1$ in the group together, suppose the MNO  will ($I_i=1$) or will not ($I_i=0$) reach an agreement with every posterior APO $i$, $i= n\ad 2,\dii,N$.
Since ${\dw}_{\langle{n}\rangle}(I_{n+1};\dii  ;I_N )$ is independent of $ I_{n+1}$, the total payoff of APOs $n$ and $n\ad 1$ (when merging together) can be written as
\begin{equation*}
\begin{aligned}
\pa_{\langle{n}\rangle}^* =
%&\textstyle \sum_{I_{n+2} }\dii \sum_{I_{N} } \big[ \sw(\x_{n \mi 1}^{\stx}, \rmkk{  x_{n}^{\stx}, x_{n+1}^{\stx}  }, I_{n+ 2}  {x_{n+ 2}^{\stx}},\dii , I_N  x_N^{\stx} ) 
%\\
%&\quad\quad\textstyle -  \sw(\x_{n \mi 1}^{\stx}, \rmkk{ 0,0 }, I_{n+ 2}  {x_{n+ 2}^{\stx}}, \dii , I_N  x_N^{\stx}) \big] \cdot 
%\frac{2}{2^{N\mi n}\cdot2}
%\\
&\textstyle \sum_{I_{n+2} }\dii \sum_{I_{N} }  
\frac{\phi_3 - \phi_0 + 
\phi_3 - \phi_0}{2^{N\mi n}\cdot2},
\end{aligned}
\end{equation*}
where 
\begin{equation*}
\begin{aligned}
\phi_3 &\textstyle \eq  \sw(\x_{n \mi 1}^{\stx}, \rmkk{  x_{n}^{\stx}, x_{n+1}^{\stx}  }, I_{n+ 2}  {x_{n+ 2}^{\stx}},\dii , I_N  x_N^{\stx} ) ,
\\
\phi_0 & \textstyle \eq \sw(\x_{n \mi 1}^{\stx}, \rmkk{ 0,0 }, I_{n+ 2}  {x_{n+ 2}^{\stx}}, \dii , I_N  x_N^{\stx}).
\end{aligned}
\end{equation*}

Now we compute the payoffs of APOs $n$, $n\ad 1$ when they bargain independently with the MNO. 
By Lemma \ref{lemma:NBS-seq-n}, we have
\begin{equation*}
\begin{aligned}
\pan^*~~ &\textstyle
 = \frac{ \adw_{n} }{2} =   \sum_{I_{n+1} }\dii \sum_{I_{N} }	 \frac{{\dw}_{n}(I_{n+1};\dii ; I_N )}{ 2^{N\mi n}\cdot2}, \\
\pa_{n+1}^* & \textstyle
 = \frac{ \adw_{n+1} }{2} =   \sum_{I_{n+2} }\dii \sum_{I_{N} }	 \frac{{\dw}_{n+1}(I_{n+2};\dii ; I_N )}{ 2^{N\mi n \mi 1}\cdot2},
\end{aligned}
\end{equation*}
where
\begin{equation*}
\begin{aligned}
{\dw}_{n}(I_{n+1};\dii  ;I_N ) = & \sw(\x_{n \mi 1}^{\stx}, \rmkk{x_{n}^{\stx}}, I_{n+ 1}  {x_{n+ 1}^{\stx}},\dii , I_N  x_N^{\stx})
\\
& - \sw(\x_{n \mi 1}^{\stx}, \rmkk{0}, I_{n+ 1}  {x_{n+ 1}^{\stx}},\dii , I_N  x_N^{\stx}),
\\
{\dw}_{n+1}(I_{n+2};\dii  ;I_N ) = & \sw(\x_{n}^{\stx}, \rmkk{x_{n+1}^{\stx}}, I_{n+ 2}  {x_{n+ 2}^{\stx}},\dii , I_N  x_N^{\stx})
\\
 & - \sw(\x_{n }^{\stx}, \rmkk{0}, I_{n+ 2}  {x_{n+ 2}^{\stx}},\dii , I_N  x_N^{\stx}).
\end{aligned}
\end{equation*}
Thus, the total payoff of APOs $n$ and $n\ad 1$ (when not merging together) can be written as
\begin{equation*}
\begin{aligned}
\pan^*+ \pa_{n+1}^* = & \textstyle  \sum_{I_{n+2} }\dii \sum_{I_{N} }	\big[
{\dw}_{n}(I_{n+1\ei0}; I_{n+2};\dii ; I_N ) 
\\
 + {\dw}_{n}(I_{n+1\ei1};& \textstyle  I_{n+2};\dii ; I_N ) 
+ 2 \cdot {\dw}_{n+1}(I_{n+2};\dii ; I_N )
\big] \cdot \frac{1}{ 2^{N\mi n}\cdot2} 
\\
= & \textstyle \sum_{I_{n+2} }\dii \sum_{I_{N} } \frac{\phi_2 - \phi_0 + \phi_3 - \phi_1 + \phi_3 - \phi_2 + \phi_3-\phi_2}{2^{N\mi n}\cdot2} 
\\
= & \textstyle \sum_{I_{n+2} }\dii \sum_{I_{N} } \frac{ \phi_3 - \phi_0  + \phi_3 - \phi_1 + \phi_3-\phi_2}{2^{N\mi n}\cdot2},
\end{aligned}
\end{equation*}
where 
\begin{equation*}
\begin{aligned}
\phi_2 &\textstyle \eq  \sw(\x_{n \mi 1}^{\stx}, \rmkk{   x_{n}^{\stx} ,0 }, I_{n+ 2}  {x_{n+ 2}^{\stx}},\dii , I_N  x_N^{\stx} ) ,
\\
\phi_1 & \textstyle \eq \sw(\x_{n \mi 1}^{\stx}, \rmkk{0,  x_{n+1}^{\stx} }, I_{n+ 2}  {x_{n+ 2}^{\stx}}, \dii , I_N  x_N^{\stx}).
\end{aligned}
\end{equation*}

By the concavity of $\sw(\cdot)$, we can easily find that
$$
\phi_3 - \phi_1 \leq \phi_2 - \phi_0.
$$
This implies that 
$$
\phi_3 - \phi_1 + \phi_3-\phi_2 \leq \phi_3 - \phi_0,
$$
and therefore $\pa_{\langle{n}\rangle}^*  \geq \pan^*+ \pa_{n+1}^*$.
That is, APOs $n$ and $n\ad1$ can achieve a higher total payoff when they merge into a group and bargain with the MNO together. 
%This implies that both APOs $n$ and $n\ad1$ have the incentive to merge together.
\end{proof}

%%%%%%%%%%%%%%%%%%%%%%%%%%%%%%%%%%%%%%%%%%%%%%%%%%
%%%%%%%%%%%%%%%%%%%%%%%%%%%%%%%%%%%%%%%%%%%%%%%%%%

\subsection{Proof for Property \ref{prop:SPE-group} in Section \ref{sec:barg:group}}\label{app:proof-prop:SPE-group}

\begin{proof}
We first study the impact of the merge of APOs $n$ and $n \ad 1$ on the payoffs of prior APOs (i.e., those bargaining before APOs $n$ and $n+1$). Consider an arbitrary APO $i $ with $i < n$. By Lemma \ref{lemma:NBS-seq-n}, its payoff is 
\begin{equation*}
\begin{aligned}
&\textstyle \qquad \pa_i^* 
 = \frac{ \adw_{i} }{2} =   \sum_{I_{i+1} }\dii \sum_{I_{N} }	 \frac{{\dw}_{i}(I_{i+1};\dii ; I_N )}{ 2^{N\mi i}\cdot2},
\end{aligned}
\end{equation*}
where
${\dw}_{i}(I_{i+1};\dii  ;I_N ) =  \sw(\x_{i \mi 1}^{\stx}, \rmkk{x_{i}^{\stx}}, I_{i+ 1}  {x_{i+ 1}^{\stx}},\dii , I_N  x_N^{\stx})
- \sw(\x_{i \mi 1}^{\stx}, \rmkk{0}, I_{i+ 1}  {x_{i+ 1}^{\stx}},\dii , I_N  x_N^{\stx})$.

Now suppose APOs $n$ and $n\ad 1$ merge together. For notation convenience, 
we denote the merged group $\{n, n\ad 1\}$ by $\langle{n}\rangle$, and introduce a dummy APO $\langle{n\ad 1}\rangle$ to keep the indexes of APOs $n\ad 2, \dii, N$ consistently.
Then, the APO $i$'s payoff under the new group structure is 
\begin{equation*}
\begin{aligned}
&\textstyle \qquad \pa_{i, \textsc{new}}^* 
 = \frac{ \adw_{i} }{2} =   \sum_{I_{i+1} }\dii \sum_{I_{N} }	 \frac{{\dw}_{i}(I_{i+1};\dii ; I_N )}{ 2^{N\mi i}\cdot2},
\end{aligned}
\end{equation*}
where ${\dw}_{i}(I_{i+1};\dii  ;I_N ) $
\begin{equation*}
\begin{aligned}
= & \sw(\x_{i \mi 1}^{\stx}, \rmkk{x_{i}^{\stx}}, I_{i+ 1}  {x_{i+ 1}^{\stx}},\dii, I_{n} x_{\langle n \rangle}^{\stx} , I_{n+1} x_{\langle n+1 \rangle}^{\stx} , \dii, I_N  x_N^{\stx})
\\
 &- \sw(\x_{i \mi 1}^{\stx}, \rmkk{0}, I_{i+ 1}  {x_{i+ 1}^{\stx}},\dii, I_{n} x_{\langle n \rangle}^{\stx} , I_{n+1} x_{\langle n+1 \rangle}^{\stx} , \dii, I_N  x_N^{\stx})
\\
= & \sw(\x_{i \mi 1}^{\stx}, \rmkk{x_{i}^{\stx}}, I_{i+ 1}  {x_{i+ 1}^{\stx}},\dii, I_{n} x_{ n}^{\stx} , I_{n} x_{ n+1 }^{\stx} , \dii, I_N  x_N^{\stx})
\\
 &- \sw(\x_{i \mi 1}^{\stx}, \rmkk{0}, I_{i+ 1}  {x_{i+ 1}^{\stx}},\dii, I_{n} x_{ n }^{\stx} , I_{n} x_{ n+1 }^{\stx} , \dii, I_N  x_N^{\stx})
\end{aligned}
\end{equation*}
The last line is because $x_{\langle n \rangle}^{\stx} = \{x_{n}^{\stx}, x_{n\ad 1}^{\stx}\}$ and $x_{\langle n \ad 1 \rangle}^{\stx} = 0$. For convenience, we introduce the following notations:
\begin{equation*}
\begin{aligned}
\delta_0  &\textstyle  =  \sw(\x_{i \mi 1}^{\stx}, x_{i}^{\stx}, I_{i+ 1}  {x_{i+ 1}^{\stx}},\dii, 0 , 0 , \dii, I_N  x_N^{\stx}) \\
& \textstyle  \quad - \sw(\x_{i \mi 1}^{\stx}, \rmkk{0}, I_{i+ 1}  {x_{i+ 1}^{\stx}},\dii, 0 , 0 , \dii, I_N  x_N^{\stx}),
\\
\delta_1  &\textstyle  =  \sw(\x_{i \mi 1}^{\stx}, x_{i}^{\stx}, I_{i+ 1}  {x_{i+ 1}^{\stx}},\dii, 0 , x_{ n+1 }^{\stx}, \dii, I_N  x_N^{\stx}) \\
& \textstyle  \quad - \sw(\x_{i \mi 1}^{\stx}, \rmkk{0}, I_{i+ 1}  {x_{i+ 1}^{\stx}},\dii, 0 , x_{ n+1 }^{\stx}, \dii, I_N  x_N^{\stx}),
\\
\delta_2  &\textstyle  =  \sw(\x_{i \mi 1}^{\stx}, x_{i}^{\stx}, I_{i+ 1}  {x_{i+ 1}^{\stx}},\dii , x_{ n }^{\stx}, 0, \dii, I_N  x_N^{\stx}) \\
& \textstyle  \quad - \sw(\x_{i \mi 1}^{\stx}, \rmkk{0}, I_{i+ 1}  {x_{i+ 1}^{\stx}},\dii , x_{ n }^{\stx},0, \dii, I_N  x_N^{\stx}),
\\
\delta_3  &\textstyle  =  \sw(\x_{i \mi 1}^{\stx}, x_{i}^{\stx}, I_{i+ 1}  {x_{i+ 1}^{\stx}},\dii, x_{ n}^{\stx}, x_{ n+1 }^{\stx}, \dii, I_N  x_N^{\stx}) \\
& \textstyle  \quad - \sw(\x_{i \mi 1}^{\stx}, \rmkk{0}, I_{i+ 1}  {x_{i+ 1}^{\stx}},\dii, x_{ n}^{\stx}, x_{ n+1 }^{\stx}, \dii, I_N  x_N^{\stx}).
\end{aligned}
\end{equation*}
Then, we can write the APO $i$'s payoff as follows:
\begin{equation*}
\begin{aligned}
\pa_i^*  & \textstyle =  \sum_{I_{i+1} }\dii\sum_{I_{n\mi 1} }\sum_{I_{n+ 2} } \dii \sum_{I_{N} }
\frac{ \delta_0 + \delta_1 + \delta_2 + \delta_3}{ 2^{N\mi i}\cdot2}
\\
\pa_{i, \textsc{new}}^*  & \textstyle =  \sum_{I_{i+1} }\dii\sum_{I_{n\mi 1} }\sum_{I_{n+ 2} } \dii \sum_{I_{N} }
\frac{ \delta_0 \cdot 2 + \delta_3\cdot 2}{ 2^{N\mi i}\cdot2}
\end{aligned}
\end{equation*}
By the concavity of $\sw(\cdot)$, we further have: $\delta_3 - \delta_2 \leq \delta_1 - \delta_0$. Therefore, we have:  $\pa_i^* \leq \pa_{i, \textsc{new}}^* $, that is, APO $i$ can achieve a higher payoff when APOs $n$ and $n\ad 1$ merge together. 

By similar analysis, we can show that 
the merge of APOs $n$ and $n\ad 1$ and has no impact on the payoff os  posterior APOs, i.e., those bargaining after APOs $n$ and $n+1$.
\end{proof}

%%%%%%%%%%%%%%%%%%%%%%%%%%%%%%%%%%%%%%%%%%%%%%%%%%%%
%%%%%%%%%%%%%%%%%%%%%%%%%%%%%%%%%%%%%%%%%%%%%%%%%%%%
%
%\subsection{Proof for Theorem \ref{theorem:NBS-group} in Section \ref{sec:barg:group}}\label{app:proof-NBS-group}
%
%\begin{proof}
%By iteratively using the result in Lemma \ref{lemma:GBS-seq}, we can easily show that all APOs  will eventually {merge into a single group}, that is, they have the incentive to merge into one group.
%
%When all APOs forming a single group, the problem degenerates to an one-to-one bargaining, and the group bargaining solution is directly given by Lemma \ref{theorem:NBS-single}
%\end{proof}
%
%
%%%%%%%%%%%%%%%%%%%%%%%%%%%%%%%%%%%%%%%%%%%%%%%%%%%%
%%%%%%%%%%%%%%%%%%%%%%%%%%%%%%%%%%%%%%%%%%%%%%%%%%%%
%

%%%%%%%%%%%%%%%%%%%%%%%%%%%%%%%%%%%%%%%%%%%%%%%%%%%
%%%%%%%%%%%%%%%%%%%%%%%%%%%%%%%%%%%%%%%%%%%%%%%%%%%

\subsection{Proof for Property \ref{prop:GBS-con} in Section \ref{sec:barg:group}}\label{app:proof-prop:GBS-con}

\begin{proof}
With a similar proof for Property \ref{prop:GBS-seq} (Appendix \ref{app:proof-prop:GBS-seq}), we can prove this property directly. 
\end{proof}

\subsection{Proof for Property \ref{prop:NOE-group} in Section \ref{sec:barg:group}}\label{app:proof-prop:NOE-group}

\begin{proof}
With a similar proof for Property \ref{prop:SPE-group} (Appendix \ref{app:proof-prop:SPE-group}), we can prove this property directly. 
\end{proof}

\newpage

\begin{center}
\Large\textbf{(IV) An Alternative Modeling Approach}
\end{center}

\vspace{-5mm}

%!TEX root = DataOffload_main_journal.tex
%SourceDoc DataOffload_main_journal.tex

%\section{Non-cooperative Game Solution}

\subsection{Non-cooperative Game Formulation and Analysis}\label{app:game}

In this section, we formulate the data offload problem as a \emph{Stackelberg game} based on the non-cooperative game theory, where the MNO acts as the game leader specifying the payments to APOs first, and then every APO acts as a game follower determining the traffic volume it is willing to deliver for the MNO.

We consider a simple \emph{linear payment}. That is, the payment $z_n$ to an APO $n$ is simply defined as a linear function of $x_n$ (i.e., the traffic offload volume to APO $n$), and denoted by
\begin{equation}
z_n(x_n) \eq p_n \cdot x_n,
\end{equation}
where $p_n$ is the unit payment to APO $n$ for one unit of traffic offload volume.
Notice that the transmission efficiency between an APO and its covered MUs is normalized to $\theta = 1$. Thus, $x_n$ also denotes the amount of APO $n$'s spectrum resource dedicate to data offload (i.e., to deliver the traffic offloaded from the MNO). In this sense,
$p_n$ can also be viewed as the unit price of APO $n$'s spectrum resource.

With the linear payment, the game process is as follows. In the first stage, the MNO (leader) specifies a price profile $\p \eq (p_1, \mbox{...}, p_N)$, each intending for one APO.
In the second stage, every APO $n$ responses with $x_n$,
the amount
of its resource for data offloading, based on the price $p_n$ and the stochastic distribution of its own traffic demand.
An Nash equilibrium (NE) is defined as such a strategy profile $\{p_n^*, x_n^*\}_{\forall n\in \N}$ such that none of th player can improve its payoff by unilitery deviating.
We solve the NE of this game by backward induction.

\subsubsection{The APO's Decision -- $x_n^*$}

First, we study the APO's optimal decision in the second stage, given the price $p_n$ specified by the MNO in the first stage. Formally, the decision problem for APO $n$ is
\begin{equation}\label{eq:APO-opt}
\begin{aligned}
\max_{x_n}& \ \Ua_n (x_n\dt p_n x_n)\\
 \mbox{s.t. }&  x_n \in [0, B_n],
\end{aligned}
\end{equation}
where $\Ua_n(\cdot,\cdot)$ is APO $n$'s payoff defined in (\ref{eq:AP-payoff}). 

The first- and second-order derivatives of $\Ua_n (x_n\dt p_n x_n)$ to $x_n$ are, respectively,
\begin{equation*}
\left\{
\begin{aligned}
\textstyle
\frac{\partial	\Ua_n }{\partial x_n} &\textstyle = -(w_n-c_n)\cdot\big[1- F_n(B_n-x_n) \big] + (p_n - c_n), \\
%&= (w_n-c_n)\cdot F_n(B_n-x_n)+ (w_n - p_n), \\
\textstyle\frac{\partial^2 \Ua_n }{\partial x_n^2} &\textstyle = -(w_n-c_n)\cdot f_n(B_n-x_n).
\end{aligned}
\right.
\end{equation*}
It is easy to see that $\frac{\partial^2 \Ua_n }{\partial x_n^2} < 0$ (since $w_n> c_n$ and $f_n(.)>0$). Thus, the  problem (\ref{eq:APO-opt}) is a convex optimization, and the optimal solution can be solved using KKT analysis.

Next we present the optimal solution $x_n^*$ analytically using the FOC analysis. The key idea is that if the FOC condition is achievable, then the optimal $x_n^*$ is given by the FOC condition in (\ref{eq:APO-foc}). Otherwise, the optimal $x_n^*$ is the lower-bound or upper-bound of $x_n$ depending on the sign of the first-order derivative.
\begin{equation}\label{eq:APO-foc}
\begin{aligned}\textstyle
\mbox{FOC:}\quad \frac{\partial	\Ua_n }{\partial x_n} = 0.
\end{aligned}
\end{equation}
For convenience, we further introduce the concept of \emph{critical price} of APO $n$, denoted by
\begin{equation}\label{eq:APO-critical}
\wc_n \eq c_n + (w_n-c_n)\cdot\big[1- F_n(B_n) \big] .
\end{equation}
It is important to note that the FOC in (\ref{eq:APO-foc}) is only achievable when $  p_n \in [\wc_n,   w_n]$.
% where $\wc_n$ is the \emph{critical price} of  APO $n$ and given by $$\wc_n \eq c_n + (w_n-c_n)\cdot\big[1- F_n(B_n) \big] .$$
%since $0\leq F_n(B_n-x_n)\leq G(B_n)$.
%, which implies that the solution given by the FOC (\ref{eq:APO-foc}) is only available when $c_n \leq p_n \leq w_n$.
When $p_n < \wc_n$ (or  $p_n > w_n$), however, the first-order derivative  $\frac{\partial	\Ua_n }{\partial x_n}$ is \emph{always} smaller (or larger) than 0 and never equals to 0, and thus the optimal solution is the lower-bound (or upper-bound) of the feasible range of $x_n$, i.e., $x_n^* = 0$ (or $x_n^* = B_n$). Formally,

\begin{lemma}[APO's Optimal Decision]\label{thrm:APO-opt}
Given the price $p_n$, every APO $n$'s optimal decision is
\begin{equation*}\label{eq:APO-solu}
x_n^*(p_n) = \left\{
\begin{aligned}
0~~~~~~~~~~~~~~~~~~~& \mbox{~~~if } p_n < \wc_n  \\
\textstyle B_n - F_n^{(\mi 1)}\left( \frac{w_n - p_n}{w_n - c_n} \right) & \mbox{~~~if } p_n \in[\wc_n, w_n]\\
B_n~~~~~~~~~~~~~~~~~ & \mbox{~~~if } p_n > w_n
\end{aligned}
\right.
\end{equation*}
where $F_n^{(\mi 1)}(\cdot)$ is the inverse function of $F_n(\cdot)$.
%, $[x]^+ = x$ if $x\geq 0$ and $[x]^+ = 0$ if $x < 0$.
\end{lemma}

The first and third cases can be referred to the previous discussion, and the second case  is derived from the FOC (\ref{eq:APO-foc}) directly.
%Due to space limit, we do not present the detailed proof.
When the price  $p_n$ falls in $ [\wc_n, w_n]$, we further have
%the following first- and second-order derivatives of $x_n^*$ to $p_n$,
\begin{equation*}
\left\{
\begin{aligned}
\textstyle
\frac{\partial x_n^*}{\partial p_n} &\textstyle = \frac{1}{f_n\big(F_n^{(\mi 1)}\big(\frac{w_n-p_n}{w_n-c_n}\big) \big)} \cdot \frac{1}{w_n-c_n} , \\
\textstyle\frac{\partial^2 x_n^*}{\partial p_n^2} &\textstyle = \frac{f_n'\big(F_n^{(\mi 1)}\big(\frac{w_n-p_n}{w_n-c_n}\big) \big)}{\big[f_n\big(F_n^{(\mi 1)}\big(\frac{w_n-p_n}{w_n-c_n}\big) \big)\big]^3} \cdot \frac{1}{(w_n-c_n)^2}.
\end{aligned}
\right.
\end{equation*}
The above formula follows because $f^{(\mi 1)\prime}(\cdot) = \frac{1}{f'(f^{(\mi 1)}(\cdot))}$. Based on above, we have the following properties for $x_n^*$.
\begin{property}\label{lemma:xn-property}
For any $p_n \in [\wc_n, w_n]$, the optimal $x_n^*$ satisfies:
\begin{enumerate}
\item[(a)]
$\frac{\partial x_n^*}{\partial p_n} > 0 $, that is, $x_n^*$ increases with $p_n$;
\item[(b)]
$\frac{\partial^2 x_n^*}{\partial p_n^2} \leq  0$, if $f_n'(\cdot) \leq 0$, and vice versa.
\end{enumerate}
\end{property}
The first condition implies that the higher price the MNO offers, the more resource the APO dedicates to data offload.
The second condition implies that $x_n^*$ is an increasing concave (or convex) function of $p_n$, if $\t_n$ has a decreasing (or increasing) PDF $f_n(\cdot)$.
For later derivational convenience, we will assume that $f_n'(\cdot) \leq 0$, and therefore $\frac{\partial^2 x_n^*}{\partial p_n^2} \leq 0$.\footnote{Note that many common distributions satisfy the condition of decreasing (non-increasing) PDF. Typical examples include uniform distributions, exponential distributions, and power distributions with negative factors, etc.}

\subsubsection{The MNO's Decision -- $\p^*$}

Now we study the MNO's best decision in the first stage, based on its prediction of every APO $n$'s optimal response $x_n^*$ in the second stage (given in Lemma \ref{thrm:APO-opt}).
%Note that $x_n^*$ is a function of $p_n$, and thus we can write it as $x_n^*(p_n)$.

Denote $\p \eq (p_1, ..., p_N)$ as the price profile for all APOs, and $\x^* \eq (x_1^*(p_1), ..., x_N^*(p_N))$ as the APOs' optimal responses.
The decision problem for the MNO is
\begin{equation}\label{eq:BS-opt}
\begin{aligned}
\max_{\p}& \ \Um (\x^*\dt \p\times\x^* )\\
 \mbox{s.t. }&  p_n \geq 0,  \quad \forall n=1,...,N, \\
& x_n^*(p_n) \leq S_n, \quad \forall n=1,...,N,
\end{aligned}
\end{equation}
where $\Um(\cdot,\cdot)$ is the MNO's payoff defined in (\ref{eq:BS-payoff}), and $\p\times\x^*$ is the pointwise product of vectors $\p$ and $\x^*$. The element in $\p\times\x^*$ denotes the payment to every APO.

We first capture some useful information from the first-order partial derivative.
Notice that $x_n^*$ is a function of $p_n$. The first-order partial derivative of $\Um (\x^*\dt \p\times\x^* )$ to $p_n$ is
\begin{equation}\label{eq:BS-first-order}
\begin{aligned}
\textstyle\frac{\partial	\Um }{\partial p_n} &\textstyle = \C'(\B(\x^*))\cdot \frac{\partial x_n^*}{\partial p_n} \cdot \frac{1}{\theta_n} - p_n\cdot \frac{\partial x_n^*}{\partial p_n}- x_n^* .
\end{aligned}
\end{equation}

By Lemma \ref{thrm:APO-opt}, we have: $\frac{\partial x_n^*}{\partial p_n} \equiv 0$, if $p_n < \wc_n$ or $p_n > w_n$.
It directly follows that: (i) $\frac{\partial	\Um }{\partial p_n} =0$ if $p_n < \wc_n$, and (ii) $\frac{\partial	\Um }{\partial p_n} =-B_n <0$ if $p_n> w_n$.
The first observation implies that any price $p_n$ lower than $\wc_n$ is indifferent to the MNO, and the second observation implies that any price $p_n$ higher than $w_n$ is dominated by $w_n$.\footnote{Intuitively, if $p_n < \wc_n$, APO $n$ always returns a zero amount of its resource for data offloading, and thus any price $p_n < \wc_n$ is indifferent to the MNO. If $p_n > w_n$, APO $n$ always returns all of its resource for data offloading, and thus a higher price (above $w_n$) cannot bring more resource (from APO $n$) for the MNO, but will definitely lead to a higher payment.}
Therefore, we can focus on the price below $w_n$.
Formally, %we have the following necessary condition for the optimal price profile $\p^* = [p_1^*,...,p_N^*]$.
\begin{lemma}\label{lemma:pn-cond-1}
For any optimal price profile $\p^*$, the following necessary condition holds:
$$
 p_n^* \leq w_n,\quad \forall n=1,...,N,
$$
and in addition, any price  $p_n^*$ below $\wc_n$ is indifferent.
\end{lemma}

From (\ref{eq:BS-first-order}), we can further find that the optimal price $p_n^*$ cannot be larger than $ \C'(\B(\x^*)) \cdot \frac{1}{\theta_n}$ (since $\frac{\partial	\Um }{\partial p_n} \geq 0$ by Lemma \ref{lemma:pn-cond-1}); otherwise, we will have $\frac{\partial	\Um }{\partial p_n}   < 0$, which implies that there must exist a price $p_n = p_n^* - \delta$ with which the MNO can achieve a higher payoff.\footnote{Here $\delta$ is an arbitrarily small positive number.}
Therefore, we can focus on the price below $\C'(\B(\x^*)) \cdot \frac{1}{\theta_n}$.
Formally,
\begin{lemma}\label{lemma:pn-cond-2}
For any optimal price profile $\p^*$, the following necessary condition holds:
$$\textstyle p_n^*  \leq  \C'(\B(\x^*)) \cdot \frac{1}{\theta_n} ,\quad \forall n=1,...,N.$$
\end{lemma}

%\begin{remark}
Note that if $\C'(\B(\x^*)) \cdot \frac{1}{\theta_n} \leq \wc_n$, then we can directly set $p_n^* $ as any price lower than $\C'(\B(\x^*)) \cdot \frac{1}{\theta_n}$, since any price below $\wc_n$ is indifferent to the MNO.
%\end{remark}

Then, we study the convexity of the optimization problem (\ref{eq:BS-opt}) by the second-partial partial derivative.
Notice that $x_n^*$ is related to $p_n$ only, while independent of $p_m$, $\forall m\neq n$. The second-order partial derivatives of $\Um (\x^*, \p\times\x^* )$ are
\begin{equation*}
\left\{
\begin{aligned}
\textstyle
\frac{\partial	^2 \Um }{\partial p_n^2} & \textstyle =  -\C''(\B(\x^*))\cdot \left(\frac{\partial x_n^*}{\partial p_n} \cdot \frac{1}{\theta_n}\right)^2
- 2\cdot \frac{\partial x_n^*}{\partial p_n} \\
&\textstyle \qquad +
\left(\C'(\B(\x^*))\cdot \frac{1}{\theta_n} - p_n\right)\cdot \frac{\partial^2 x_n^*}{\partial p_n^2},\ \forall n,
\\
\textstyle \frac{\partial	^2 \Um }{\partial p_m \partial p_n}  & \textstyle = -\C''(\B(\x^*))\cdot \frac{\partial x_n^*}{\partial p_n} \cdot  \frac{\partial x_m^*}{\partial p_m} \cdot \frac{1}{\theta_n  \theta_m},\ \forall n\neq m.
\end{aligned}
\right.
\end{equation*}

It is easy to see that (i) $\frac{\partial	^2 \Um }{\partial p_m \partial p_n} \leq 0$, since $\C''(b) \geq 0$ and $\frac{\partial x_n^*}{\partial p_n} \geq 0$;
and (ii)
$\frac{\partial	^2 \Um }{\partial p_n^2} \leq 0$, since $\C'(\B(\x^*))\cdot \frac{1}{\theta_n} \geq p_n$ (by Lemma \ref{lemma:pn-cond-2}) and $\frac{\partial^2 x_n^*}{\partial p_n^2} \geq 0$ (by the assumption of $f_n'(\cdot) \leq 0$).
Thus, $\Um $ is concave in $\p$.
Furthermore, the constraint set of problem (\ref{eq:BS-opt}) is obviously a convex set.
Therefore,
\begin{lemma}\label{lemma:BS-convexity-x}
The problem (\ref{eq:BS-opt}) is a convex optimization.
\end{lemma}

By the convexity of problem (\ref{eq:BS-opt}), we can solve the problem using classic KKT analysis.
Similar to Section \ref{sec:barg-xo}, we capture some useful properties of the optimal price profile $\p^*$ using the FOC analysis.
Suppose all constraints of (\ref{eq:BS-opt}) are strictly satisfied under the optimal solution. Then, the optimal $\p^*$ must satisfy the FOC condition:
\begin{equation}\label{eq:BS-FOC}
\begin{aligned}\textstyle
\mbox{FOC:}\quad \frac{\partial \Um }{\partial p_n} = 0, \quad \forall n = 1,...,N,
\end{aligned}
\end{equation}
which leads to the following optimality condition immediately,
\begin{theorem}[Optimality]\label{thrm:BS-opt}
Suppose all constraints of (\ref{eq:BS-opt}) are  not binding.
The  optimal price profile $\p^*$ for the MNO satisfies the following conditions: $\forall n\in \N$,
\begin{equation}\label{eq:BS-FOC-eq}
\textstyle
 \C'(\B(\x^*)) =  x_n^* \cdot \theta_n \cdot\frac{\partial p_n}{\partial x_n^*} + p_n\cdot \theta_n.
\end{equation}
\end{theorem}

Now we capture some insight behind the above optimal $\p^*$ given in Theorem \ref{thrm:BS-opt}.
One one hand, the left hand side of (\ref{eq:BS-FOC-eq}) is the \emph{marginal cost} ($\MC$) of the MNO.
On the other hand, the right hand side of (\ref{eq:BS-FOC-eq}) is the \emph{marginal payment} ($\MP_n$) to APO $n$, i.e., the increase of the MNO's payment induced by offloading $\theta_n$ additional units of traffic to APO $n$.
Specifically, to increase the traffic offload volume by $\theta_n$, the MNO has to increase the price $p_n$ by $\Delta p_n \eq  \theta_n \cdot  \frac{\partial p_n}{\partial x_n^*}$, which will introduce an additional payment $x_n^*  \cdot \Delta p_n $ for the existing $x_n^*$ units of offloaded traffic volume, and a new payment
$p_n \cdot \theta_n $ for the coming $\theta_n$ units of offloaded traffic volume.
The equation (\ref{eq:BS-FOC-eq}) suggests that in an optimal solution $\p^*$, \textbf{the $\MC$ equals to the $\MP_n$ to every APO $n$}.
Intuitively, if the $\MC$ is larger (or smaller) than the $\MP_n$ to APO $n$, then the MNO can immediately improve its payoff by offloading more (or less) traffic to APO $n$ through increasing (or decreasing) $p_n$.

By (\ref{eq:BS-FOC-eq}), we further have the following property.
\begin{property}\label{thrm:BS-opt-2}
Suppose all constraints of (\ref{eq:NBS-multi-swm}) are not binding.
The optimal price profile $\p^*$ satisfies:
\begin{equation}\label{eq:BS-FOC-xxx}
\MP_m = \MP_n, \quad \forall m,n\in \N,
\end{equation}
where $\MP_n =  x_n^* \cdot \theta_n \cdot\frac{\partial p_n}{\partial x_n^*} + p_n\cdot \theta_n$, $\forall n\in \N$.
\end{property}

Property \ref{thrm:BS-opt-2} states that under the optimal $\p^*$, the $\MP$s to all APOs would be the same. Intuitively, if the $\MP_n<\MP_m$, then the MNO can immediately increase its payoff by increasing the traffic volume offloaded to APO $n$ and decreasing the traffic volume offloaded  to APO $m$.

\end{document}